\documentclass[11pt,onecolumn]{article}

\usepackage[margin=2.5cm]{geometry}

\usepackage[T1]{fontenc}
\usepackage{graphicx}
\usepackage[utf8]{inputenc}
\usepackage{hyperref}
\usepackage{graphicx}

\usepackage{mathtools}
\usepackage{amsfonts}

\usepackage{url}

\usepackage{float}

\usepackage[vlined]{algorithm2e}

\usepackage{authblk}

\newcommand{\normal}{\mathcal{N}}
\newcommand{\obs}{\text{obs}}

\newcommand{\ABC}{\mathrm{ABC}}

\newcommand{\dy}{\mathop{dy}}

\newcommand{\R}{\mathbb{R}}

\newcommand{\E}{\mathbb{E}}

\title{A Wall-time Minimizing Parallelization Strategy for Approximate Bayesian Computation}

\date{}

\author[1,$*$]{Emad Alamoudi}
\author[1,$*$]{Felipe Reck}
\author[2]{Nils Bundgaard}
\author[2,3,4]{Frederik Graw}
\author[5]{Lutz Brusch}
\author[1,6,7,$\dagger$]{Jan Hasenauer}
\author[1,6,7]{Yannik Schälte}

\affil[1]{
	University of Bonn, Life and Medical Sciences Institute, 53113 Bonn, Germany}
\affil[2]{
	Heidelberg University, BioQuant - Center for Quantitative Biology, 69120 Heidelberg, Germany}
\affil[3]{
	Heidelberg University, Interdisciplinary Center for Scientific Computing, 69120 Heidelberg, Germany}
\affil[4]{
	Friedrich-Alexander-University Erlangen-N\"{u}rnberg, Department of Medicine 5, 91054 Erlangen, Germany}
\affil[5]{
	Technische Universität Dresden, Center of Information Services and High Performance Computing (ZIH), 01062 Dresden, Germany}
\affil[6]{
	Helmholtz Zentrum München, Institute of Computational Biology, 85764 Neuherberg, Germany}
\affil[7]{
	Technische Universität München, Center for Mathematics, 85748 Garching, Germany
}
\affil[$*$]{
	These authors contributed equally to this work
}
\affil[$\dagger$]{
	To whom correspondence should be addressed (jan.hasenauer@uni-bonn.de)
}

\begin{document}
	
	\maketitle
	
	\begin{abstract}
		
		\noindent Approximate Bayesian Computation (ABC) is a widely applicable and popular approach to estimating unknown parameters of mechanistic models.
		As ABC analyses are computationally expensive, parallelization on high-performance infrastructure is often necessary.
		However, the existing parallelization strategies leave resources unused at times and thus do not optimally leverage them yet.
		
		\noindent We present look-ahead scheduling, a wall-time minimizing parallelization strategy for ABC Sequential Monte Carlo algorithms, which utilizes all available resources at practically all times by proactive sampling for prospective tasks.
		Our strategy can be integrated in e.g.\ adaptive distance function and summary statistic selection schemes, which is essential in practice.
		Evaluation of the strategy on different problems and numbers of parallel cores reveals speed-ups of typically 10-20\% and up to 50\% compared to the best established approach.
		Thus, the proposed strategy allows to substantially improve the cost and run-time efficiency of ABC methods on high-performance infrastructure.
		
	\end{abstract}
	
	
	\section{Introduction}
	
	The ultimate goal of systems biology is a holistic understanding of biological systems \cite{Kitano2002,KarrSan2012}.
	Mechanistic models are important tools towards this aim, allowing to describe and understand underlying mechanisms \cite{Gershenfeld1999}.
	Usually, such models have unknown parameters that need to be estimated by comparing model outputs to observed data \cite{Tarantola2005}.
	For complex stochastic models, in particular multi-scale models used to describe the complex dynamics of multi-cellular systems, evaluating the likelihood of data given parameters however becomes quickly computationally infeasible \cite{TavareBal1997,HasenauerJag2015}.
	For this reason, simulation-based methods that circumvent likelihood evaluation have been developed, such as approximate Bayesian computation (ABC), popular for its simplicity and wide applicability \cite{PritchardSei1999,BeaumontZha2002}.
	
	ABC generates samples from an approximation to the true Bayesian posterior distribution.
	While asymptotically exact, a known disadvantage of ABC is its reliance on repeated simulation, often hundred thousands to millions of times. Therefore, methods to efficiently explore the search space have been developed \cite{SissonFan2018Handbook}.
	In particular, ABC is frequently combined with a Sequential Monte Carlo scheme (ABC-SMC), which over several generations successively refines the posterior approximation via importance sampling while maintaining high acceptance rates \cite{DelMoralDou2006,SissonFan2007}.
	Furthermore, in ABC-SMC the sampling for each generation can be parallelized, enabling the use of high-performance computing (HPC) infrastructure. This has in recent years enabled tackling increasingly complex problems via ABC \cite{JagiellaRic2017,ImleKum2019,DursoKum2021,AlamoudiSch2023fitmulticell}.
	
	It would be desirable if available computational resources were perfectly exploited at all times, to minimize both the wall-time until results become available to the researcher, and the cost associated with allocated resources.
	However, the problem is that established parallelization strategies to distribute ABC-SMC work over a set of workers leave resources idle at times and thus fall short of this aim.
	The parallelization strategy used in most established HPC-ready ABC implementations is \textit{static scheduling (STAT)}, which defines exactly as many parallel tasks as accepted particles are required \cite{DuttaSch2017,KangasraasioLin2016}. While it minimizes the active compute time and consumed energy, typically a substantial amount of workers become idle towards the end of each generation.
	\textit{Dynamic scheduling (DYN)} mitigates this problem and reduces the overall wall-time by continuing sampling on all workers until sufficiently many particles have been accepted \cite{KlingerRic2018}. It was shown to reduce the wall-time substantially.
	However, also in this strategy at the end of each generation workers become idle, waiting until all simulations have finished.
	
	In this manuscript, we present a novel ABC-SMC parallelization strategy for multi-core and distributed systems, called \textit{look-ahead scheduling (LA)}. The strategy builds upon dynamic scheduling, but in addition, as workers get idle at the end of each generation, preemptively formulates tentative sampling tasks for the next generation. By this, it makes use of all available workers at almost all times.
	We show that by appropriate sample reweighting we obtain an unbiased Monte Carlo sample.
	We provide an HPC-ready implementation and test the method on various problems, demonstrating both efficiency and accuracy.
	Moreover, we show that the strategy can be integrated with adaptive algorithms for e.g.\ summary statistics, distance functions, or acceptance thresholds.
	
	
	\section{Methods}
	
	
	\subsection{ABC}
	
	We consider a mechanistic model described via a generative process of simulating data $y\sim\pi(y|\theta)\in\R^{n_y}$ for parameters $\theta\in\R^{n_\theta}$.
	Given observed data $y_\obs$, in Bayesian inference the likelihood $\pi(y_\obs|\theta)$ is combined with prior information $\pi(\theta)$ to the posterior distribution $\pi(\theta|y_\obs) \propto \pi(y_\obs|\theta)\cdot\pi(\theta).$
	We assume that evaluating the likelihood is computationally infeasible, but that it is possible to simulate data $y\sim\pi(y|\theta)$ from the model.
	Then, classical ABC consists in the 3 steps of first sampling parameters $\theta\sim\pi(\theta)$, second simulating data $y\sim \pi(y|\theta)$, and third accepting $\theta$ if $d(y,y_\obs) \leq \varepsilon$, for a distance metric $d:\R^{n_y}\times\R^{n_y}\rightarrow\R_{\geq 0}$ and acceptance threshold $\varepsilon>0$.
	This is repeated until sufficiently many particles, say $N$, are accepted. The population of accepted particles constitutes a sample from an approximation of the posterior distribution,
	\begin{equation}\label{eq:pyabc}
		\pi_{\ABC,\varepsilon}(\theta|y_\obs) \propto \int I[d(y,y_\obs)\leq\varepsilon]\pi(y|\theta)\dy\cdot\pi(\theta).
	\end{equation}
	Under mild assumptions, $\pi_{\ABC,\varepsilon}(\theta|y_\obs)$ converges to the actual posterior $\pi(\theta|y_\obs)$ as  $\varepsilon\rightarrow 0$ \cite{Wilkinson2013,SchaelteAla2021}. Commonly, ABC operates not directly on the measured data, but summary statistics thereof, capturing relevant information in a low-dimensional representation \cite{FearnheadPra2012}. Here, for notational simplicity we assume that $y$ already incorporates summary statistics, if applicable.
	
	\subsection{ABC-SMC}
	
	The vanilla ABC formulation exhibits a trade-off between reducing the approximation error induced by $\varepsilon$, and high acceptance rates.
	Thus, ABC is frequently combined with a Sequential Monte Carlo scheme (ABC-SMC) \cite{ToniWel2009,Beaumont2010}.
	In ABC-SMC, a series of particle populations $P_{t}=\{(\theta_{t}^i,w_{t}^i)\}_{i\leq N}$ is generated, constituting samples of successively better approximations $\pi_{\ABC,\varepsilon_{t}}(\theta|y_\obs)$ of the posterior, for generations $t=1,\ldots,n_t$, with acceptance thresholds $\varepsilon_t>\varepsilon_{t+1}$.
	In the first generation ($t=1$), particles are sampled directly from the prior, $g_1(\theta) = \pi(\theta)$.
	In later generations ($t>1$), particles are sampled from proposal distributions $g_t(\theta)\gg\pi(\theta)$ based on the last generation's accepted weighted population $P_{t-1}$, e.g.\ via a kernel density estimate.
	The importance weights $w_t^{i}$ are the Radon-Nikodym derivatives $w_t(\theta) = \pi(\theta)/g_t(\theta)$.
	This is precisely such that the weighted parameters are samples from the distribution
	\begin{equation}\label{eq:pyabc_weighted}
		\int w_t(\theta)I[d(y,y_\obs)\leq\varepsilon_t]\pi(y|\theta)\dy\cdot g_t(\theta) = \int I[d(y,y_\obs)\leq\varepsilon_t]\pi(y|\theta)\dy\cdot \pi(\theta),
	\end{equation}
	i.e.\ the target distribution \eqref{eq:pyabc} for $\varepsilon = \varepsilon_t$.
	
	Common proposal distributions first select an accepted parameter from the last generation and then perturb it, in which case $g_t$ takes the form
	$g_t(\theta) = \sum_{i=1}^Nw_{t-1}^iK(\theta|\theta_{t-1}^i) / \sum_{i=1}^Nw_{t-1}^i$,
	with e.g.\ $K(\theta|\theta_{t-1}^i) = \normal(\theta|\theta_{t-1}^i,\Sigma_{t-1})$ a normal distribution with mean $\theta^i_{t-1}$ and covariance matrix $\Sigma_{t-1}$.
	The performance of ABC-SMC algorithms relies heavily on the quality of the proposal distribution, on its ability to efficiently explore the parameter space.
	Methods that adapt to the problem structure, e.g.\ basing $\Sigma_{t-1}$ on the previous generation's weighted sample covariance matrix and potentially localizing around $\theta_i$, have shown superior \cite{BeaumontCor2009,FilippiBar2013,KlingerHas2017}.
	The steps of ABC-SMC are summarized in Algorithm~\ref{alg:abcsmc}.
	
	\begin{algorithm}[t]
		\caption{ABC-SMC algorithm.}
		\For{$t=1,\ldots,n_t$}{
			set $g_t$ and $\varepsilon_t$\\
			\While{less than $N$ acceptances}{
				sample parameter $\theta \sim g_t(\theta)$\\
				simulate data $y \sim p(y|\theta)$\\
				accept $\theta$ if $d(y, y_\obs) \leq \varepsilon_t$\\
			}
			compute weights $w_t^{i} = \frac{\pi(\theta_t^i)}{g_t(\theta_t^i)}$, for accepted parameters $\{\theta_t^{i}\}_{i\leq N}$
		}
		\label{alg:abcsmc}
	\end{algorithm}
	
	The output of ABC-SMC is a population of weighted parameters 
	\[P_{n_t}=\{(\theta_{n_t}^{i},w_{n_t}^{i})\}_{i\leq N} \sim \pi_{\ABC,\varepsilon_{n_t}}(\theta|y_\obs).\]
	For a statistic $f:\R^{n_\theta}\rightarrow\R$, the expected value under the posterior is then approximated via the self-normalized importance estimator
	\[\E_{\pi_{\ABC,\varepsilon_{n_t}}(\theta|y_\obs)}[f] \approx \hat f = \sum_{i=1}^{N}W_{n_t}^{i}f(\theta_{n_t}^{i}),\]
	which is asymptotically unbiased.
	Here, $W^i_t := w^i_t / \sum_{j=1}^N w^j_t$ are self-normalized weights.
	This is necessary because the weights $w_t(\theta) = \pi(\theta) / g_t(\theta)$ are not normalized in the joint sample space $(\theta,y)$, therefore effectively another Monte Carlo estimator is employed for the normalization constant (for details see the Supplementary Information, Section 1.1).
	
	In importance sampling, samples are assigned different weights, such that some impact estimates more than others. This can be quantified e.g.\ via the \textit{effective sample size (ESS)} \cite{Liu1998,SissonFan2018Handbook}:
	\begin{equation}\label{eq:ess}
		\operatorname{ESS}(\{w_i\}_{i\leq N}) = \frac{(\sum_{i\leq N}w_i)^2}{\sum_{i\leq N}w_i^2}
	\end{equation}
	
	
	\subsection{Established parallelization strategies}
	
	In ABC, often hundred thousands to millions of model simulations need to be performed, which is typically the computationally critical part. To speed up inference, parallelization strategies have been developed that exploit the independence of the $N$ particles constituting the $t$-th population. Suppose we have $W$ parallel workers, each worker being a computational processor unit e.g.\ in an HPC environment. There are two established techniques to parallelize execution over the workers:
	
	\begin{figure}[t]
		\centering
		\includegraphics[width=\textwidth]{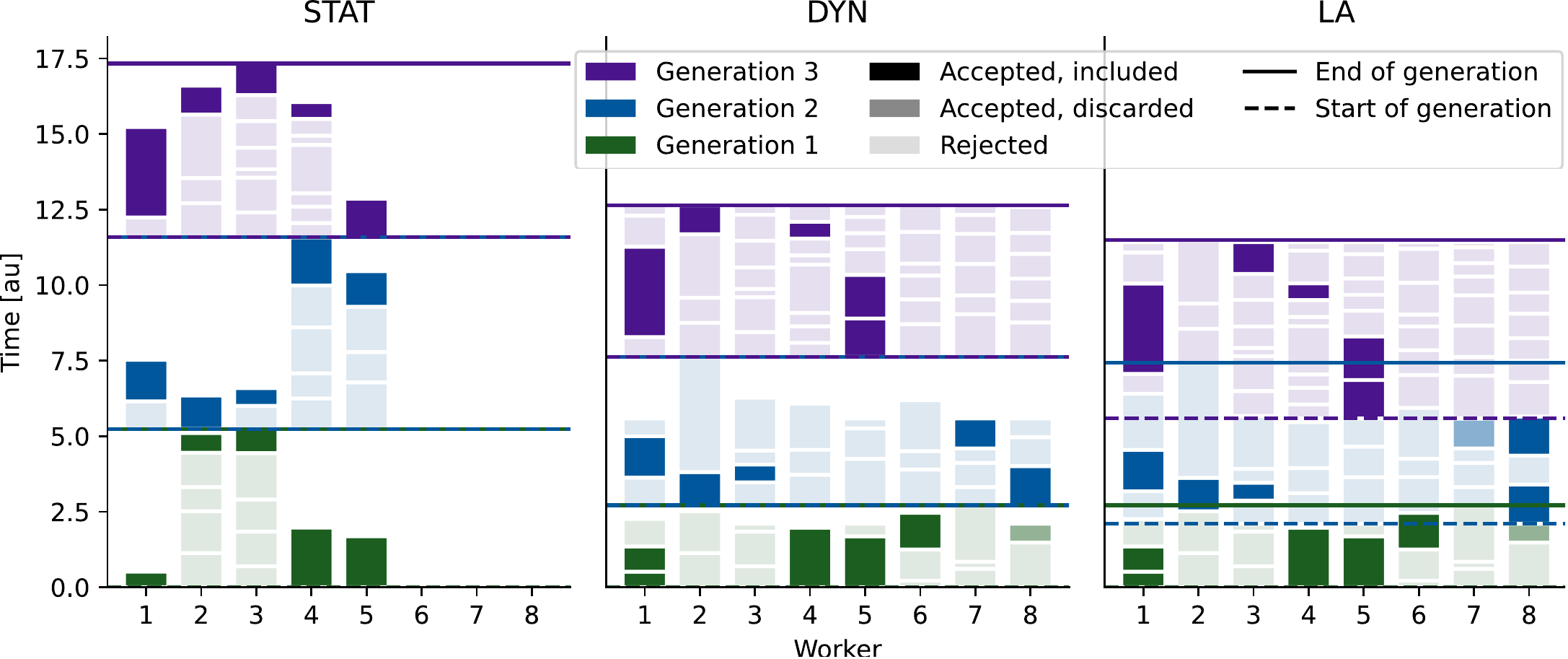}
		\caption{Illustration of core usage over run-time for static (STA), dynamic (DYN) and look-ahead (LA) scheduling for a population size $N=5$ on $W=8$ workers, over 3 generations (colors). The shading indicates whether a sample satisfies the acceptance criterion and is included in the final population (dark), satisfies the acceptance criterion but is discarded because enough earlier-started accepted samples exist (medium, for DYN+LA), or does not satisfy the acceptance criterion and is rejected (light). Solid lines indicate the end of a generation, dashed lines indicate the (preliminary) beginning of a generation (different from solid lines only for LA).}
		\label{fig:strategies}
	\end{figure}
	
	In \textit{static scheduling (STAT)}, given a population size $N$, $N$ tasks are defined and distributed over the workers. Each task consists in sampling until one particle gets accepted (Figure~\ref{fig:strategies}A).
	The tasks are queued if $N\geq W$. STAT minimizes the active computation time and number of simulations and is easy to implement, only requiring basic pooling routines available in most distributed computation frameworks. However, even for $W>N$ only $N$ workers are employed, although the number of required simulations is usually substantially larger than $N$. In addition, at the end of every generation the number of active workers decreases successively, most workers idly waiting for a few to finish their tasks. STAT is available in most established ABC-SMC implementations \cite{KangasraasioLin2016,DuttaSch2017}.
	
	In \textit{dynamic scheduling (DYN)}, sampling is performed continuously on all available workers until $N$ particles have been accepted (Figure~\ref{fig:strategies}B).
	However, simply taking those first $N$ particles as the final population would bias the population towards parameters with short-running simulations. 
	Instead, in DYN all workers are waited for to finish, and out of the then $\tilde N\geq N$ accepted particles, only the $N$ that started earliest are considered accepted \cite{KlingerRic2018}. This ensures that acceptance probability of a particle is in accordance with the target distribution, independent of later events and thus its run-time.
	
	
	\subsection{Parallelization using look-ahead dynamic scheduling}
	
	DYN allows to exploit the available parallel infrastructure to a higher degree than STAT and therefore already substantially decreases the wall-time (by a factor of between 1.4 and 5.3 in test scenarios, see \cite{KlingerRic2018}).
	Nonetheless, some workers remain idle at the end of each generation while waiting for others to complete.
	This fraction increases as the number of workers increases relatively to the population size.
	Additionally, the idle time increases if simulation times are heterogeneous, which is often the case, e.g.\ with estimated reaction rates determining the number of simulated events (Supplementary Information, Section 3.6).
	In case of fast model simulations, also the time between generations, e.g.\ to post-process and store results, may be relatively long.
	
	
	\begin{figure}[t]
		\centering
		\includegraphics[width=\textwidth]{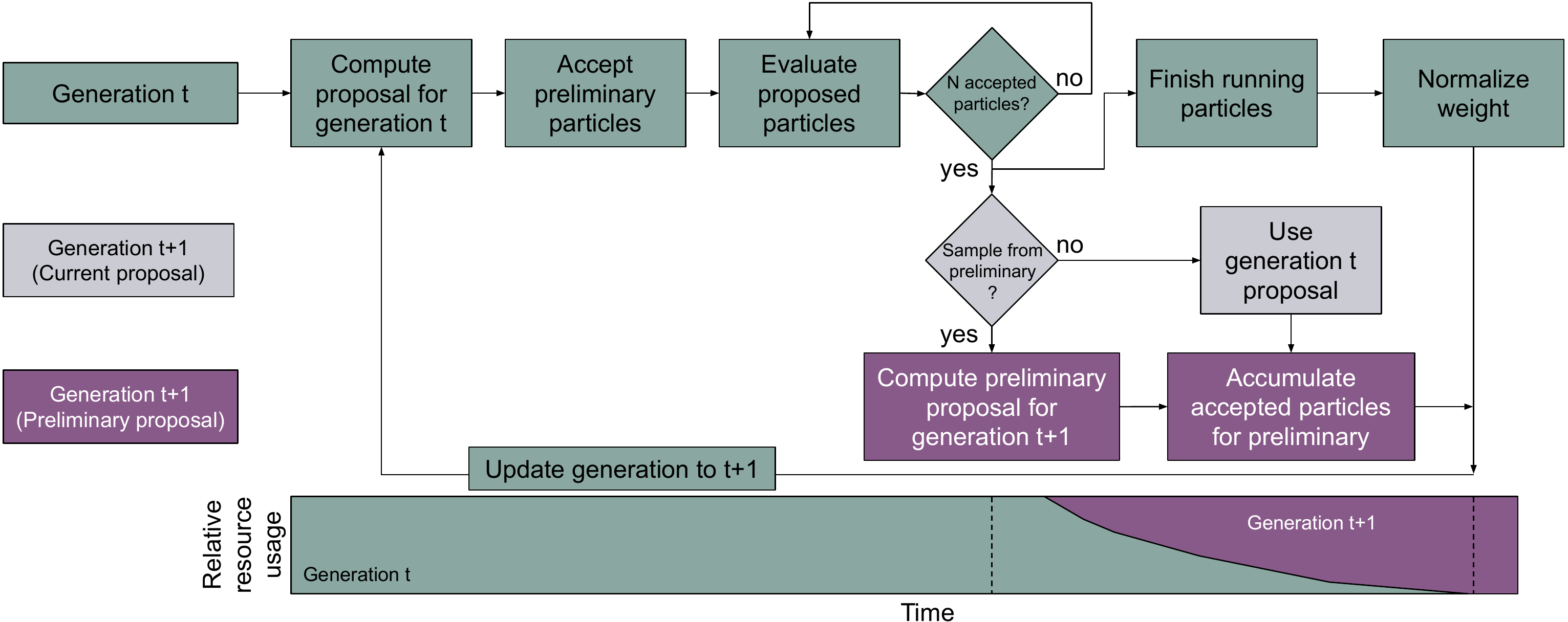}
		\caption{Concept visualization of look-ahead scheduling (LA). As soon as no more simulations are required for generation $t$ (green), a preliminary simulation task for generation $t+1$ is formulated either based on generation $t$ (dark purple) or $t+1$ (light purple). Resulting simulations are considered when evaluating the next generation ($t=t+1$), and suitable weight normalization is applied to all samples (top right). Over time, the number of workers dedicated to generation $t$ decreases, while that for generation $t+1$ increases (bottom).}
		\label{fig:concept}
	\end{figure}
	
	\subsubsection{Proposed algorithm}
	
	We propose to extend dynamic scheduling by using the free workers at the end of each generation to proactively sample for the next generation (Figure~\ref{fig:concept}):
	As soon as $N$ acceptances have been reached in generation $t-1$ and workers thus start to get idle, we construct a preliminary proposal $\tilde g_t$, based on which particles for generation $t$ are generated and simulations performed on the free workers.
	$\tilde g_t$ can be based on a preliminary population of accepted particles $\hat P_{t-1} = \{(\hat\theta_{t-1}^i,\hat w_{t-1}^i)\}_{i\leq N}$ based on these first $N$ acceptances.
	However, $\hat P_{t-1}$ may introduce a practical bias (in a finite sample sense) towards particles with faster simulations times. This can in particular occur when computation time is highly parameter-dependent. To address this issue, the preliminary proposal can alternatively be based on $P_{t-2}$ (such that $\tilde g_{t} = g_{t-1}$), giving inductively practically unbiased proposals.
	If a particle $\tilde\theta_t\sim\tilde g_t$ gets accepted according to the acceptance criteria of generation $t$, its non-normalized weight is calculated as $\tilde w_t(\tilde\theta_t) = \frac{\pi(\tilde\theta_t)}{\tilde g_t(\tilde\theta_t)}$.
	As soon as all simulations for generation $t-1$ have finished and thus the actual $P_{t-1}$ is available, all workers are updated to continue working with the actual sampling task based on proposal $g_t$. As the time-critical part of typical ABC applications is the model simulation, the cost of generating the preliminary sampling task is usually negligible.
	
	The assessment of acceptance of preliminary samples depends on whether everything is pre-defined:
	If the acceptance components, including distance function $d$ and acceptance threshold $\varepsilon_t$ for generation $t$ are defined a-priori, then acceptance can be checked directly on the workers without knowledge of the complete previous population $P_{t-1}$.
	If however any component of the algorithm is adaptive and hence based on $P_{t-1}$ (e.g. the acceptance threshold is commonly chosen as a quantile of $\{d(y_{t-1}^i,y_\obs)\}_{i\leq N}$), the acceptance check must be delayed until the actual $P_{t-1}$ is available.
	This serves to use one common acceptance criterion across all particles within a generation, so that all particles target the same distribution.
	
	The population of generation $t$ is then, corrected for run-time bias as in DYN by only considering the $N$ accepted particles that were started first, given as 
	\begin{equation}\label{eq:la_population}
		P_t = \{\{(\tilde\theta_t^i,\tilde w_t^i)\}_{i\leq \tilde N}, \{(\theta_t^i,w_t^i)\}_{\tilde N<i\leq N}\},
	\end{equation}
	with $0\leq \tilde N\leq N$ particles based on the preliminary proposal $\tilde g_t$, and $N-\tilde N$ on the final $g_t$. The weights need to be normalized appropriately, as explained in the following section.
	We call this parallelization strategy, which during generation $t-1$ already looks ahead to generation $t$, \textit{Look-ahead (dynamic) scheduling (LA)}.
	
	
	\subsubsection{Weights and unbiasedness}
	
	A key property of ABC methods is that they provide an asymptotically unbiased Monte Carlo sample from $\pi_{\ABC,\varepsilon_{n_t}}(\theta|y_\obs)$, with $\pi_{\ABC,\varepsilon_{n_t}}(\theta|y_\obs) \rightarrow \pi(\theta|y_\obs)$ for $\varepsilon\rightarrow 0$. The sample \eqref{eq:la_population} obtained via LA conserves this property:
	The point is that each subpopulation on its own gives an asymptotically unbiased estimator, since the weights $\tilde w_t(\tilde\theta) = \pi(\tilde\theta) / \tilde g_t(\tilde\theta)$, $w_t(\theta) = \pi(\theta) / g_t(\theta)$ are exactly the Radon-Nikodym derivatives w.r.t.\ the respective proposal distributions.
	These estimates are then combined, which decreases the Monte Carlo error due to the larger sample size. Instead of simply tossing all samples together, it is preferable to first normalize the weights relative to their subpopulation, $\tilde W^i_t := \tilde w^i_t/\sum_{i=1}^{\tilde N}\tilde w^j_t$, $W^i_t := w^i_t/\sum_{i=\tilde N+1}^Nw^i_t$ (Supplementary Information, Section 1.3). This is because both weight functions are non-normalized, with generally different normalization constants, which renders them not directly comparable.
	A joint estimate based on the full population can then be given as
	\begin{equation}\label{eq:jointestimate}
		\E_{\pi_{\ABC,\varepsilon_t}(\theta|y_\obs)}[f] \approx \beta\sum_{i=1}^{\tilde N}\tilde W^i_t f(\tilde\theta^i_t) + (1-\beta)\sum_{i=\tilde N+1}^N W^i_t f(\theta^i_t)
	\end{equation}
	with $\beta\in[0,1]$ a free parameter.
	A straightforward choice is $\beta=\tilde N/N$, rendering the contribution of each subpopulation proportional to the respective number of samples. Instead, we propose to choose it to maximize the overall effective sample size \eqref{eq:ess}, rendering the Monte Carlo estimate more robust. This is a simple constrained optimization problem with solution
	\begin{equation*}
		\beta = \frac{\operatorname{ESS}(\{\tilde W^i_t\}_{i\leq \tilde N})}{\operatorname{ESS}(\{\tilde W^i_t\}_{i\leq \tilde N}) + \operatorname{ESS}(\{W^i_t\}_{\tilde N <i\leq N})}
	\end{equation*}
	i.e.\ the contribution of each subpopulation is proportional to its effective sample size (Supplementary Information, Section 1.4).
	Supposing that for $N\rightarrow\infty$, $\tilde N/N\rightarrow\alpha\in[0,1]$, \eqref{eq:jointestimate} converges to the left-hand side, as required.
	A more detailed derivation and extension to more than two proposal distributions is given in the Supplementary Information, Section 1.
	
	
	\subsection{Implementation and availability}
	
	We implemented LA in the open-source Python tool pyABC \cite{SchaelteKli2022}, which already provided STAT and DYN. We employ a Redis low-latency server to handle the task distribution.
	If all components are pre-defined, we perform evaluation of the ``look-ahead'' samples $(\tilde\theta,\tilde y)$ directly on the workers.
	If there are adaptive components, the delayed evaluation is performed on the main process.
	To avoid generating unnecessarily many preliminary samples in the presence of some very long-running simulations, we limited the number of preliminary samples to a default value of 10 times the number of samples in the current iteration.
	To not start preliminary sampling unnecessarily, we employed schemes predicting whether any termination criterion will be hit after the current generation.
	The code underlying this study can be found at \url{https://github.com/EmadAlamoudi/Lookahead_study}. A snapshot of code and data can be found at \url{https://doi.org/10.5281/zenodo.7875905}.
	
	
	\section{Results}
	
	Wall-time superiority of DYN over STAT has already been established in prior work \cite{KlingerRic2018}.
	To study the performance of LA and compare it to DYN, we applied both to several parameter estimation problems and in various scenarios of population size $N$ and workers $W$.
	We distinguish between ``LA Pre'' using the preliminary $\hat P_{t-1}$ to generate $\tilde g_t$, and ``LA Cur'' using $P_{t-2}$ instead.
	
	
	\subsection{Test problems}
	
	\begin{table}[t]
		\centering
		\caption{Overview of application examples}
		\label{tab:problems}
		{%
			\begin{tabular}{l|l|l|l}
				ID & Description & Implementation & $n_\theta$\\
				\hline
				T1 & Bimodal run-time-skewed model & Python & 1\\ 
				T2 & Conversion reaction ODE model & Python & 2\\ 
				M1 & \begin{tabular}[c]{@{}l@{}}Tumor spheroid growth \cite{JagiellaRic2017} \end{tabular} & C++ & 7 \\
				M2 & Liver tissue regeneration \cite{MeyerMor2020} & Morpheus & 14\\
			\end{tabular}%
		}
	\end{table}
	
	We considered four problems (Table~\ref{tab:problems}): Problems T1-T2 are simple test problems, while M1-M2 are realistic application examples.
	
	Problem T1 is a bimodal model $y \approx \theta^2$, in which simulations from one mode have an artificially longer run-time, reflecting parameter dependence of run-times.
	Problem T2 is an ordinary differential equation (ODE) model with 2 parameters describing a conversion reaction $x_1\leftrightarrow x_2$, with observables obscured by random multiplicative noise.
	To analyze sampler behavior under simulation run-time heterogeneity, we added random log-normally distributed delay times $t_{\text{sleep}}$ of various variances atop the ODE simulations. 
	For this model, run-times are fast, permitting repeated analysis to check correctness of the method, quantify stochastic effects and assess average behavior.
	Problem M1 describes the growth of a tumor spheroid using a hybrid discrete-continuous approach, modeling single cells as stochastically interacting agents and extracellular substances via PDEs \cite{JagiellaRic2017}.
	Problem M2 describes the metabolic status of mechano-sensing during liver generation, describing the reaction network dynamics by a set of ODEs \cite{MeyerMor2020}.
	
	\begin{figure}[t]
		\centering
		\includegraphics[width=0.8\textwidth]{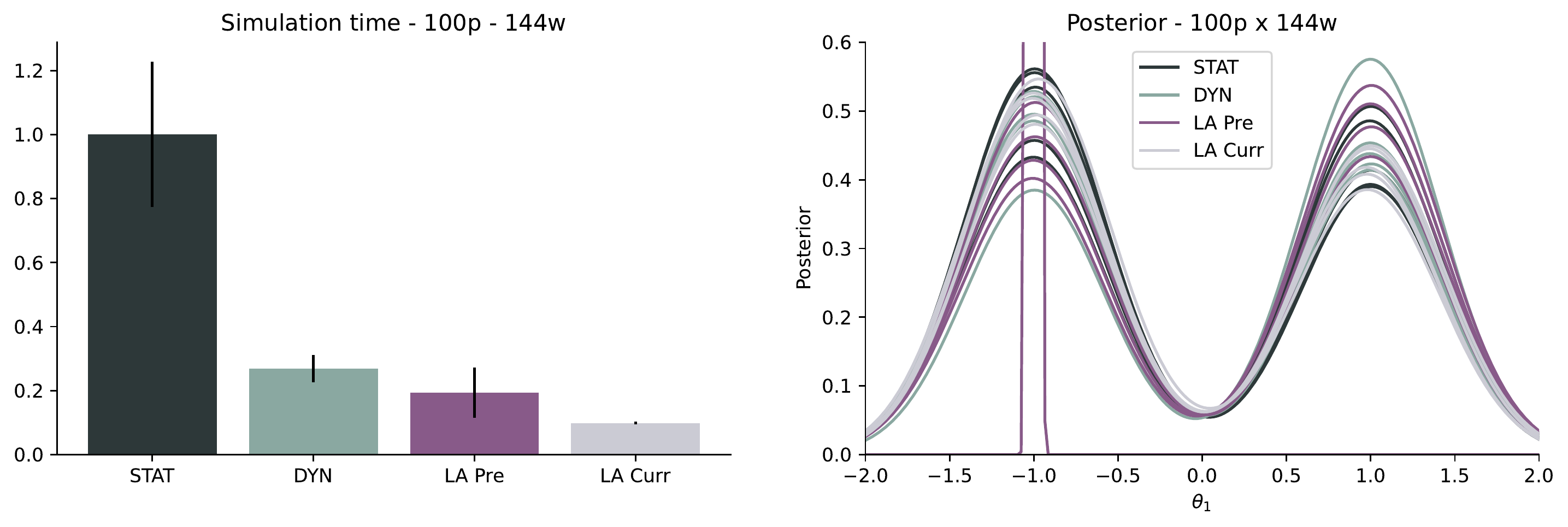}
		\includegraphics[width=0.8\textwidth]{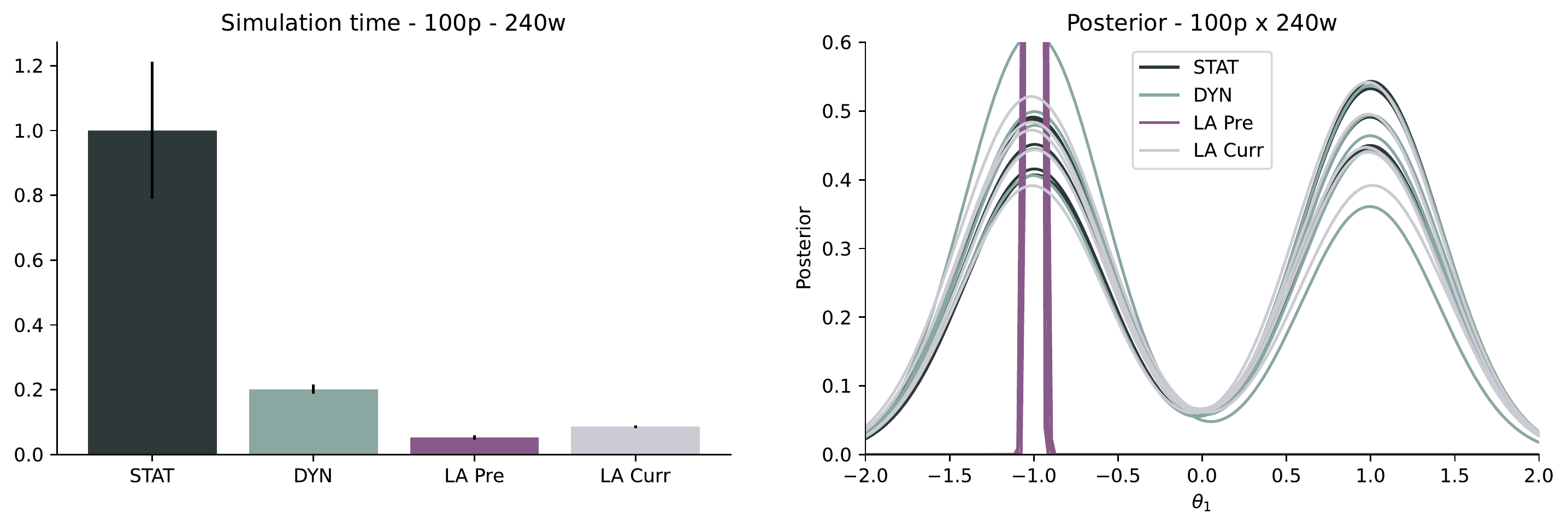}
		\includegraphics[width=0.8\textwidth]{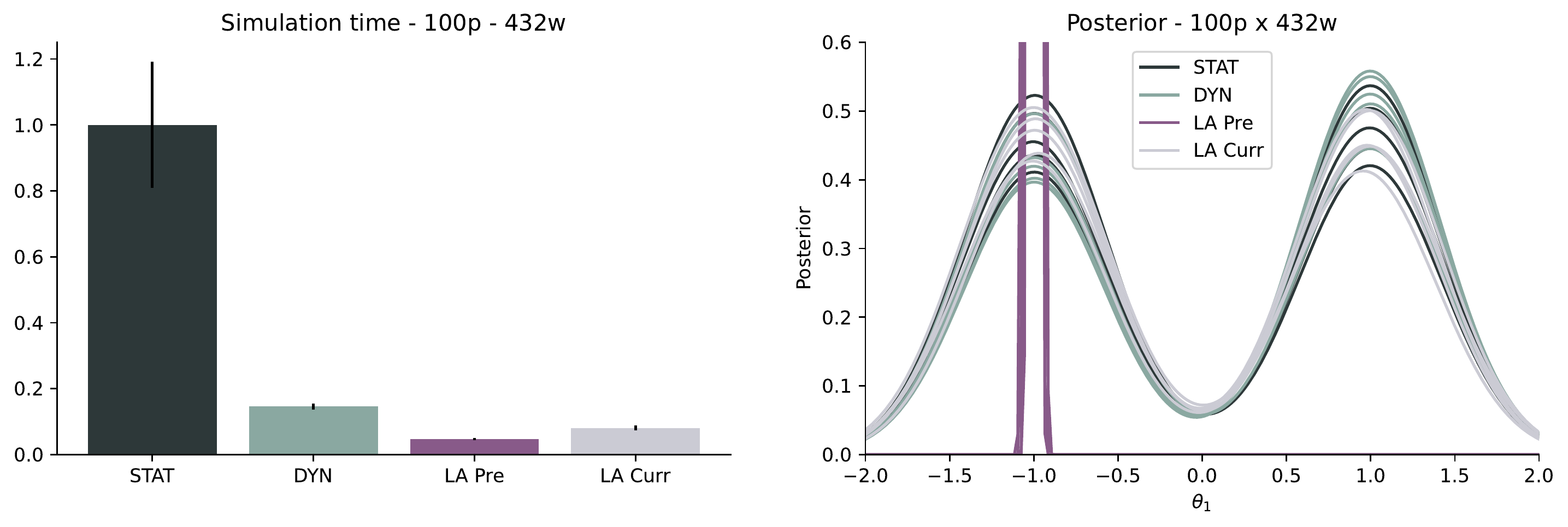}
		
		\caption{Run-time and posterior approximation for 5 different runs of model T1 with STAT, DYN, LA Pre and LA Cur, with population size $N=100$ on $W=144, 240, 432$ workers.}
		
		\label{fig:T1}
	\end{figure}
	
	\subsection{Biased proposal can induce practical bias in accepted population}
	
	The analysis of test model T1 revealed that for small population sizes $N$ relative to the number of workers $W$, together with high acceptance rates, LA Pre can indeed lead to a bias towards short-running simulations (Figure~\ref{fig:T1} right). This can happen when $\hat P_{t-1}$ is only based on short-running simulations, and only proposes particles from that regime, enough of which are then accepted to form $P_t$. For $N\geq W$, this effect no longer arose, likely because given large population sizes and sampling from other modes with associated high importance weights eventually happened, and because low acceptance rates make it improbably that acceptances in $P_t$ only stem from $\hat P_{t-1}$.
	
	\subsection{Sampling from unbiased proposal solves bias}
	
	When replacing LA Pre by LA Cur, i.e. sampling from $P_{t-2}$ instead of $\hat P_{t-1}$, the bias no longer occurred (Figure~\ref{fig:T1} right).
	This is as expected, because $\tilde g_{t-2}$ is has no run-time bias.
	In practice, we did not encounter any such problems on the considered application examples, where results from DYN, LA Pre and LA Cur were highly consistent.
	Yet, LA Pre may fail in some situations, which also demonstrates that ABC-SMC algorithms are sensitive to potential bias in the proposal distribution.
	Thus, in the following, we focus on the stable LA Cur, showing pendants for LA Pre in the Supplementary Information.
	
	\subsection{Look-ahead sampling gives accurate results}
	
	We used problem T2 to analyze different scenarios, with population sizes $N$ from 20 to 1280, worker numbers $W$ from 32 to 256, and log-normally distributed simulation times of variances $\sigma^2$ from 1.0 to 4.0. We ran each scenario 13 times to obtain stable average statistics.
	We considered means and standard deviations as point and uncertainty measures.
	
\begin{figure}[t]
		\centering
		\begin{minipage}{0.63\textwidth}
			\includegraphics[width=\textwidth]{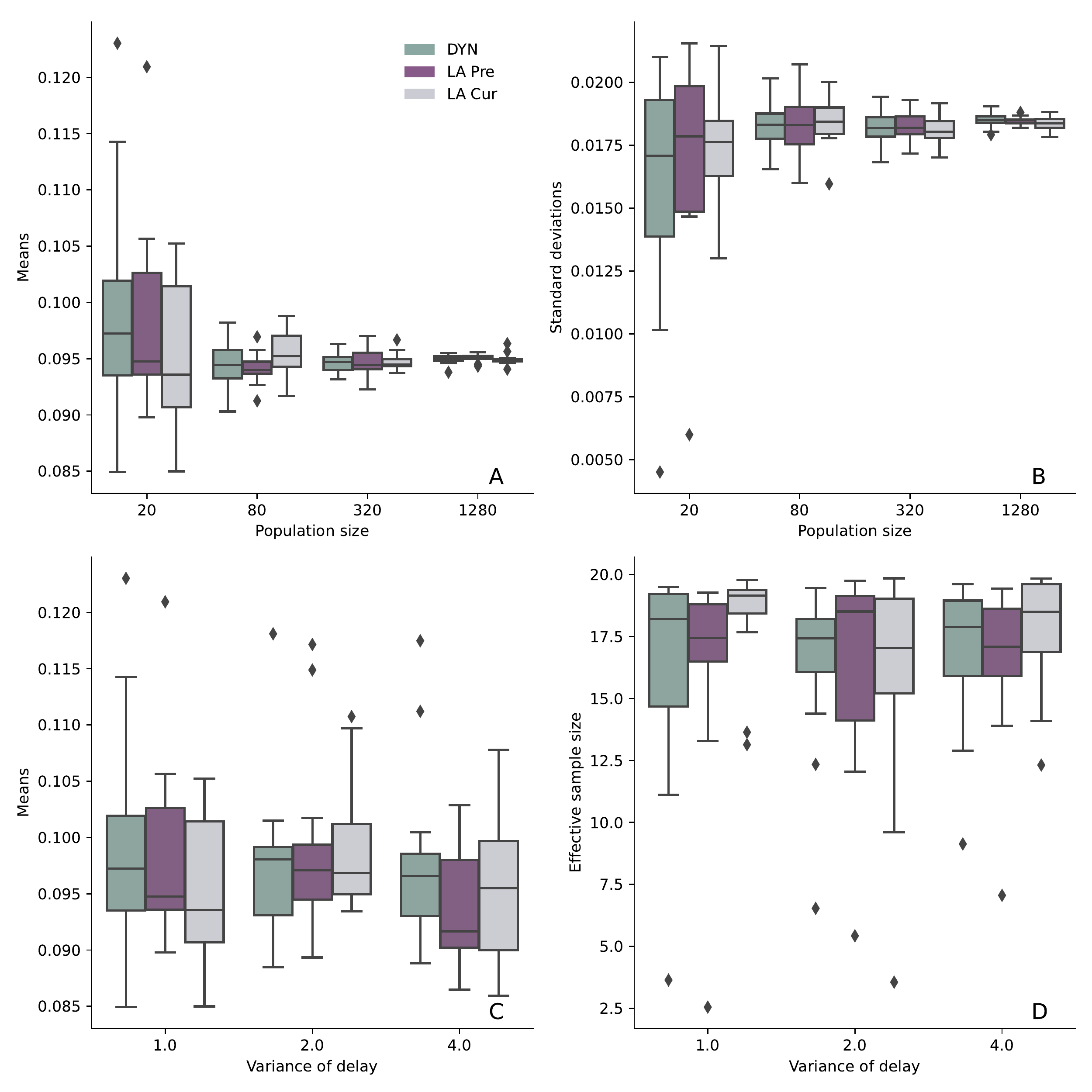}        
		\end{minipage}
		\begin{minipage}{0.34\textwidth}
			\includegraphics[width=\textwidth]{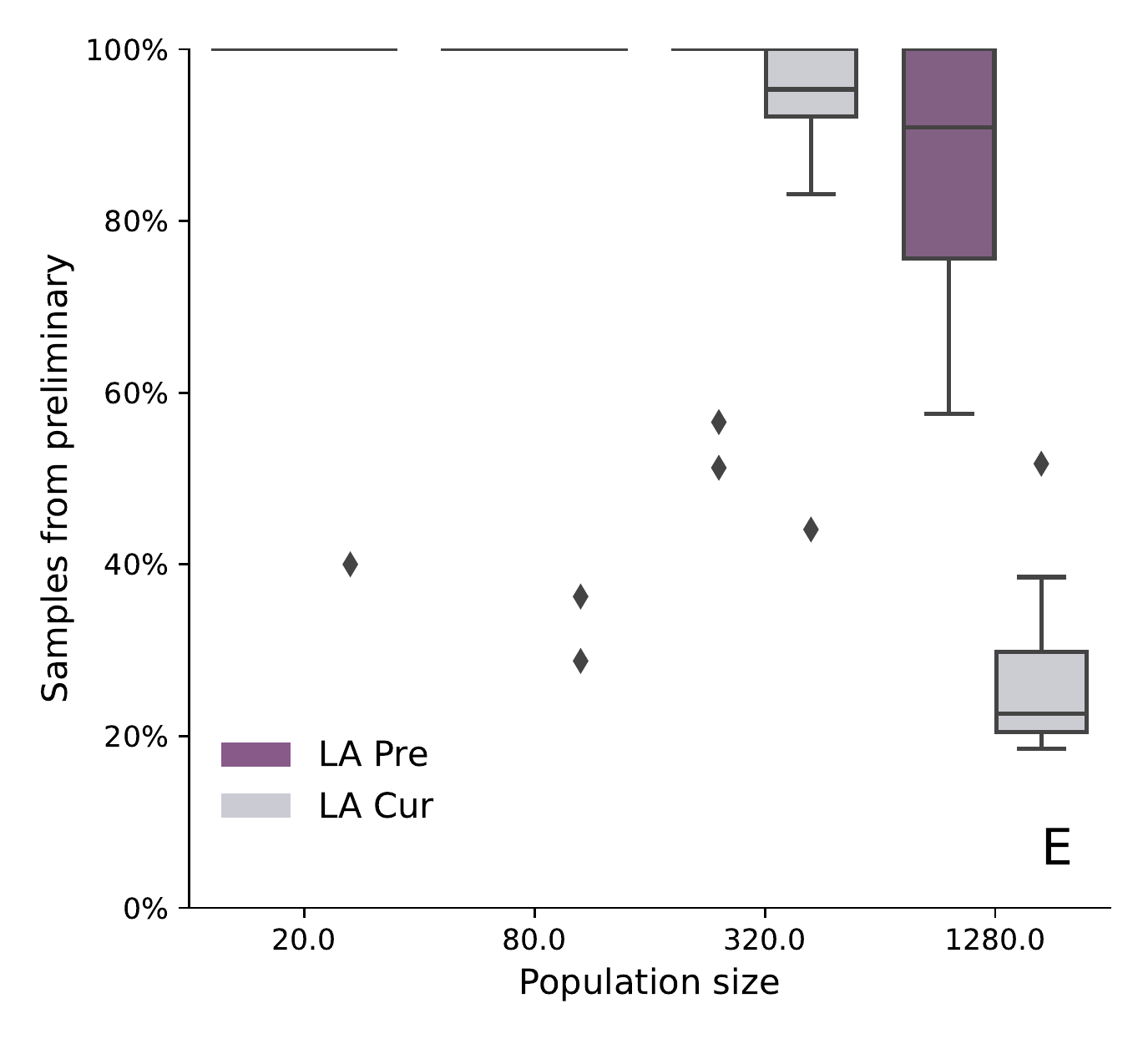}
			\includegraphics[width=\textwidth]{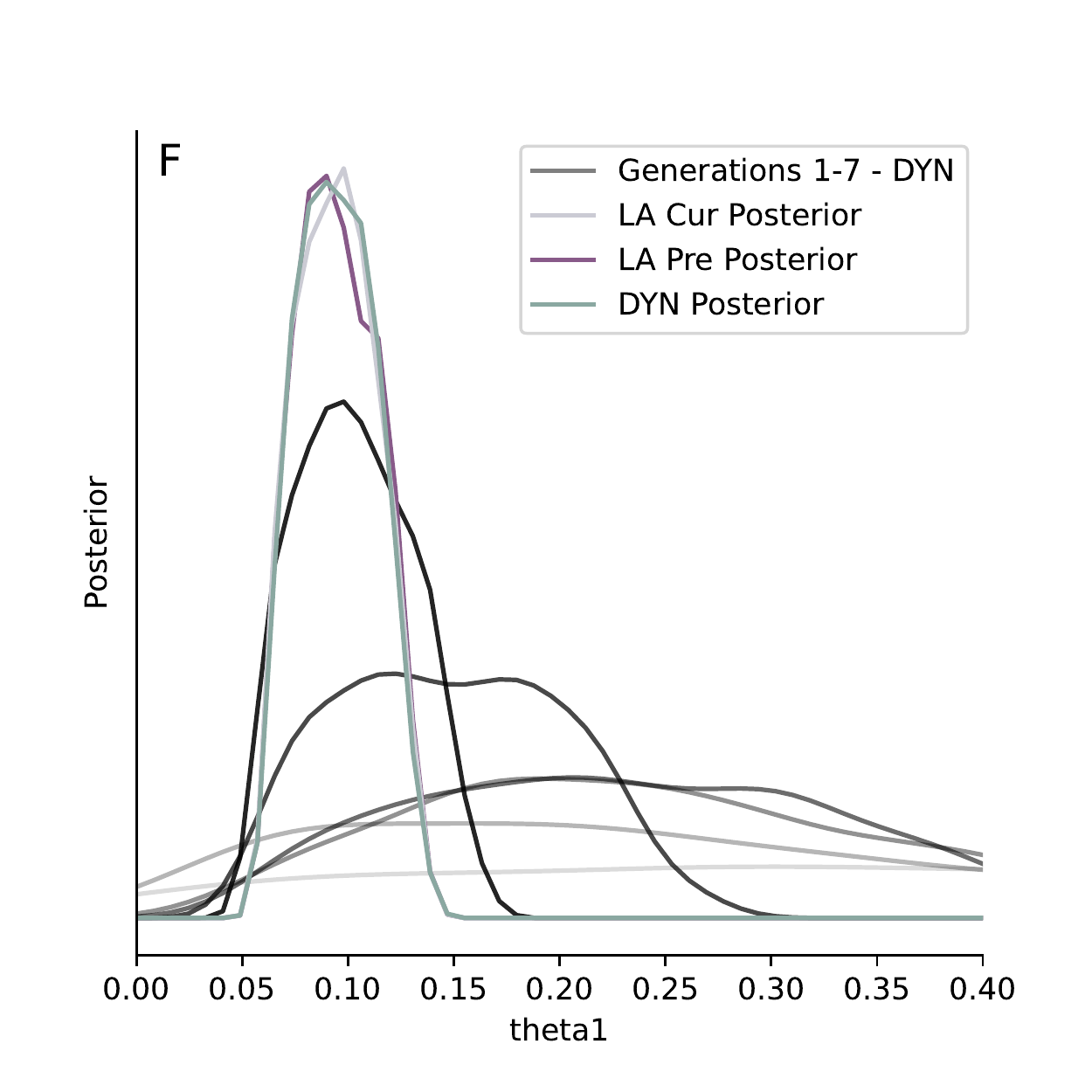}
		\end{minipage}
		
		\caption{
			Results for problem T2 for different population sizes $N$, worker numbers $W$, and sleep time variances $\sigma^2$. Unless otherwise specified, we used $N=1280$, $W=256$, and a log-normally distributed sleep time $t_\text{sleep}$ of variance $\sigma^2=1$. To increase comparability, the $\varepsilon_t$ values over $n_t=8$ generations were pre-defined. (A) and (B): Mean and variance of the posterior approximation $\pi_{\ABC,\varepsilon_{n_t}}(\theta|y_\obs)$. Box-plot over 13 repetitions. (C): Posterior mean for different sleep time variances, for $W=20$. (D): Effective sample size across different sleep time variances, for $N=256$ and $W=20$, in which case it is likely that several generations are sampled completely from the preliminary proposal. (E): Fraction $\tilde N / N$ of accepted samples in the final population $t=n_t$ that originate from the preliminary proposal $\tilde g_{n_t}(\theta)$ for LA Pre and LA Cur. (F): Exemplary visualization of 1d posterior approximation marginals for single runs.
		}
		\label{fig:T2ODEBoxPlots}
		
	\end{figure}

	Point estimates for DYN and LA converged to the same values across population sizes (Figure~\ref{fig:T2ODEBoxPlots} A+B).
	The proportion of accepted LA samples in the final population originating from the preliminary distribution ranged from nearly 100\% to 50\% (LA Pre) and 20\% (LA Cur), as expected decreasing for larger population sizes (Figure~\ref{fig:T2ODEBoxPlots} E+F). The more pronounced decrease for LA Cur than LA Pre is reasonable because void of bias, $\hat P_{t-1}$ provides a better sampling distribution than $P_{t-2}$.
	Effective sample sizes were stable across DYN and LA (Figure~\ref{fig:T2ODEBoxPlots}~D).
	A higher run-time variance lead to an increase in accepted samples originating from the preliminary proposal distribution (Supplementary Information, Figure S1), which is expected, because greater heterogeneity in run-times raises the chance of encountering exceptionally long-running simulations, which DYN has to wait for, while LA proceeds already.
	
	\subsection{Considerable speed-up towards high worker numbers}
	
	To analyze the effect of scheduling strategy on the overall wall-time, we ran model T2 systematically for different population sizes and numbers of workers .
	We considered population sizes $32 \leq N \leq 1280$ and numbers of parallel workers $32 \leq W \leq 256$, which covers typical ranges used in practice.
	Each scenario was repeated between $13$ times to assess average behavior, here we report mean values.
	
\begin{figure}[t]
		\centering
		\includegraphics[width=\textwidth]{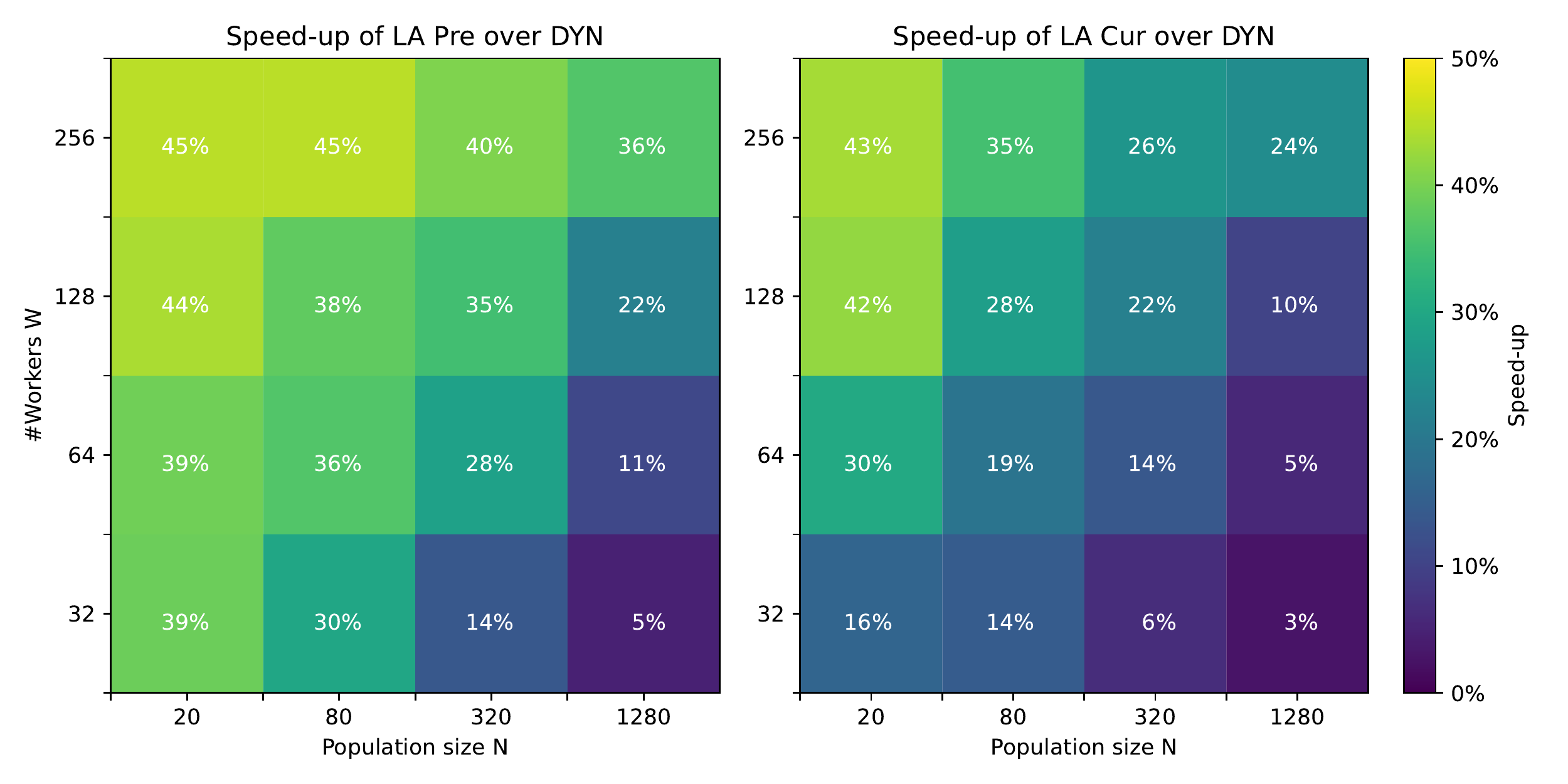}
		\caption{
			Speed-up ($1 - \{\text{Wall-time LA}\} / \{\text{Wall-time DYN}\}$) of LA Pre (left) and LA Cur (right) over DYN for various population sizes and numbers of workers, for a model sleep time variance of $\sigma^2=1$.
		}
		\label{fig:T2ODEParEff}
		
	\end{figure}

	As a general tendency, the wall-time speed-up of LA over DYN became larger with increasing ratio of the number of workers by the population size.
	For a model sleep time variance of $\sigma^2=1$ (Figure \ref{fig:T2ODEParEff}), e.g.\ for $N=20$ and $W=256$, the average wall-time got reduced by a factor of almost 1.8. In most scenarios, a wall-time reduction by a factor of between $1.11$ and $1.8$ was observed.
	Only when the population size was large compared to the number of workers, was the speed-up comparably small.
	
	For a sleep time variance of $\sigma^2=2$ (Supplementary Information, Figure S2), we observed similar behavior. There, the acceleration was generally more pronounced with up to a factor of roughly $1.9$ and many factors in the range $1.2$ to $1.9$. This indicates that indeed the advantage of LA over DYN is more pronounced in the presence of highly heterogeneous model simulation times.
	
	Also on T1, the comparison of run-times (Figure~\ref{fig:T1} left) revealed a speed-up of LA over DYN. Further, we could confirm on both T1 and T2 (Figure~\ref{fig:T1} and Supplementary Information, Figure S3)the substantial speed-up DYN already provides over STAT, as reported in \cite{KlingerRic2018}, on which we here improved further.
	
	\subsection{Scales to realistic application problems}
	
	Given the high simulation cost of the application problems M1-2, we only performed selected analyses to compare LA and DYN.
	A reliable comparison of run-times in real-life application examples is challenging, because the total wall-time varies strongly due to stochastic effects, and computations are too expensive to perform inference many times.
	
	For the two models, the parameter estimates obtained using LA (both LA Per and LA Cur) and DYN are consistent, up to expectable stochastic effects (Figures S4+5 and S9-11).
	Together with the previous analyses, this indicates that for practical applications, the multi-proposal approach of LA allows for stable and accurate inference, similar to the single proposal used by DYN.
	In early generations, a considerable part of the accepted particles was based on the preliminary proposal distribution (near 100\%), which then decreased in later generations (Supplementary Information, Figures S6 and S12).
	This is consistent with the decrease in acceptance rate and thus the relative time during which the preliminary and not the final proposal distribution is used.
	
\begin{figure}[t]
		\centering
		\includegraphics[width=0.8\textwidth]{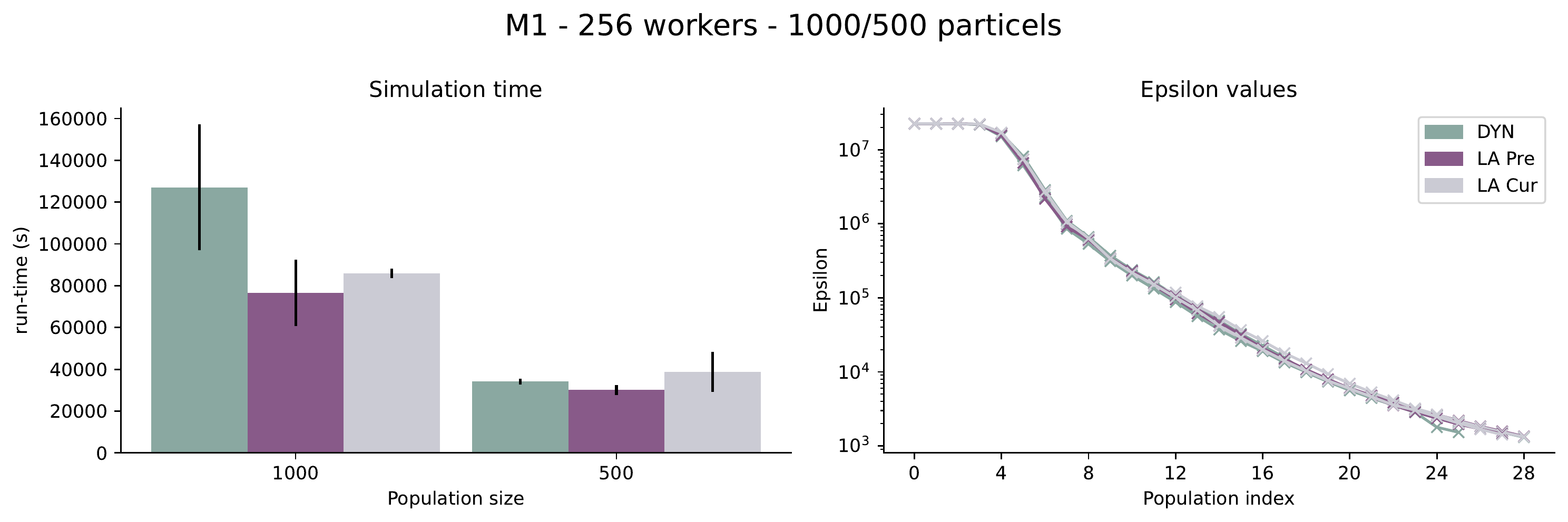}
		\includegraphics[width=0.8\textwidth]{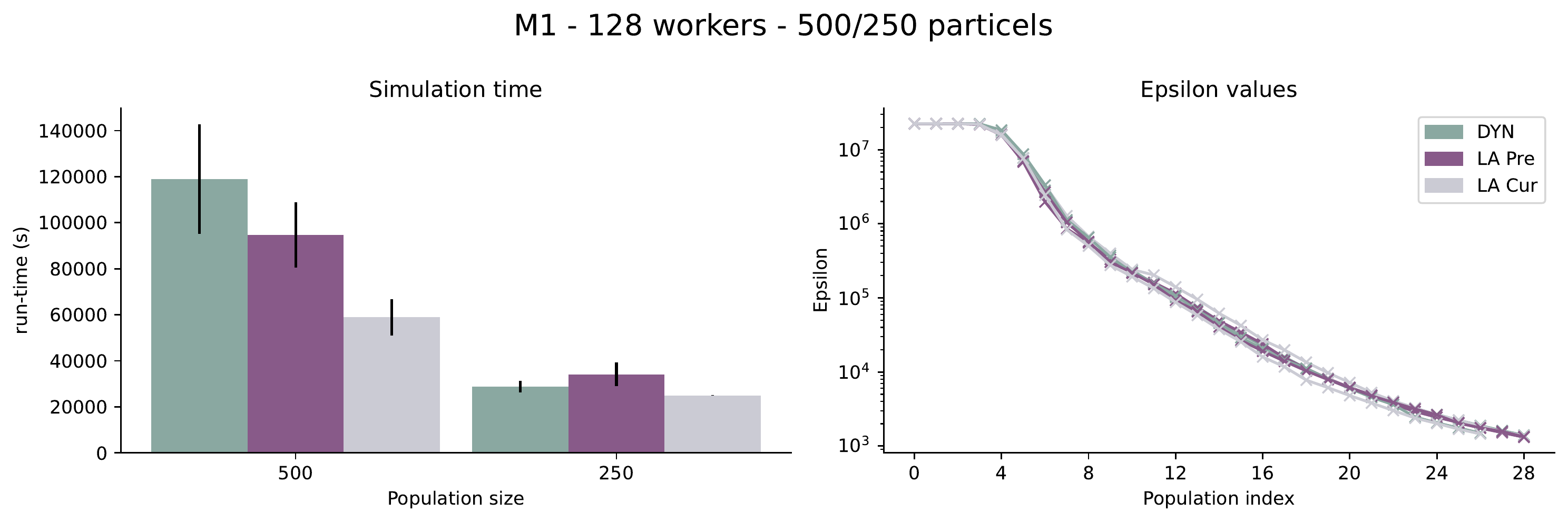}
		
		\caption{The run-time and posterior distributions for 2 different runs of the model (M1) with population size 1000, 500, 250 on 128 and 256 workers.}
		\label{img:M1_time_eps}
	\end{figure}

	For the tumor model M1, we used an adaptive quantile-based epsilon threshold schedule \cite{DrovandiPet2011}, with DYN, LA Pre and LA Cur, population sizes $N\in\{250,500,1000\}$, and $W\in\{128, 256\}$ workers. For each considered configuration we performed $2$ replicates (in total 8) to assess average behavior. Reported run-times are until a common final threshold was hit by all runs.
	The speed-up of LA over DYN varied depending on the ratio of population size and number of workers, similar to what we observed for T1+2. For high ratios, LA was consistently faster up to 35\%.
	However, for low ratios, less improvement was observed. In some runs, LA was slightly slower than DYN (Figure~\ref{img:M1_time_eps}).
	Over the 8 runs, we observed a mean speed-up of 21\% (13\%) and a median of 23\% (16\%) for LA Cur (LA Pre).
	This indicates expected speed-ups of 13-23\%, however it should be remarked that large run-time differences and volatility could be traced back to single generations taking vast amounts of time (Supplementary Information, Figure S7).
	Those long generations occurred in all scheduling variants and exist most likely because the epsilon for
	that generation was chosen too optimistically, indicating a weakness of the used epsilon scheme, rather than the parallelization strategy.
	
\begin{figure}[t]
		\centering
		\includegraphics[width=0.8\textwidth]{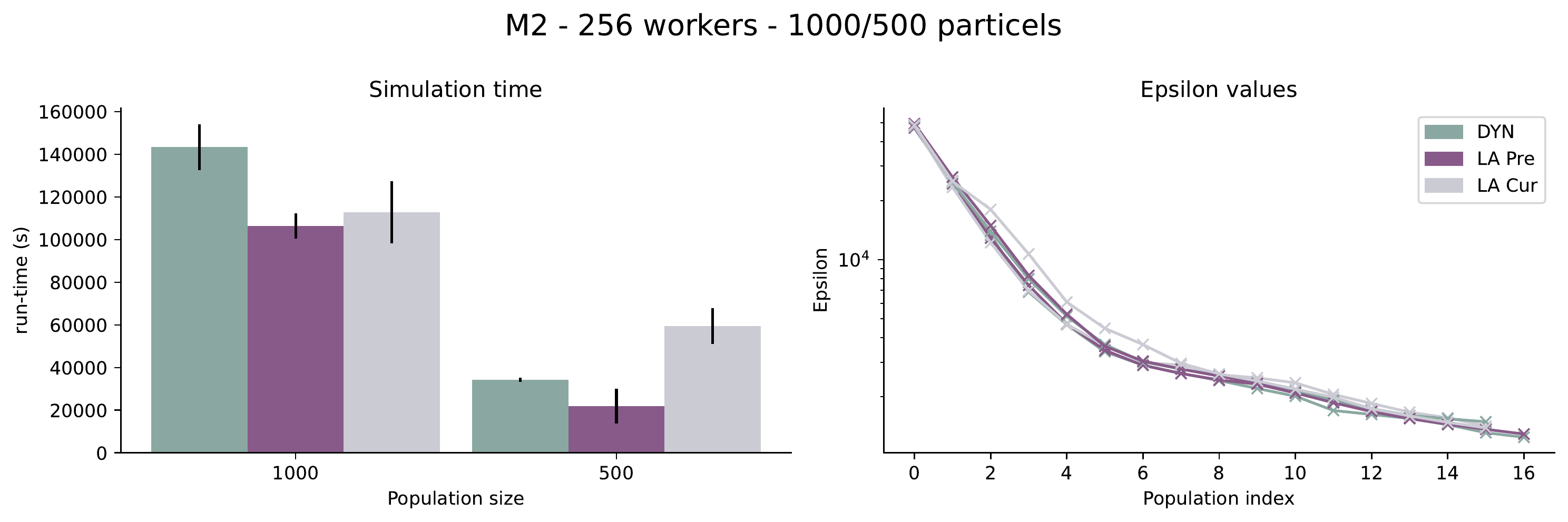}
		\includegraphics[width=0.8\textwidth]{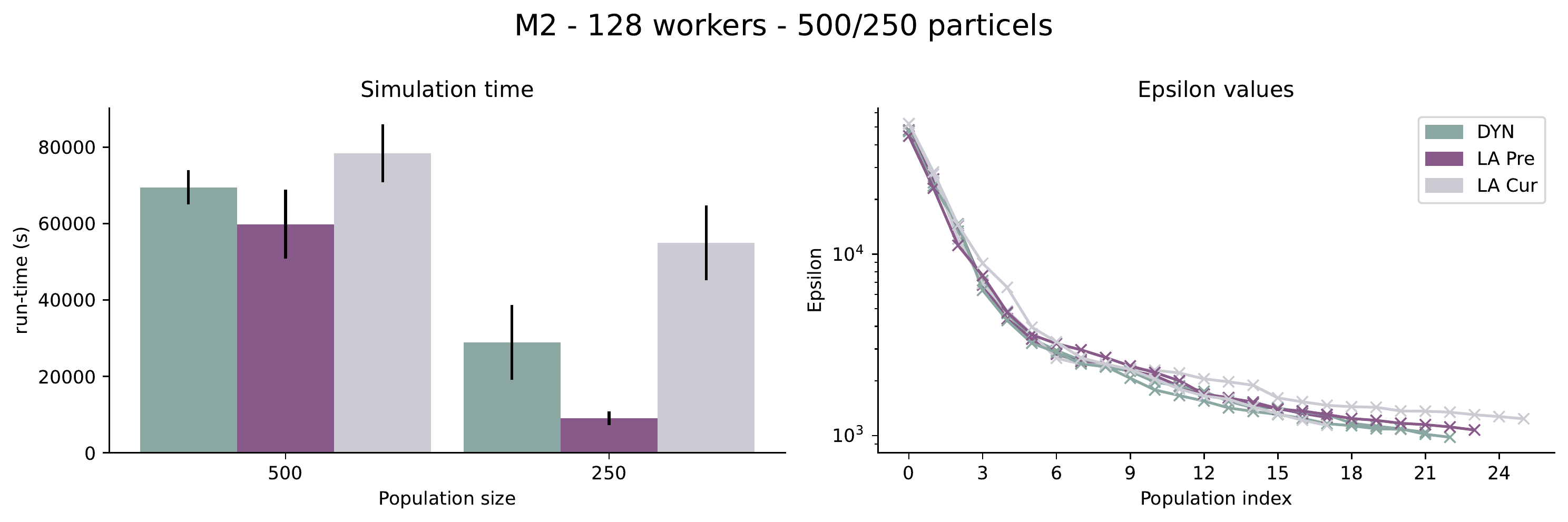}
		
		\caption{The run-time and posterior distribution of 2 different runs of the model (M2) with population size 1000, 500, 250 on 128 and 256 workers.}
		\label{img:M2_time_eps}
	\end{figure}

	For the liver regeneration model M2, we performed similar analyses, with adaptive quantile-based epsilon threshold schedules, population sizes $N\in\{250,500,1000\}$ and $W\in\{128,256\}$ workers, with 2 replicates per configuration.
	Similar to model M1, we observed a faster performance of up to 35\%. However, with a smaller ratio between population size and the number of workers, a slightly lower performance improvement was achieved (Figure~\ref{img:M2_time_eps}). 
	Similarly to M1, the acceleration varied quite strongly. For LA Pre we observed a mean speed-up over all 8 runs of 36\% (median 31\%) over DYN. However, for LA Cur we observed contrarily a mean slow-down of 39\% (median 43\%) over DYN.
	It is not clear what caused this stark difference, which is again subject to high fluctuations. Further tests would be needed to assess the reasons for this specific model.
	
	\section{Discussion}
	
	Simulation-based ABC methods have made parameter inference increasingly accessible even for complex stochastic models, are however limited by computational costs. Here, we presented ``look-ahead'' sampling, a parallelization strategy to minimize wall-time by using all available high-performance computing resources at near-all times.
	On various test and application examples, we verified the accuracy and robustness of the novel approach in typical settings. Depending on model simulation run-time heterogeneity, and the relation of population size and the number of cores, we observed a speed-up of up to 45\% compared to the previously most efficient strategy, dynamical scheduling. On typical application examples, we observed a speed-up of roughly 20-30\%, however with some variability, assessing which would require further tests with considerable computational resources.
	We would also like to remark that ABC-SMC is sensitive to the choice of proposal distribution. Finite samples can induce a practical bias, as we observed here for parameter-dependent run-times of models -- a problem that occurred in extreme cases but could only be solved by using look-ahead sampling with the previous, and not the preliminary, proposal distribution.
	
	Conceptually and aside implementation details, the presented strategy provides the minimal wall-time among all parallelization strategies, as all cores are made use of at practically all times.
	We observed that look-ahead sampling using preliminary results (LA Pre) provided a performance speed-up over re-using the previous generation (LA Cur), however at the cost of practical bias.
	Were it possible to construct an unbiased proposal using those preliminary results, e.g.\ via reweighting or imbalance detection, we could thus increase the speed-up with robust performance.
	When using delayed evaluation, it would be possible to parallelize the evaluation as well, which we have not done here. If evaluation times are long relative to simulation times, e.g.\ if (adaptive) summary statistics involve complex operations, this would be beneficial. In order to reduce a potential bias in the preliminary proposal distribution towards fast-running simulations, it may be beneficial to update it as soon as more particles finish. This would imply the use of more than two importance distributions, the theory of which we have already provided in the Supplementary Information.
	
	In conclusion, we showed how we can minimize wall-time and associated computing cost of ABC samplers with substantial performance gains over established methods. This concept is generally applicable for sequential importance sampling methods, thus we are certain that it will find widespread use.
	
	
	\section*{Acknowledgments}
	
	The authors acknowledge the Gauss Centre for Supercomputing e.V. (www.gauss-centre.eu) for funding this project by providing computing time on the GCS Supercomputer JUWELS at J\"ulich Supercomputing Centre (JSC).
	This work was supported by the German Federal Ministry of Education and Research (BMBF) (FitMultiCell/031L0159C and EMUNE/031L0293C) and the German Research Foundation (DFG) under Germany's Excellence Strategy (EXC 2047 390873048 and EXC 2151 390685813 and the Schlegel Professorship for JH).
	YS acknowledges support by the Joachim Herz Foundation. FG was supported by the Chica and Heinz Schaller Foundation.
	
	
	\section*{Author contributions}
	
	JH and YS conceived the idea. YS, EA and FR developed and implemented the algorithms. EA and FR performed the case studies with assistance of NB and FG. EA, FG and LB developed the application problems. All authors discussed the results and jointly wrote and approved the manuscript.
	
	\clearpage
	\newpage
	
	\setcounter{figure}{0}
	\renewcommand{\thefigure}{S\arabic{figure}}
	\clearpage
	\newpage
	
	\bibliographystyle{unsrt}
	\bibliography{Database}
\end{document}


\maketitle
	
	\tableofcontents
	
	\section{Multiple proposal importance sampling}
	
	Let $\theta\in\R^{n_\theta}$ denote the latent parameters, $\pi(\theta)$ the prior, and $\pi(y|\theta)$ the likelihood which we assume to be able to sample data $y\in\R^{n_y}$ from, while direct evaluation is not possible. Let $y_\obs\in\R^{n_y}$ be the observed data. By Bayes' Theorem, we get the posterior distribution
	\begin{equation*}
		\pi(\theta|y_\obs) = \frac{\pi(y_\obs|\theta)\pi(\theta)}{\pi(y_\obs)} \propto \pi(y_\obs|\theta)\pi(\theta)
	\end{equation*}
	of parameters given observed data.
	
	Let $\varepsilon>0$ and $d: \R^{n_y}\times\R^{n_y}\rightarrow\R_{\geq 0}$ be a distance metric. ABC methods can be interpreted to generate samples from the joint distribution
	\begin{equation}\label{eq:abcdistribution}
		(\theta, y) \sim \pi_{\ABC,\varepsilon}(\theta,y|y_\obs) \propto I[d(y,y_\obs)\leq \varepsilon]\pi(y|\theta)\pi(\theta),
	\end{equation}
	such that the target marginal of interest, 
	\begin{equation*}
		\pi_{\ABC,\varepsilon}(\theta|y_\obs) \propto \int I[d(y,y_\obs)\leq\varepsilon]\pi(y|\theta)\dy \cdot \pi(\theta)
	\end{equation*}
	can be obtained by simple projection to the first component. $\pi_{\ABC,\varepsilon}(\theta|y_\obs)$ is an approximation of the actual parameter posterior, with $\pi(\theta|y_\obs) = \lim_{\varepsilon\searrow 0}\pi_{\ABC,\varepsilon}(\theta|y_\obs)$ under mild assumptions.
	
	\subsection{Standard importance sampling}
	
	We employ importance sampling (ABC-IS) embedded in a sequential Monte-Carlo scheme (ABC-SMC). For a proposal distribution $g(\theta)\gg\pi(\theta)$ and a target population size $N$, we use an ABC-IS scheme as shown in Algorithm \ref{alg:abcis}. In ABC-SMC, Algorithm \ref{alg:abcis} is then iterated multiple times for successively refined thresholds $\varepsilon$ and proposals $g$. Here we focus on a single such iteration.
	
	\begin{algorithm}[hbt]\label{alg:abcis}
		\caption{Importance ABC algorithm.}
		\While{less than $N$ acceptances}{
			sample parameter $\theta \sim g(\theta)$\\
			simulate data $y \sim \pi(y|\theta)$\\
			accept $\theta$ if $d(y, y_\obs) \leq \varepsilon$\\
		}
		compute weights $w_{i} = \frac{\pi(\theta^i)}{g(\theta^i)}$, for accepted parameters $\{\theta^{i}\}_{i\leq N}$\\
		output: weighted samples $\{(\theta_i,w_i)\}_{i\leq N}$
	\end{algorithm}
	
	This form of ABC-IS generates samples from the distribution
	\begin{equation}\label{eq:isdistribution}
		(\theta,y) \sim G(\theta,y) \propto I[d(y,y_\obs)\leq\varepsilon]\pi(y|\theta)g(\theta).
	\end{equation}
	Here, the $g(\theta)$ is because we sample $\theta\sim g(\theta)$, further we simulate data $y$ from the likelihood and discard particles not satisfying the distance condition, such that
	$$y|\theta \sim I[d(y,y_\obs)\leq\varepsilon]\pi(y|\theta).$$
	Therefore, the importance weights, Radon-Nikodym derivatives, of the proposal \eqref{eq:isdistribution} against the target \eqref{eq:abcdistribution} are given by
	\begin{equation*}v(\theta,y) = \frac{\pi_{\ABC,\varepsilon}(\theta,y|y_\obs)}{G(\theta,y)} = \frac{C\cdot I[d(y,y_\obs)\leq\varepsilon]\pi(y|\theta)\pi(\theta)}{C_G\cdot I[d(y,y_\obs)\leq\varepsilon]\pi(y|\theta)g(\theta)} = \frac{C}{C_G}\frac{\pi(\theta)}{g(\theta)} = \frac{C}{C_G} w(\theta)
	\end{equation*}
	with normalization constants $C, C_G$. If all normalizations were known exactly, we could calculate the unbiased importance estimator of a test function $f$ over the ABC posterior as 
	$$\E_{\pi_{\ABC,\varepsilon}(\theta,y|y_\obs)}[f] = \E_{G(\theta,y)}[v f] \approx \frac 1 N\sum_i v_if(\theta_i,y_i).$$
	In general, we however do not know $C$ and $C_G$ and thus $v$ only up to normalization, and therefore need to employ the self-normalizing estimate 
	\[\E_{\pi_{\ABC,\varepsilon}(\theta,y|y_\obs)}[f] = \frac{\E_{G(\theta,y)}[wf]}{\E_{G(\theta,y)}[w]} \approx \frac{\frac 1 N\sum_i w_if(\theta_i,y_i)} {\frac 1 N \sum_i w_i} = \sum_i W_if(\theta_i,y_i)\]
	with self-normalized weights 
	\[W_i:=\frac{w_i}{\sum_jw_j},\]
	which uses another Monte-Carlo approximation for the normalization constant. It is only asymptotically unbiased, converging almost surely as $N\rightarrow\infty$.
	
	Remark: Were we instead to accept all particles irrespective of the distance, as some implementations do, we would need to regard $I[(y,y_\obs)\leq\varepsilon]$ as part of the weighting, as in that case $G(\theta,y) \propto \pi(y|\theta)g(\theta)$ s.t.\ $v(\theta,y) = C\cdot I[d(y,y_\obs)\leq\varepsilon] \pi(\theta) / g(\theta)$. In general, this interchangeability of importance and rejection sampling occurs in various places in ABC algorithms. Further note that the here presented scenario based on a uniform acceptance kernel $I[d(\cdot,y_\obs)\leq\varepsilon]$, which is the most widely used one, naturally generalizes to arbitrary acceptance kernels $K_\varepsilon(\cdot,y_\obs)$.
	
	\subsection{Effective sample size}
	
	Importance sampling allows to e.g.\ better explore high-density regions by tailored proposals. However, as the particles need to be weighted, some contribute more to estimates than others, which can be interpreted as having a lower \textit{effective sample size (ESS)} than a sample of the same size directly from the target distribution. A common definition of the ESS can be motivated as follows: Consider a linear combination
	\begin{equation*}
		S_N = \frac{\sum_{i\leq N}w_iX_i}{\sum_{i\leq N}w_i}
	\end{equation*}
	of i.i.d.\ random variables $X_i$ of variance $\sigma^2>0$, with weights $w_i\geq 0$. The unweighted mean $\frac 1 N_e \sum_{i\leq N_e}X_i$ of $N_e$ variables has variance $\sigma^2/N_e$. Equating, we obtain
	\begin{equation}\label{eq:ess}
		\frac{\sigma^2}{N_e} \overset{!}{=} \operatorname{Var}(S_N) = \frac{\sum_{i\leq N}w_i^2}{(\sum_{i\leq N}w_i)^2}\sigma^2 \Rightarrow \operatorname{ESS} := N_e = \frac{(\sum_{i\leq N}w_i)^2}{\sum_{i\leq N}w_i^2}
	\end{equation}
	as a scale-invariant quantification of the ESS of the weighted sum.
	
	\subsection{Importance sampling with multiple proposal distributions}
	
	Now assume we have multiple proposal distributions $\{g_l(\theta)\}_{l\leq L}$ instead of just one, and have generated $N_l>0$ accepted particles $\{(\theta^l_i,w^l_i)\}_{i\leq N_l}$ from each proposal, with $N = \sum_lN_l$, giving a total population $P = \{\{(\theta^l_i,w^l_i)\}_{i\leq N_l}\}_{l\leq L}$ of size $N$. This generalizes the scenario presented in the main manuscript, where we have two proposal distributions, a preliminary one and a final one.
	
	Assume $N_l/N\rightarrow\alpha_l\in[0,1]$ for $N\rightarrow\infty$. Denote the normalized importance densities $v^l(\theta,y) = \pi_{\ABC,\varepsilon}(\theta,y|y_\obs) / G_l(\theta,y) = C_lw^l(\theta,y)$ with normalization constants $C_l>0$.
	
	The goal is to for a test function $f$ define a robust Monte-Carlo estimate making use of all samples. One possible estimator could be obtained by just ignoring the fact that multiple proposals were used and just throwing all samples and weights together:
	\begin{equation}\label{eq:joint1}
		\begin{split}
			\E_{\pi_{\ABC,\varepsilon}(\theta,y|y_\obs)}[f] &\xleftarrow{N\rightarrow\infty} \frac{\frac 1 N\sum_l\sum_{i\leq N_l}w^l_if(\theta^l_i, y^l_i)}{\frac 1 N\sum_l\sum_{i\leq N_l}w^l_i} = \frac{\sum_l\frac{N_l}{N}\frac{1}{N_l}\sum_{i\leq N_l}w^l_if(\theta^l_i,y^l_i)}{\sum_l\frac{N_l}{N}\frac{1}{N_l}\sum_{i\leq N_l}w^l_i}\\
			&\xrightarrow{N\rightarrow\infty} \frac{\sum_l\alpha_l\E_{G_l(\theta,y)}[w^lf]}{\sum_l\alpha_l\E_{G_l(\theta,y)}[w^l]} = \frac{\sum_l\alpha_lC_l^{-1}\E_{\pi_{\ABC,\varepsilon}(\theta,y|y_\obs)}[f]}{\sum_l\alpha_lC_l^{-1}}\\
			&= \E_{\pi_{\ABC,\varepsilon}(\theta,y|y_\obs)}[f].
		\end{split}
	\end{equation}
	Equation \eqref{eq:joint1} shows that we obtain an asymptotically unbiased estimate, however the subpopulations are weighted by the normalization constants $C_l$, which is in general not meaningful, unless these carry an appropriate interpretation.
	Further, this joint estimator does not account for the difference in importance estimator quality of the various $G_l$.
	
	Instead of \eqref{eq:joint1}, we suggest to first consider each subpopulation separately, and obtain a joint estimate as a weighted sum of estimators
	\begin{equation}\label{eq:joint2}
		\begin{split}
			\E_{\pi_{\ABC,\varepsilon}(\theta,y|y_\obs)}[f] &= \sum_l\beta_l\E_{G_l(\theta,y)}[v^lf]
			= \sum_l\beta_l\frac{\E_{G_l(\theta,y)}[w^lf]}{\E_{G_l(\theta,y)}[w^l]}\\
			&\xleftarrow{N\rightarrow\infty} \sum_l \beta_l\sum_{i\leq N_l}W^l_if(\theta^l_i,y^l_i)
		\end{split}
	\end{equation}
	with coefficients $\beta_l$ s.t.\
	\begin{equation}\label{eq:beta}
		\sum_l\beta_l=1,
	\end{equation}
	and
	\[W^l_i := \frac{w^l_i}{\sum_{j\leq N_l} w^l_j}\]
	subpopulation-wise self-normalized weights.
	
	\subsection{Subpopulation weighting}
	
	The $\beta_l$ are free parameters and should be chosen to yield a robust estimator. A straightforward choice is
	\begin{equation}\label{eq:beta1}
		\beta_l = \alpha_l\approx \frac{N_l}{N},
	\end{equation}
	i.e.\ to normalize the contribution of each proposal by the number of samples generated from that proposal.
	
	As the proposal quality can vary, this weighting may not be ideal. Instead of \eqref{eq:beta1}, we suggest to choose the $\beta_l$ to e.g.\ maximize the overall ESS, in order to obtain a robust estimator. Here, \eqref{eq:ess} takes the shape
	\begin{equation}\label{eq:ess_multiprop}
		\operatorname{ESS} = \frac{(\sum_l\sum_{i:l}\beta_lW^l_i)^2}{\sum_l\sum_{i:l}(\beta_lW^l_i)^2} = \frac{1}{\sum_l\beta_l^2\sum_{i\leq N_l}(W^l_i)^2}.
	\end{equation}
	Denote for short $Q_l:=\sum_{i\leq N_l}(W^l_i)^2$, the diagonal matrix $Q:=\operatorname{diag}(Q_1,\ldots,Q_L)\in\R^{L\times L}$, and the column vectors $\beta:=(\beta_1,\ldots,\beta_L)^T\in\R^L$ and $\mathbbm{1}:=(1,\ldots,1)^T\in\R^{L}$. Without loss of generality, assume that $Q_l>0$ for all $l$, otherwise just set $\beta_l=0$. Then, maximizing \eqref{eq:ess_multiprop} subject to \eqref{eq:beta} is equivalent to the quadratic form with linear equality constraint
	\begin{equation}\label{eq:quadform}
		\begin{matrix*}[l]
			\text{minimize} & \beta^TQ\beta \\
			\text{subject to} & \mathbbm{1}^T\beta = 1.
		\end{matrix*}
	\end{equation}
	Evaluating $\nabla_{\beta,\lambda}[\beta^TQ\beta + \lambda(\mathbbm{1}^T\beta - 1)]$, using a Lagrange multiplier $\lambda\in\R$, readily shows that the unique solution of \eqref{eq:quadform} is given by the positive-definite linear system
	\begin{equation*}
		\begin{pmatrix}2Q & \mathbbm{1} \\ \mathbbm{1}^T & 0\end{pmatrix} \cdot
		\begin{pmatrix}\beta \\ \lambda\end{pmatrix} = \begin{pmatrix}0 \\ 1\end{pmatrix} \Leftrightarrow
		\begin{pmatrix}\beta\\\lambda\end{pmatrix} = 
		\begin{pmatrix}
			\frac{Q^{-1}\mathbbm{1}}{\mathbbm{1}^TQ^{-1}\mathbbm{1}}\\
			\frac{-2}{\mathbbm{1}^TQ^{-1}\mathbbm{1}}.
		\end{pmatrix}
	\end{equation*}
	Written out, we obtain
	\begin{equation*}
		\beta_l = \frac{\frac{1}{\sum_{i\leq N_l}(W^l_i)^2}}{\sum_k\frac{1}{\sum_{i\leq N_k}(W^k_i)^2}}
		= \frac{\operatorname{ESS}(\{W^l_i\}_{i\leq N_l})}{\sum_k\operatorname{ESS}(\{W^k_i\}_{i\leq N_k})},\quad l=1,\ldots,L,
	\end{equation*}
	i.e.\ the overall ESS is maximized by setting the contribution of sub-population $l$ proportional to its ESS, which intuitively makes sense.
	
	\section{Implementation}
	
	\subsection{Algorithm implementation in pyABC}
	
	The ABC-SMC algorithm and all samplers used and developed in this work have been implemented and made available as part of the python-based open-source package pyABC (\url{https://github.com/icb-dcm/pyabc}). The sampler implementation uses the \textit{Redis} package (\url{https://redis.io}) as broker between main process and workers on a distributed high-performance computing (HPC) architecture.
	
	Before the analysis, a Redis server is set up. To this server the desired number of workers is then connected. Workers can also be added interactively during the analysis. In addition, there is a process for the main analysis, the ABC-SMC algorithm, which submits in each generation simulation tasks to the server, which broadcasts them to the workers and collects results, to be read in again by the main process. The whole process of environment setup and analysis has been fully automated for HPC infrastructure.
	
	Simulation tasks are broadcast by the server via dedicated port channels, which workers listen on if they are idle. Workers return simulation results via a queue, which is continually collected from by the main process. Start times are tracked via shared variables to ensure the main process is aware which tasks need to be waited for.
	
	\subsection{Technical specifications}
	
	For our tests we used Anaconda 3, Python 3.8, with package versions pyABC 0.12.07, tumor2d 1.0.0, Morpheus 2.2.0 \cite{StarrussBac2014}.
	Analyses of model T1 was performed on the Juelich Supercomputing Center (JSC), Juwels cluster, standard compute nodes, specification of which are 2 $\times$ Intel Xeon Platinum 8168 CPU, 2 $\times$ 24 cores, 2.7 GHz 96 ($12 \times 8$) GB DDR4, 2666 MHz.
	
	Analyses of model T2, M1, and M2 were performed on the Bonna cluster at the University of Bonn. A standard node has the specification of 2 $\times$ 16 cores, 6 GB RAM per core, and 2x480GB SSD.
	For the different models, we used 1-9 nodes for T1 and 1-8 nodes for T2, M1 and M3.
	
	\section{Test models}
	
	To test the correctness and the performance of the new concept, we used 4 models as described in the Main Manuscript, Table 1, which we describe in further detail in this section.
	First, we used LA on some basic toy models (T1 and T2) to examine the general behavior and some specific properties.
	Later, we also performed tests on more complex application examples (M1 and M2) to test the performance in realistic settings.

	\subsection{(T1) Unbalanced Modes}
	
	Usually, the run-time of a single simulation will not be independent of the parameter candidate but instead, be somehow correlated to the region of the parameter space.
	High variance of the run-time paired with correlation between the simulation time and the candidate could possibly lead to a strong imbalance in the preliminary proposal.
	This raises the question of what happens when it is likely that the preliminary population is strongly biased.
	
	To analyze this phenomenon, we constructed a two-mode scenario by considering a model that simply squares its only parameter, i.e. $y=\theta^2$ and then adds additive noise to that.
	Assuming the observed data consist of one observation with value 1 and that the prior is uniformly distributed on $[-2,4]$, we get two parameters that could equally be true, $-1$ and $+1$.
	We added a shorter idle time sampled from a log-normal distribution with mean 0.05 and $\sigma^2$=0.025 to the negative mode and a larger one (mean=1 and $\sigma^2$=0.5) to the positive mode.
	Given a large number of parallel workers and a comparably small population sizes, it is thus possible that the preliminary population $\hat P_{t-1}$ consists entirely of samples from the negative side.
	Depending on the shape of the thus created preliminary proposal $\tilde g_t$, it is then possible that it only proposes samples from the negative side. It can happen that enough of these get accepted, such that the subsequent accepted population $P_t$ consists only of samples from the negative side.
	
	\subsection{(T2) ODE Model}
	
	While ODE based models are not the main application area of ABC, it is fairly simple to create an example for which we can easily verify the correct result.
	Parameter inference for such a model is also not as time intensive as real life application examples.
	Therefore, we can perform it several hundred times with the new concept and compare the results to the ones of established approaches, making this a good test model to perform sanity checks on the new algorithm.
	
	Our basic ODE-model with two parameters describes the inter-conversion between two species $x_1, x_2$, with rates $\theta_1,\theta_2$.
	It is given by
	
	\begin{align}
		\frac{d}{dt}x_1 =& -\theta_1 x_1+\theta_2 x_2\\
		\frac{d}{dt}x_2 =& \phantom{-}\theta_1 x_1 -\theta_2 x_2.
	\end{align}
	
	Using this ODE, we can immediately verify that the results converge towards the same value as the original version for which the asymptotical correctness is known.
	In order to not have a fully deterministic system and to consider that in reality measurements are not perfectly accurate, we added multiplicative normal noise $\sim\mathbb{N}(1,0.03)$ to each model evaluation.
	
	To analyse the run-times and the effect of heterogeneous simulation times, we used the same ODE and added some idle time to each model call. This idle time was chosen in a way that imitates some single simulations taking vastly longer than the average value.
	As distribution for $t_{idle}$ we employed a log-normal one $t_{idle}\sim \operatorname{lognormal}(\mu_n,\sigma_n^2)$, where $\mu_n,\sigma_n^2$ are the mean and variance of the underlying normal distribution.
	$\mu_n,\sigma_n^2$ were chosen such that the real variance $\sigma^2$ takes values between 0.25 and 4 and the real mean $\mu$ is constantly equal to 1 for the different values of $\sigma^2$, i.e.
	\begin{align}
		\sigma_n^2 = 2\log\left(\frac{\sqrt{1^2+\sigma^2}}{1}\right),\ \ \mu_n = \log(1)-\frac{\sigma_n^2}{2}.
	\end{align}
	
	We used as prior a uniform distribution on the unit square $\pi\sim U([0,1]^2)$, initial values $(x_1,x_2)(0) = (1,0)$, and as true parameters $\theta_1=e^{-2.5}\approx0.0821,\ \theta_2=e^{-2}\approx0.135$, so here a point close to one corner of the prior distribution.
	As observed data $y_{obs}$ we considered the trajectory of species $x_2$ at time points $0,1,\ldots,10$.
	As distance function we used an $l_1$ norm.
	Candidates are accepted if the distance between their simulated data and the observation falls below a fixed $\varepsilon$-threshold for each generation, with pre-defined thresholds of $\varepsilon_1,...,\varepsilon_8=8,4,2,1,0.7,0.5,0.33,0.25$. 
	
	\subsubsection*{Results -- Correctness}
	Using (T2) we generated 13 replicas of different configurations of both using DYN and LA scheduling for each population size on 32 to 256 workers.
	
	Both scheduling approaches seem to return equal results (see Figure 4) when taking into account that any two runs will always slightly differ due to the innate random nature of the sampling, the noise and the run-time. 
	
\begin{figure}[H]
		\centering
		\includegraphics[width=0.8\textwidth]{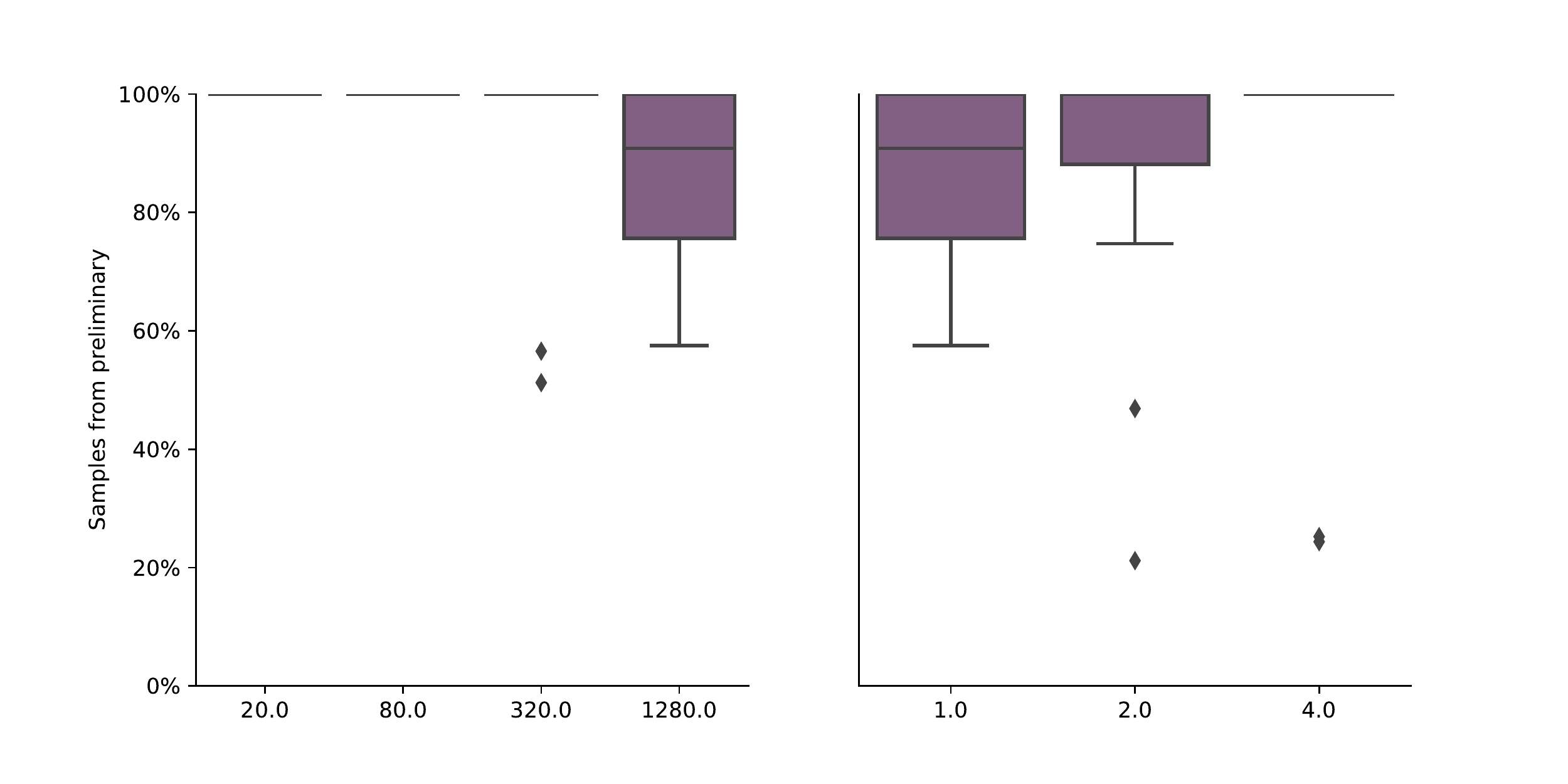}    
		\includegraphics[width=0.8\textwidth]{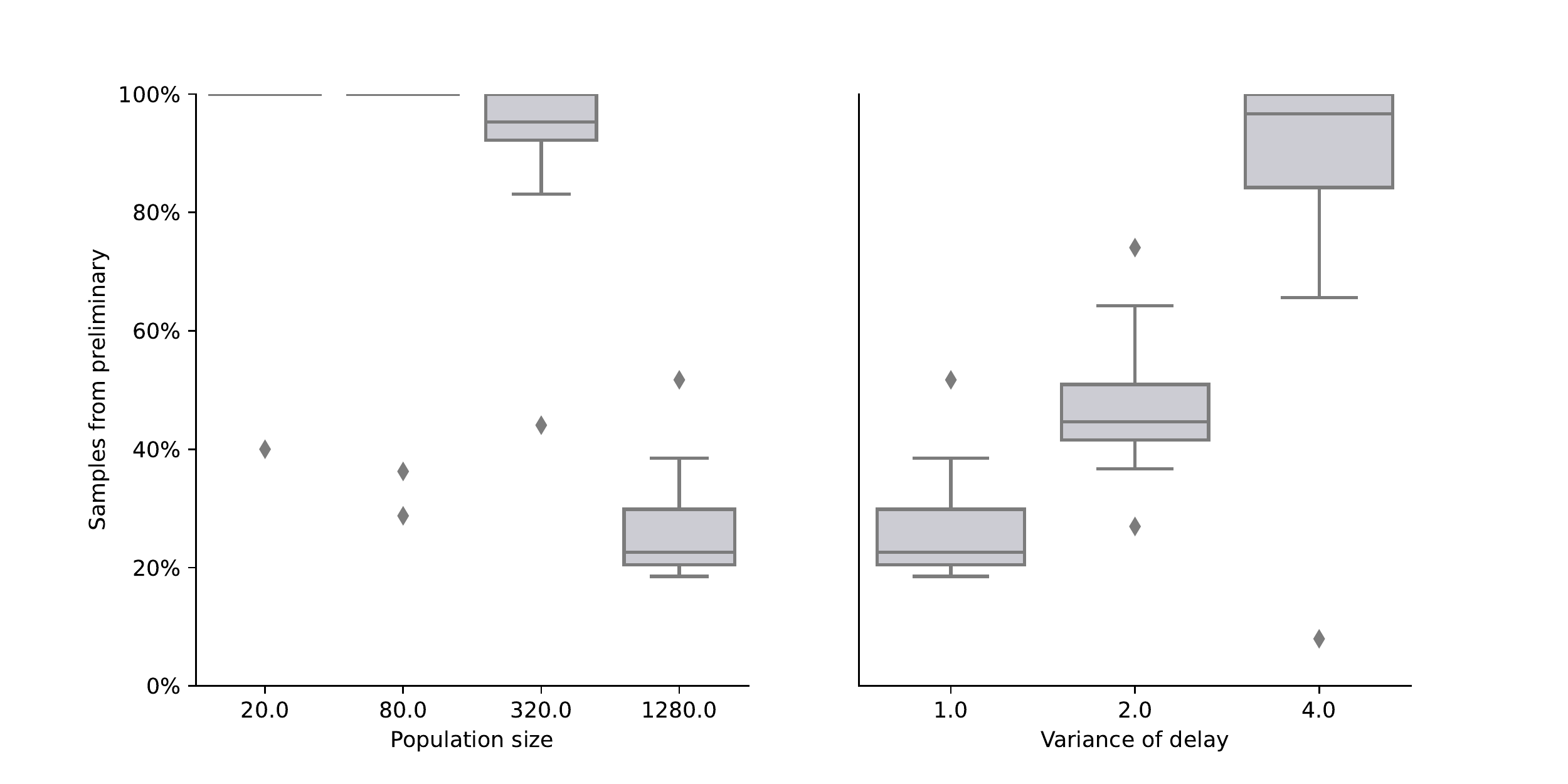}    
		\caption{Fraction of particles sampled from the preliminary proposal in the last generation (for $\sigma^2=1$ (left) and $N=1280$ (right)) for LA Pre (upper row) and LA Cur (second row).}.
		\label{fig:ODEprelfrac}
	\end{figure}

	(A) demonstrates how the mean for both DYN and LA converges to the same value as we increase the population size.
	At the same time, (B) shows that also the standard deviation within individual distributions decreases, meaning that, as $N$ increases, both posteriors peak more sharply at the same value.
	(C) makes it clear that a stronger variances for $t_{idle}$ does not affect the mean, even though there is a larger fraction of preliminary particles expected.
	In (D), one can see that LA does not seem to substantially affect the effective sample size either even though we here used $N=20$, which, on 256 workers, makes it likely that several generations are sampled completely from the preliminary.
	In E and F, we can see the fraction $\tilde N / N$ of accepted samples in the final population at $t=n_t$ that were generated using the preliminary proposal $\tilde g_{n_t}(\theta)$. 
	In G, we can see a comparison of the posterior distribution of the different scheduling. One can appreciate the nice agreement in the shape of the posterior from different scheduling strategies. 
	By showing that the preliminary and the final proposal both considerably contribute to the posterior distribution, we get a strong indication that everything works as intended.
	And indeed, in many of the relevant scenarios the amount of particles that were based on the preliminary in the last generation ends up making up roughly half of the population, which is returned as output by the ABC-SMC algorithm (see Figure~\ref{fig:ODEprelfrac}).
	
	In Figure~\ref{fig:ODEprelfrac}, one can also observe how the fraction of particles sampled from the preliminary decreases as the population size increases.
	This makes sense, since keeping the number of workers constant results in a more or less constant amount of resources used for the preliminary sampling between the $N$-th acceptance in a generation and the last worker to finish working on that generation.
	So, there should be a roughly constant number of preliminary acceptances opposite to an increasing size of the total population.
	
	Similarly, a higher run-time variance increases the fraction of preliminary particles, as this results in more time between the $N$-th acceptance and the last simulation of a generation.
	
	\subsubsection*{Results -- Run-Time}
	
\begin{figure}[H]
		\centering
		\begin{minipage}{\textwidth}
			\includegraphics[width=\textwidth]{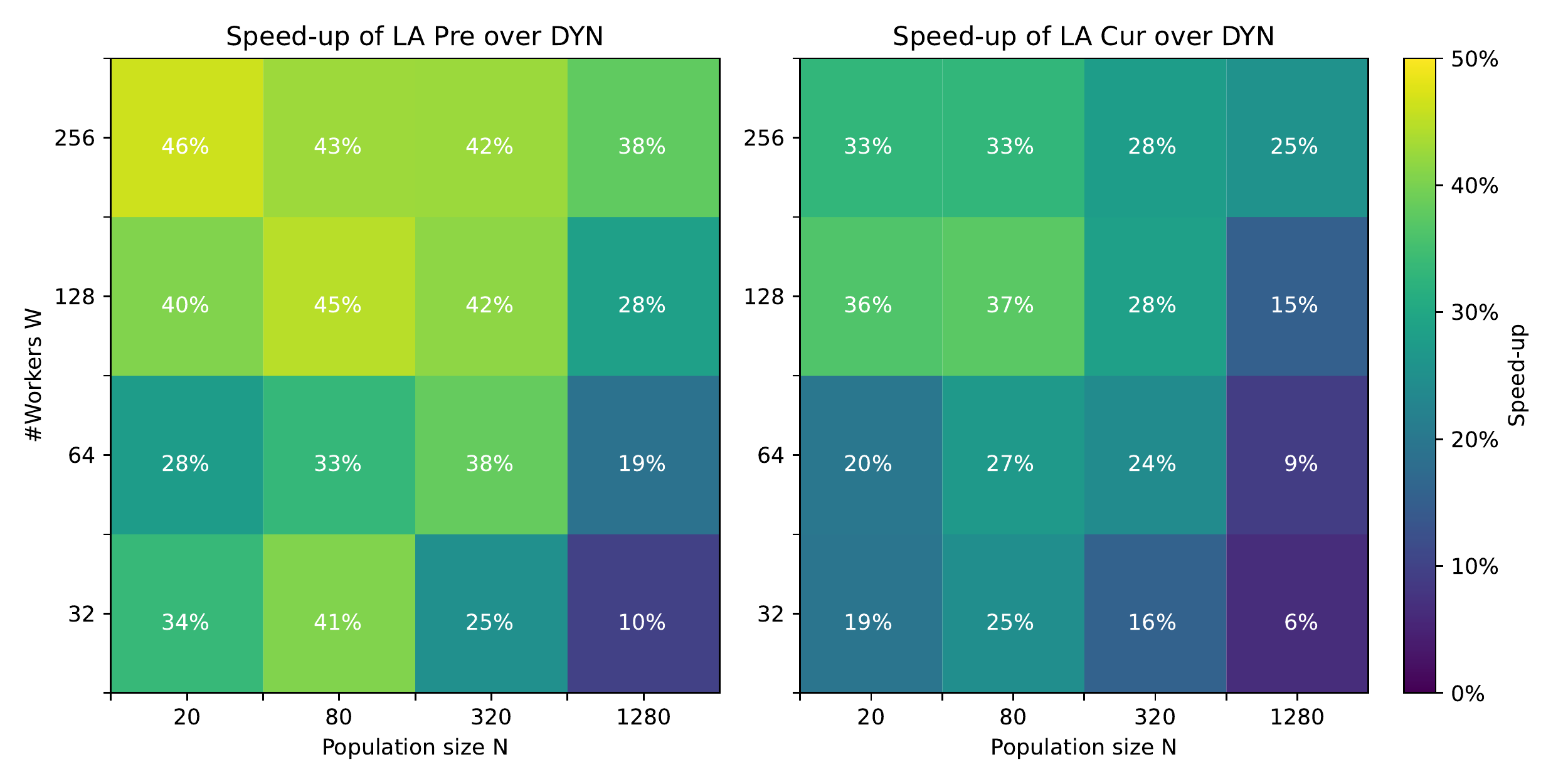}
		\end{minipage}
		\caption{The speed-up for LA Pre (left) and LA Cur (right) with $\sigma^2 = 2$}
		\label{fig:ODESpeedVar2}
	\end{figure}

	To analyze how much effect the new scheduling can have on the wall time, we ran the same model on different population size to worker ratios.
	First, we used an idle time with variance $\sigma^2=1$, the same population sizes as above, i.e. between 20 and 1280 and ran each of those on 1, 2, 4, and 8 nodes with 32 workers each, so on 32 to 256 workers.
	For a run-time variance of $\sigma^2=1$, we started with 13 repetitions of each scheduling approach (Except for STAT, where we used only 3 repetitions for time-preserving purposes) for each population size on 256 workers (see Figure 5).
	For $\sigma^2=2$, run-times were generally higher (see Figure~\ref{fig:ODESpeedVar2}).
	
	We can observe a substantial speed-up when using LA instead of DYN scheduling (see Figure 5 and \ref{fig:ODESpeedVar2}), whenever the population size is about as large or slightly larger than the amount of workers.
	In the best case, the LA approach decreased the wall time by nearly a factor of two when compared to the dynamic scheduling runs.
	On the other hand, the difference in efficiency is much less apparent, or even almost trivial, in case the two factors, population and worker size, are vastly different.
	
	In the case where the amount of workers is substantially larger than our population size, only few simulations remain for each worker.
	This can go as far as even having enough preliminary acceptances to complete the next generation even before the previous one is finished leaving all further simulations retrospectively useless, which results in LA scheduling also having a poor parallel efficiency.
	Nonetheless, in that case there is still a substantial acceleration when compared to established approaches as LA can almost complete two generations in the time it takes to complete one using DYN.
	
	When the population size is multiple times larger than the amount of available workers, dynamic scheduling already performed very well, leaving little room for improvement.
	However, even in that scenario we do not have any changes to the negative, so any additional computation necessary to enable the sampling from the preliminary population appears always worth the effort.
	
	Further, using the same setup, we examined the effect of the run-time variance of the single evaluations on the speed-up.
	For that we ran the same tests as before with an idle time variance of $\sigma^2=2$; and indeed results (see Figure~\ref{fig:ODESpeedVar2}) seem to indicate that the achieved acceleration is even slightly higher than for the less varying run-time with $\sigma^2=1$.
	
\begin{figure}[H]
		\centering
		\includegraphics[width=0.8\textwidth]{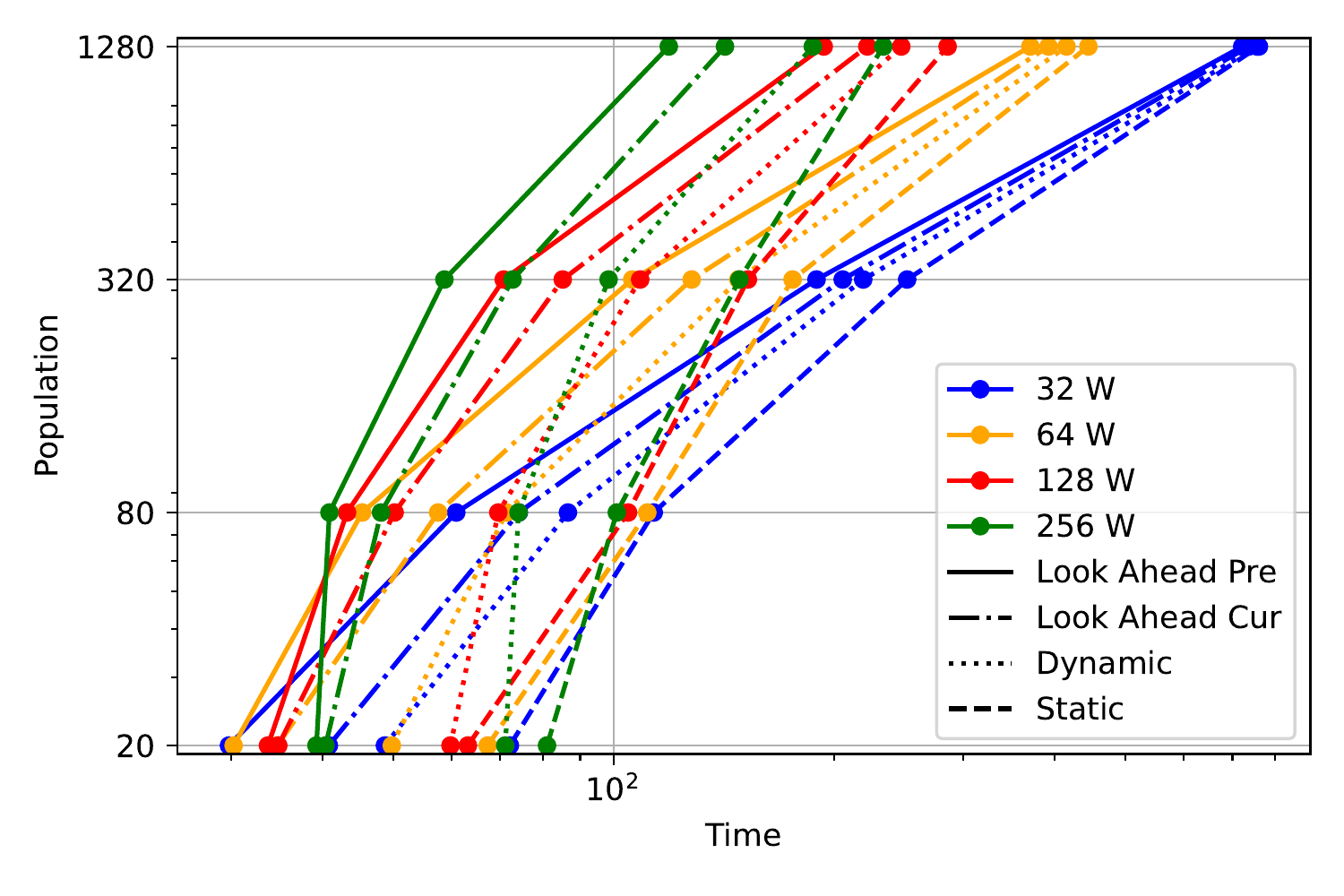}
		
		\caption{The time taken by different scheduling methods on different population sizes and numbers of workers.}
		\label{img:T2Cores}
	\end{figure}

	Finally, both DYN and LA scheduling scale better than STAT when using more computational resources (See Figure~\ref{img:T2Cores})
	
	\subsection{(M1) Tumor Growth Model}
	
	\begin{minipage}{0.6\textwidth}
		The tumor growth model (M1) is our first test instance based on a real life example.
		The model was developed mainly in \cite{jagiella2012} and aims to describe the growth of a tumor spheroid while taking into account its spacial structure.
		For example, the proliferating cells are almost exclusively the ones in the outer rim of the spheroid whereas the ones contained in the core are mostly necrotic.
		
		As hybrid discrete-continuous model it uses different mathematical tools trying to accurately depict a real life biological system. 
	\end{minipage}
	\begin{minipage}{0.05\textwidth}
		\hfill
	\end{minipage}
	\begin{minipage}{0.32\textwidth}
		\begin{figure}[H]
			\centering
			\includegraphics[width=0.85\textwidth]{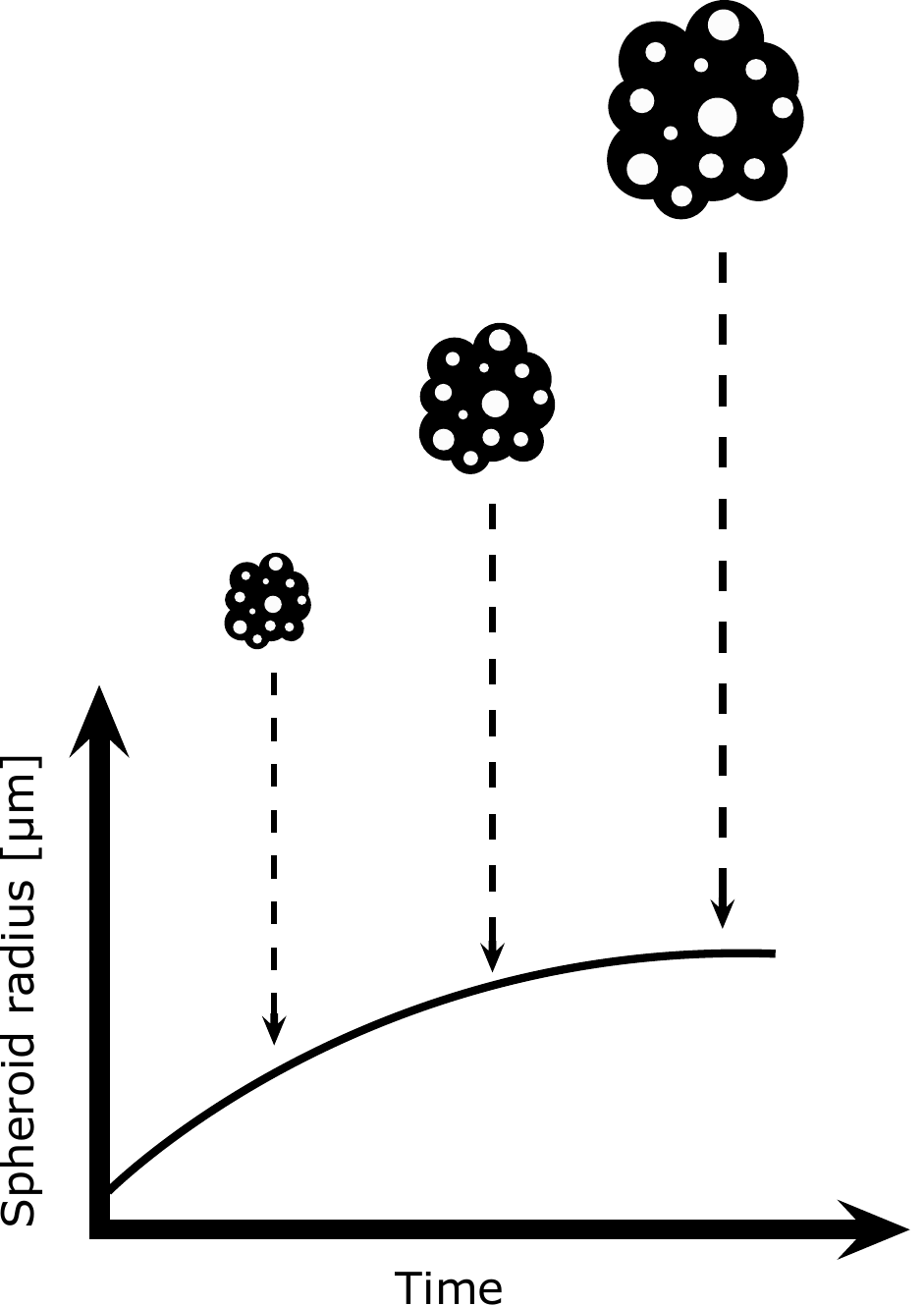}
		\end{figure}
	\end{minipage}
	\smallskip
	
	An agent-based approach with stochastic interactions between separate particles in a system, is used to model the relations between the individual cells, while a system of PDEs describes the extracellular matrix.
	At the same time, mechanisms like cell division and cell death are modeled using a continuous time Markov process.
	For further details about the model see \cite{jagiella2012, JagiellaRic2017} and for some biological background see e.g. \cite{CarverMin2014, KwapiszewskaMic2014}.
	
	The version we used as test instance is a two dimensional implementation of the tumor growth model with seven parameters.
	It is available in the python package \textit{tumor2d} (\url{https://github.com/ICB-DCM/tumor2d}).
	Even using several hundred workers and a population size of e.g.\ 1000, the parameter inference for the tumor growth model takes hours to days, making it close to impossible to run the simulation on anything but large-scale parallelized infrastructure and strictly necessary to employ an efficient workload distribution scheme.
	Yet, the run-time is not yet as exorbitantly high as for some other models, such that it was possible to perform several runs to obtain at least some reliability in the results.
	This was particularly important as the trajectories of the tumor growth model underlie stochastic fluctuations, especially in the early phases, as there the number of cells is still very limited. 
	Towards the end, the cell numbers are far higher and the fluctuations are much less severe due to the averaging effect of the Law of Large Numbers.
	
	\subsubsection*{Results -- Correctness}
	
	We ran model M1 with population sizes ranging from 250 to 1000, employing two different amounts of workers 128 and 256.
	The acceptance criteria for candidates of a generation are adaptively chosen after the previous one finishes as the 50\% quantile of the previous accepted distances. 
	The run finishes after a generation $t$ in which the threshold $\varepsilon_t$ falls below a certain value, in our runs $\varepsilon_{n_t}\leq700$.
	Every time one scenario was run, one instance with dynamic scheduling, one with LA Pre, and one with LA Cur scheduling was performed with the same setup to directly compare the results.
	
	During the runs, no substantial deviations in the quality of results were observed (see Figure~\ref{img:M1_CI} and Figure~\ref{fig:TumorPosteriors}; detailed results of the other runs can be found in the supplementary code).
	
	\begin{figure}[H]
		\centering
		\includegraphics[width=1\textwidth]{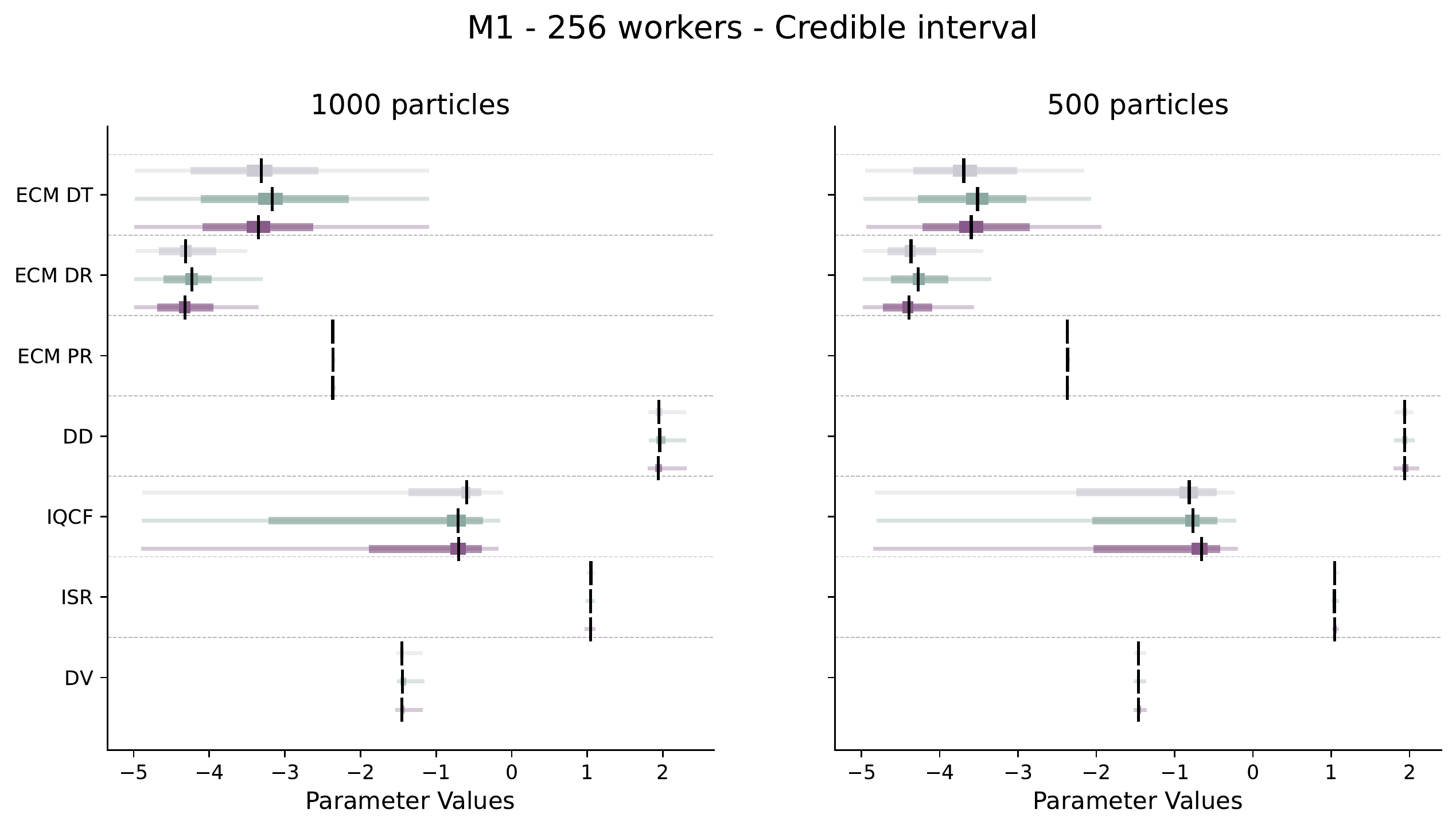}
		\includegraphics[width=1\textwidth]{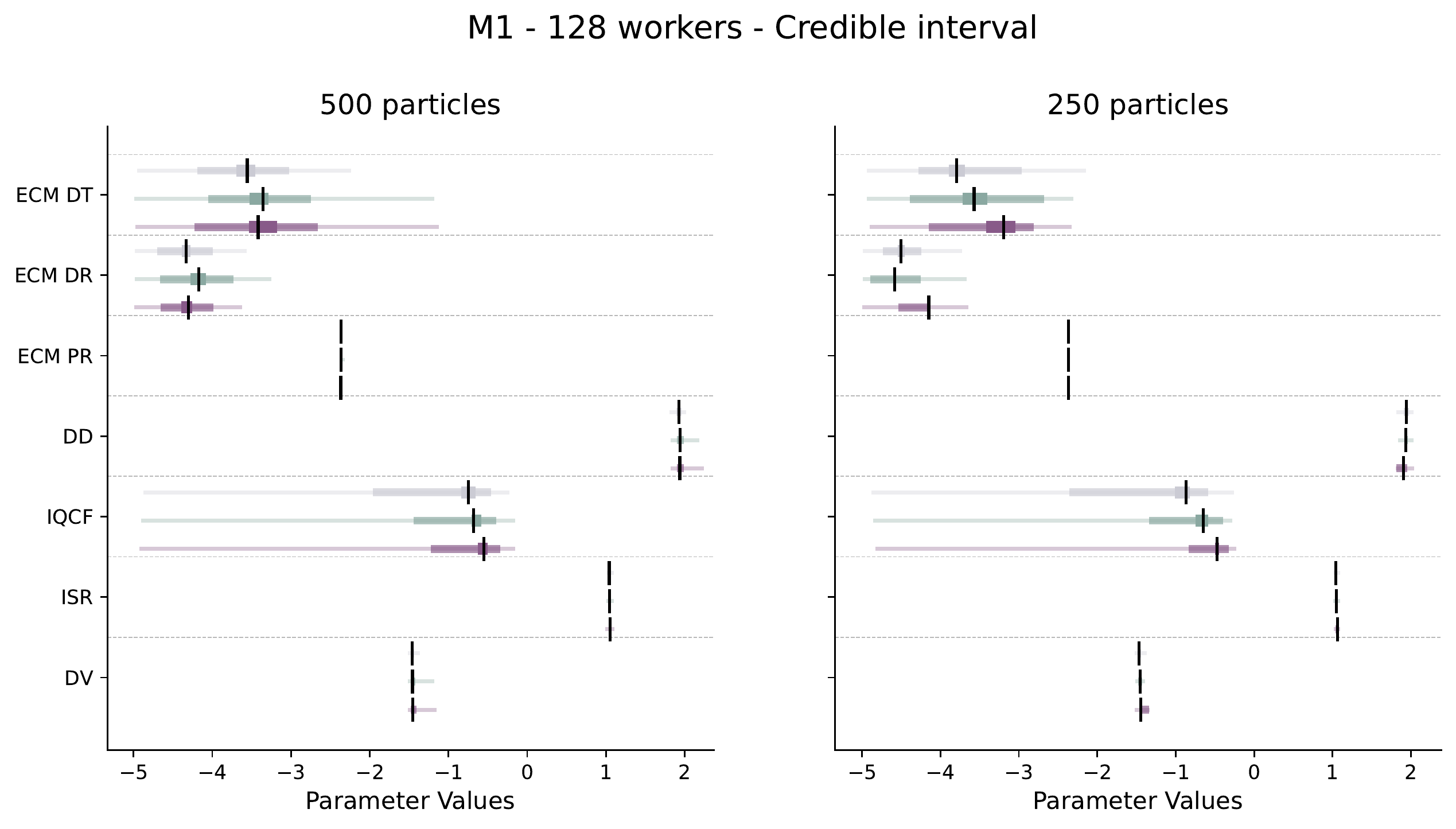}
		
		\caption{The credible interval of the model M1 with population size 1000, 500, 250 on 128 and 256 workers.}
		\label{img:M1_CI}
	\end{figure}
	
	\begin{figure}[H]
		\centering
		\includegraphics[width=0.8\textwidth]{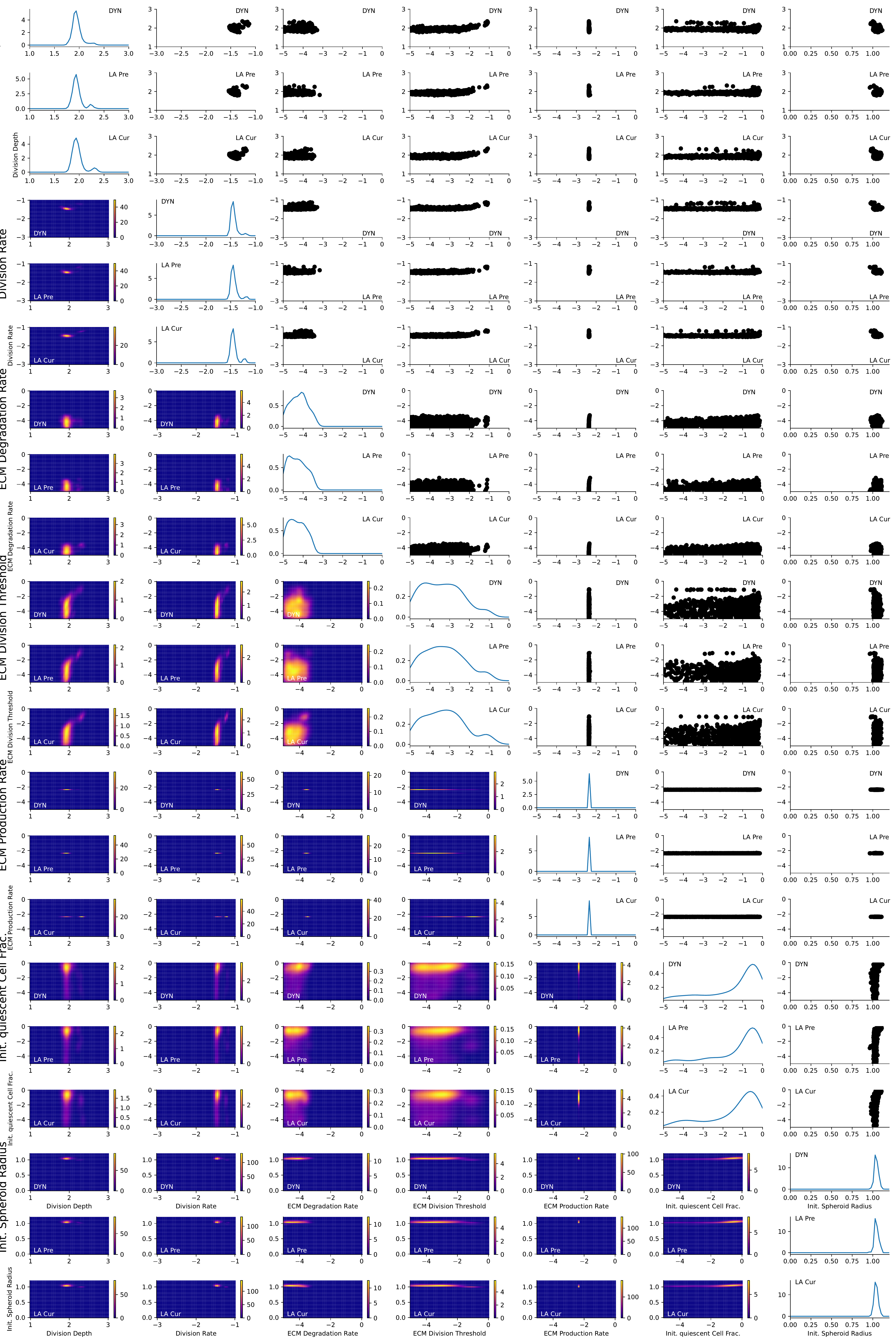}
		\caption{Direct comparison of the posterior distributions for all seven parameters of (M1) for a run with $N=1000$ on $W=256$ workers using DYN, LA Pre, and LA Cur scheduling. The parameters with sharp peaks are visibly at the same location and if the inference returned a broader distribution, it did so in all cases.}
		\label{fig:TumorPosteriors}
	\end{figure}
	
\begin{figure}[H]
		\centering
		\includegraphics[width=0.7\textwidth]{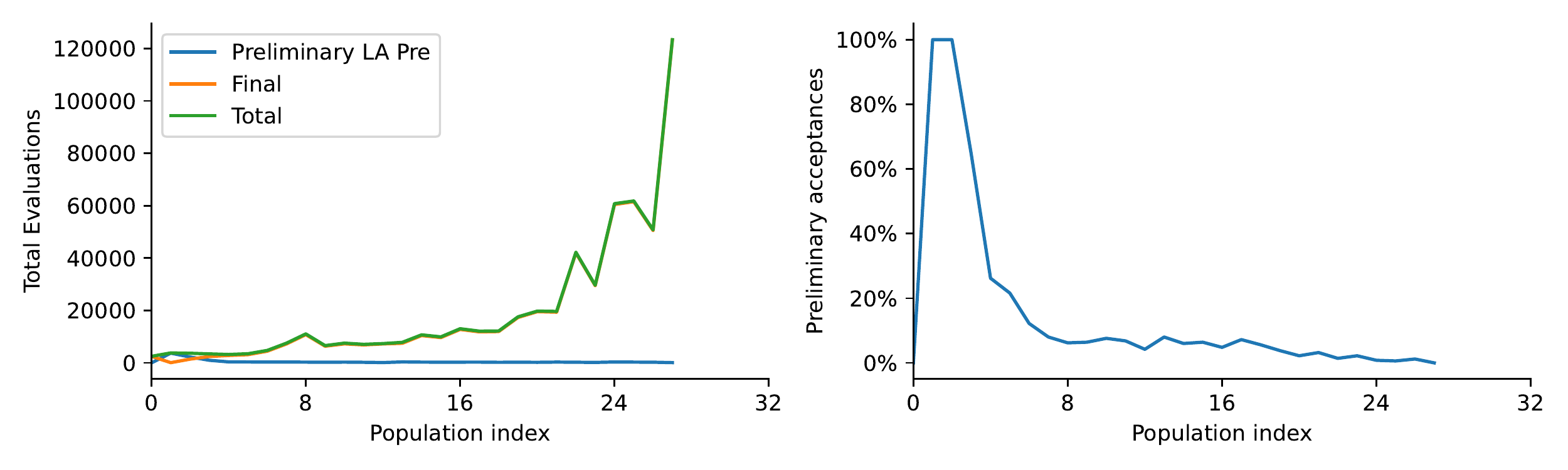}
		\includegraphics[width=0.7\textwidth]{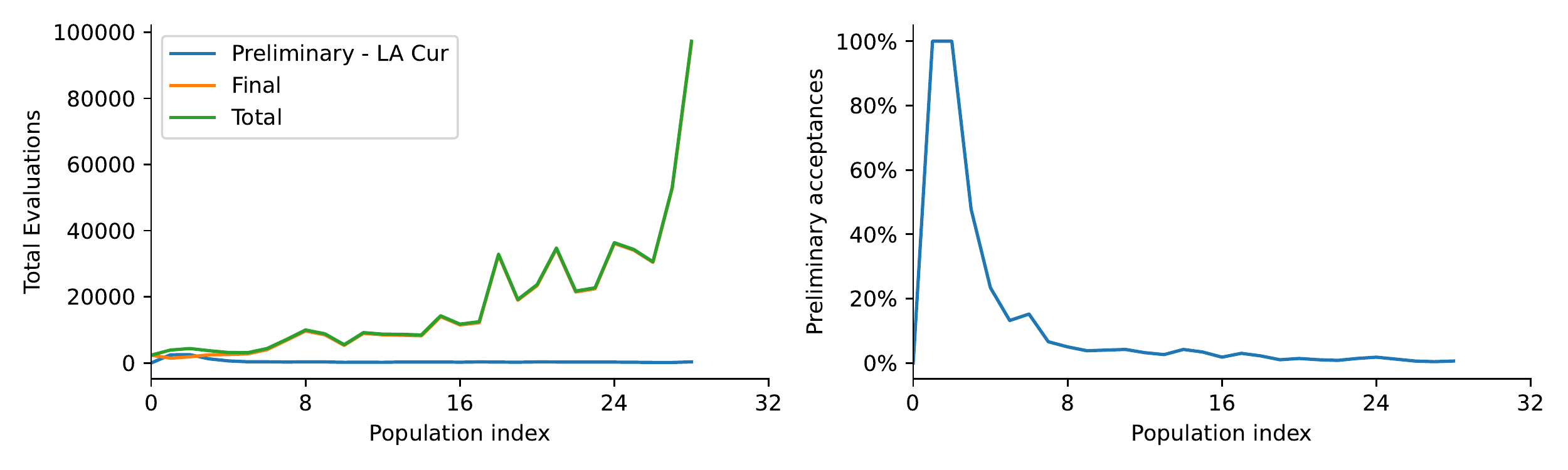}
		
		\caption{Total evaluations and fraction of accepted particles of M1 based on the preliminary population in each generation in the LA Pre (top) and LA Cur (bottom) runs of the same run as in Figure~\ref{fig:TumorPosteriors}}
		\label{fig:TumorPrelAcc}
	\end{figure}

	While the fraction of preliminary particles tends to decrease over the course of the inference, it varies strongly from generation to generation (see Figure~\ref{fig:TumorPrelAcc}).
	As the time between the $N$-th acceptance in generation $t$ and the last worker to finish is expected to be more or less constant, we have a similar amount of evaluations from the preliminary proposal in each generation.
	Generally, the acceptance rate decreases with the epsilon threshold however, and this also holds for the preliminary proposal based candidates. 
	So, as more total evaluations are necessary to reach the desired number of accepted particles $N$, a smaller fraction of those should end up being sampled from the preliminary as the generations progress and the acceptance rate decreases.

	\subsubsection*{Results -- Run-Time}
	\label{sec:TumorRT}
	
	For more complex models, it is usually far more effective to use the adaptive epsilon schedule mentioned in the previous section.
	That however, leads to a different final acceptance threshold in every run, which strongly affects the run-time.
	Together with the very stochastic start of the tumor growth model, this yields a high variance of the wall times, making a direct comparison more difficult. 
	Nonetheless, we can for example observe the time point after which each epsilon value is achieved.
	
\begin{figure}[H]
		\centering
		\begin{minipage}{0.3\textwidth}
			\includegraphics[width=\textwidth]{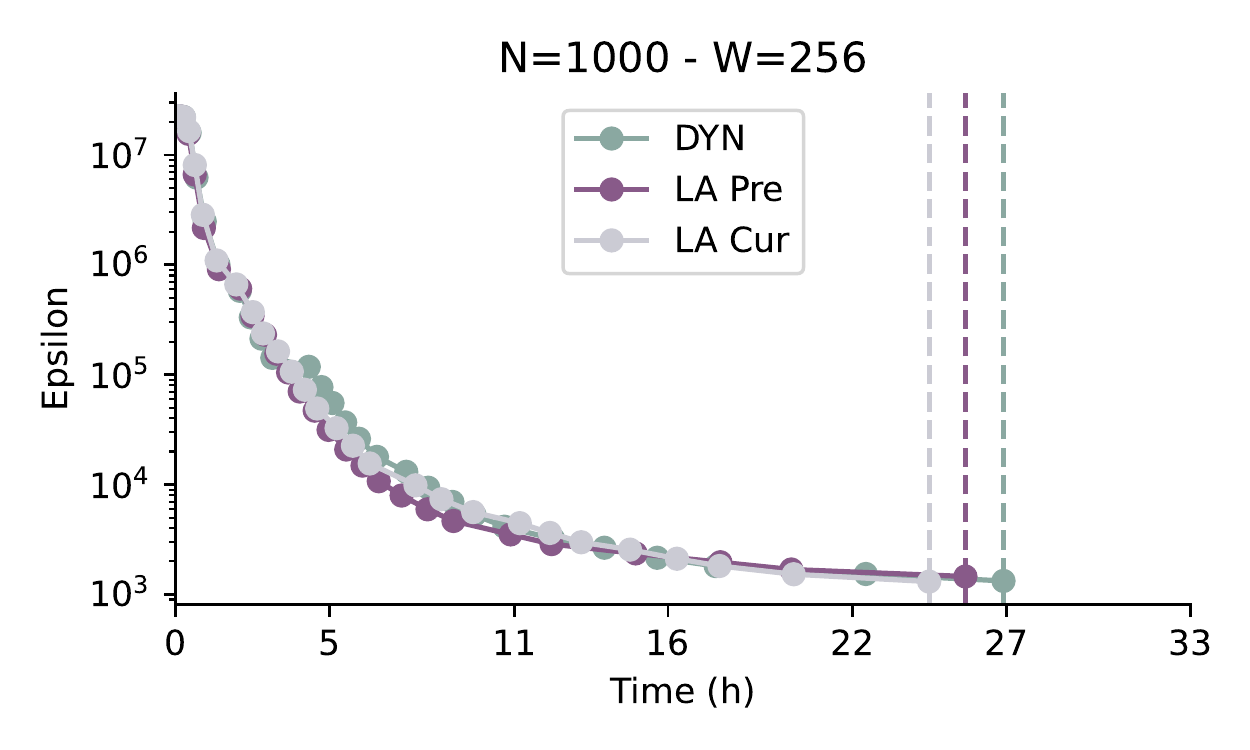}
			\includegraphics[width=\textwidth]{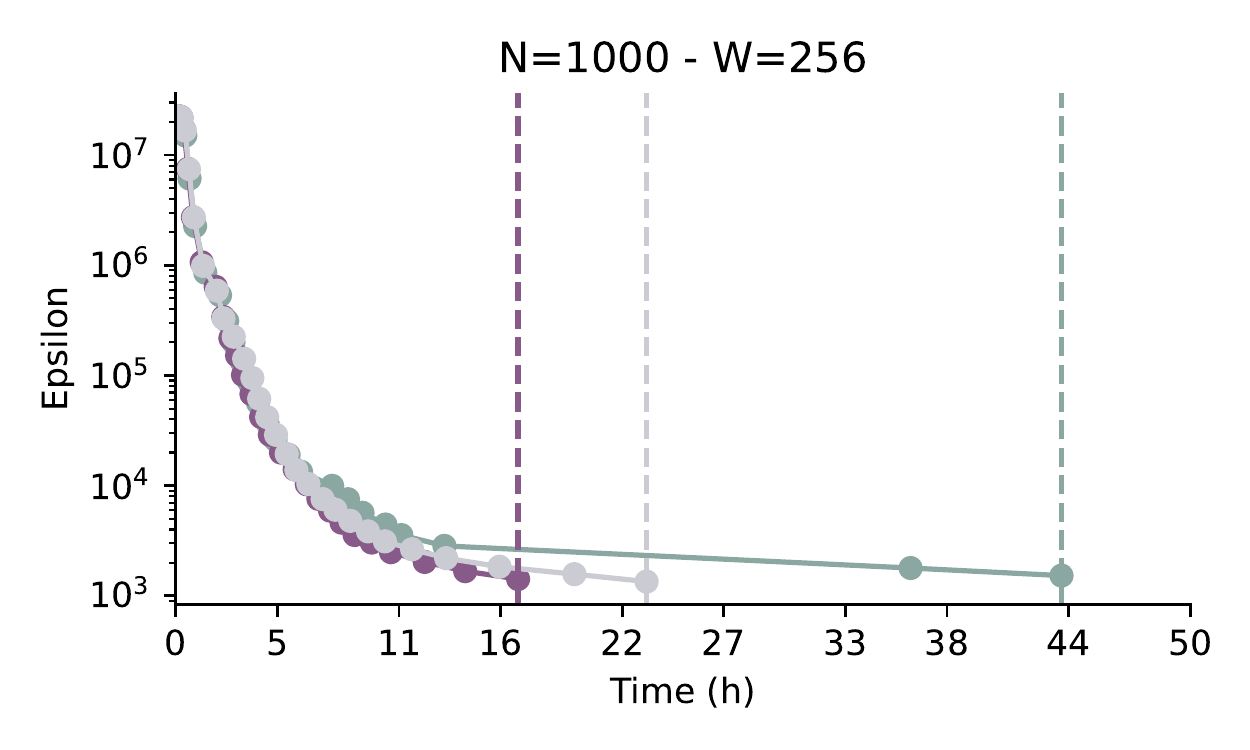}
			\includegraphics[width=\textwidth]{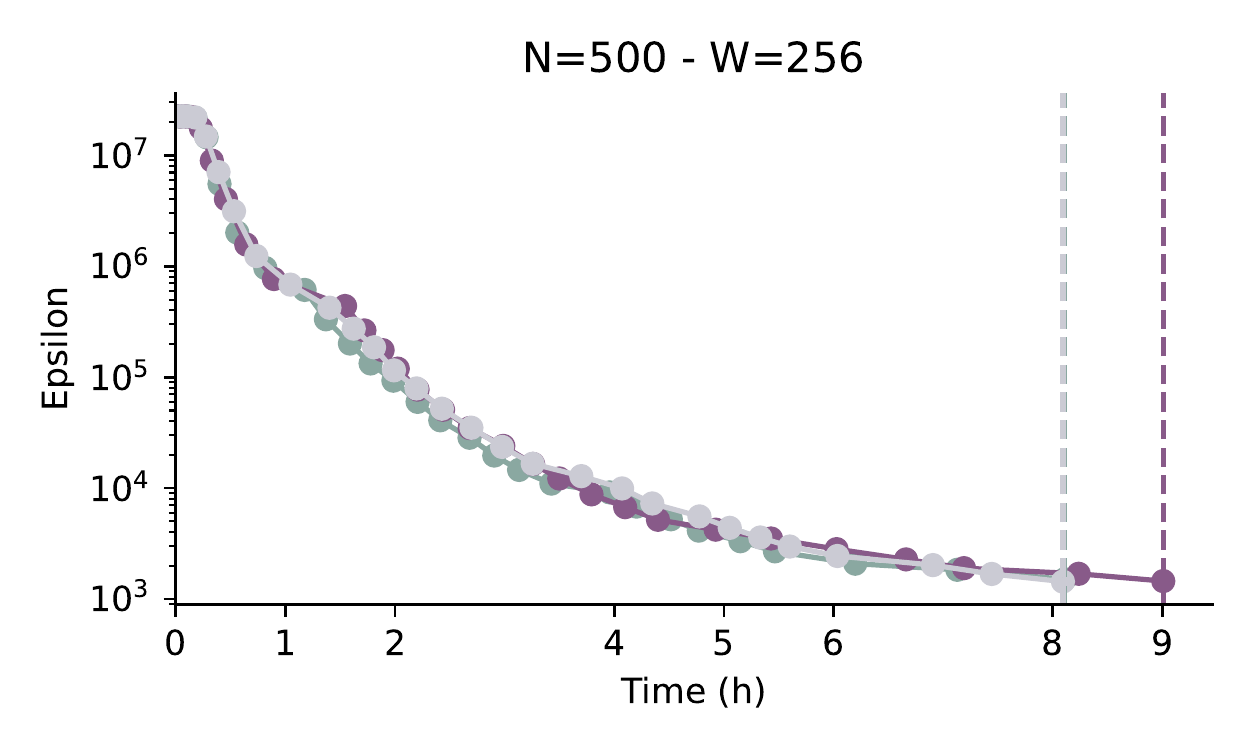}
		\end{minipage}
		\begin{minipage}{0.3\textwidth}
			\includegraphics[width=\textwidth]{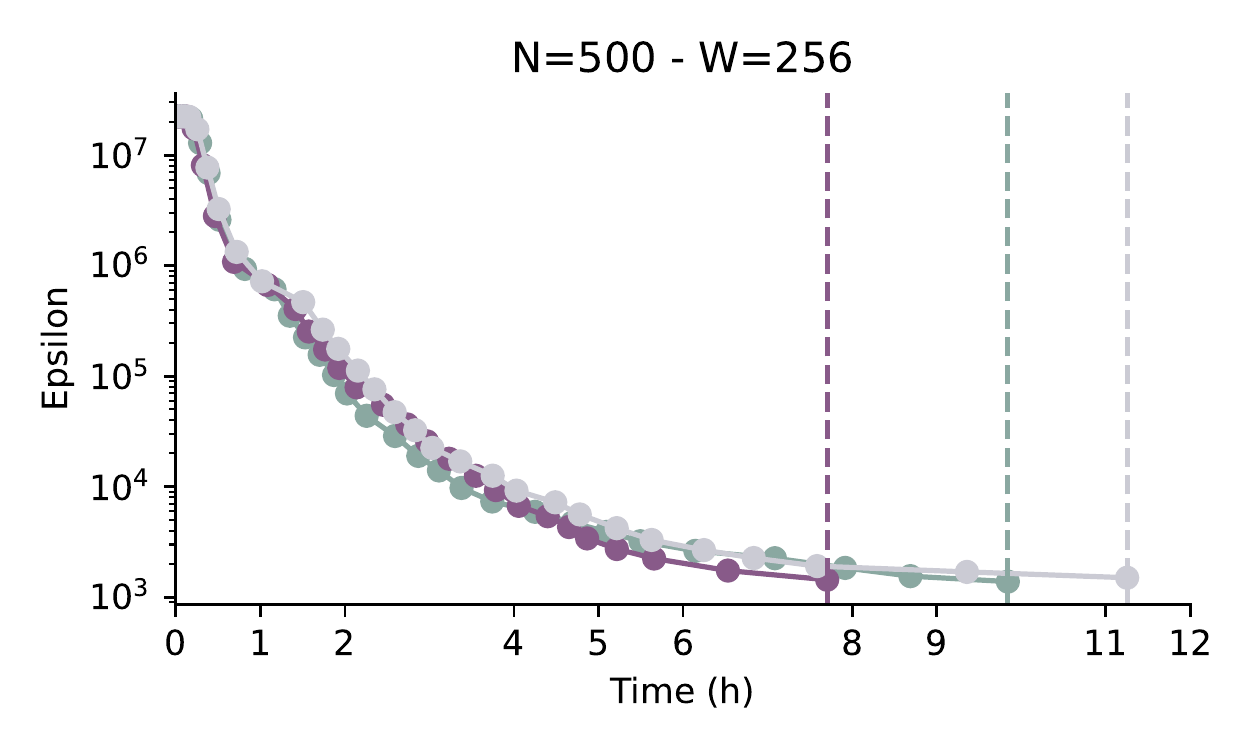}
			\includegraphics[width=\textwidth]{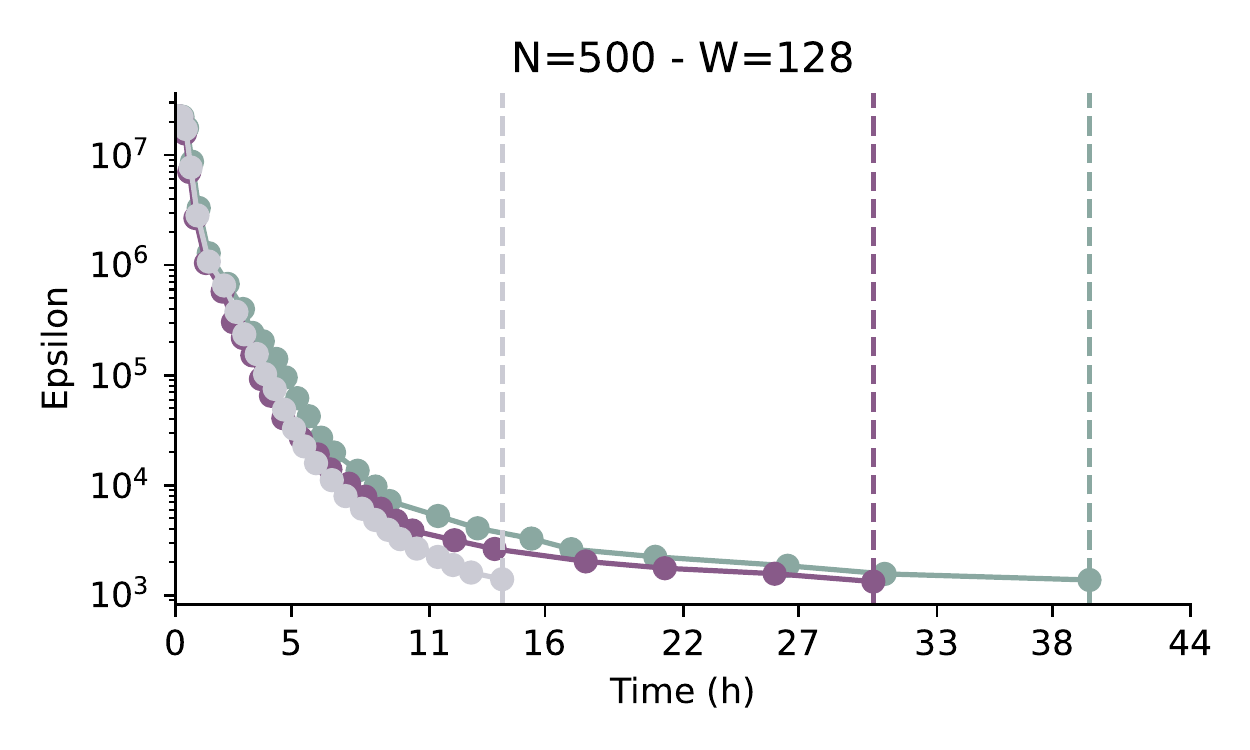}
			\includegraphics[width=\textwidth]{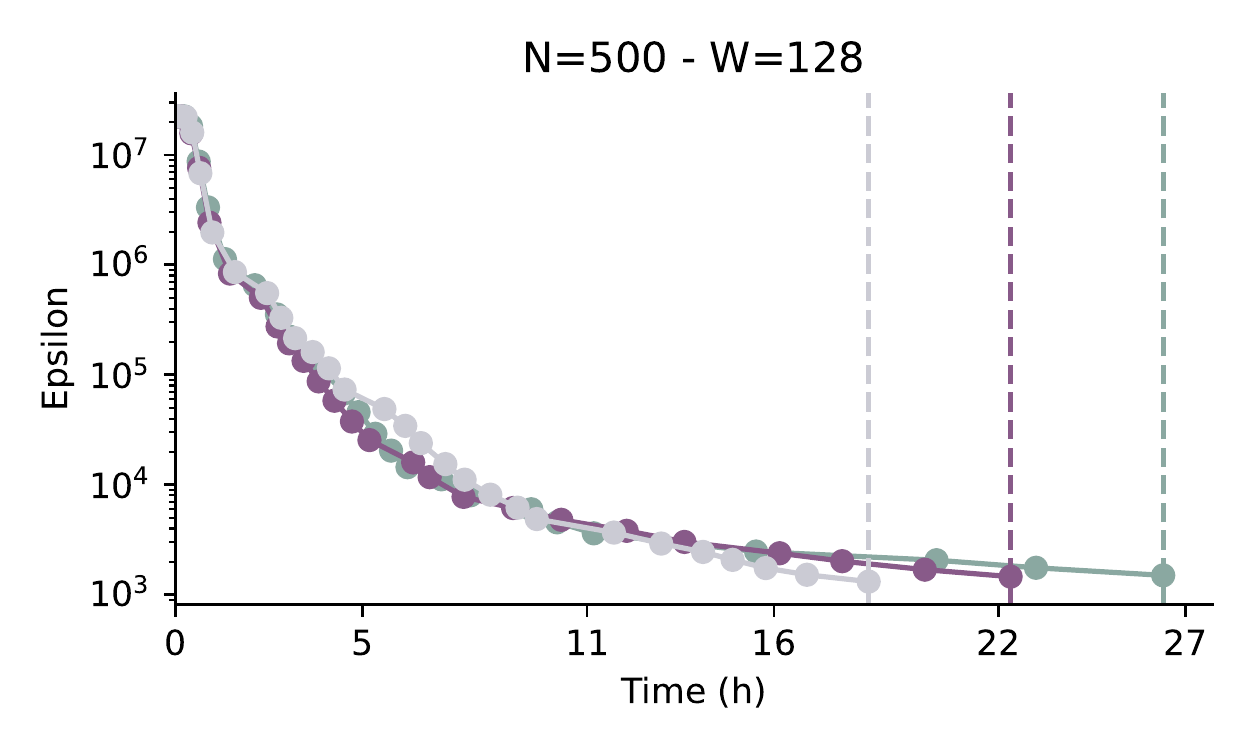}
		\end{minipage}
		\begin{minipage}{0.3\textwidth}
			\includegraphics[width=\textwidth]{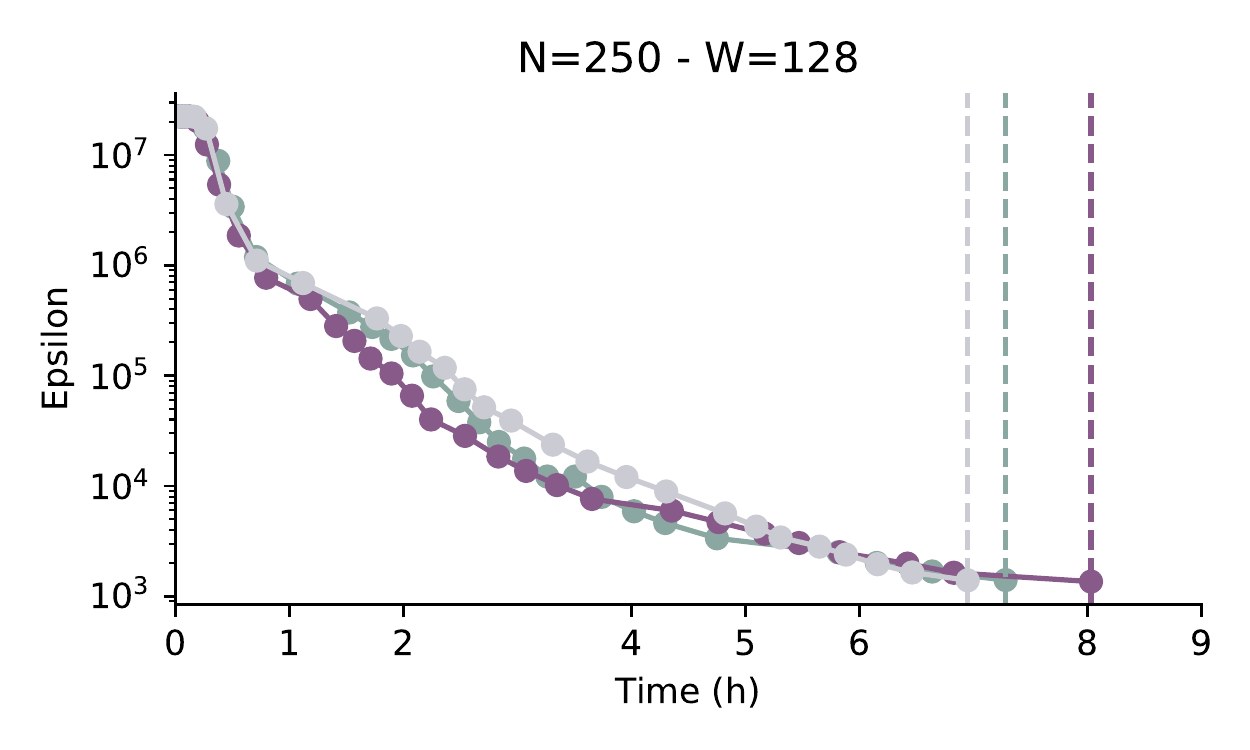}
			\includegraphics[width=\textwidth]{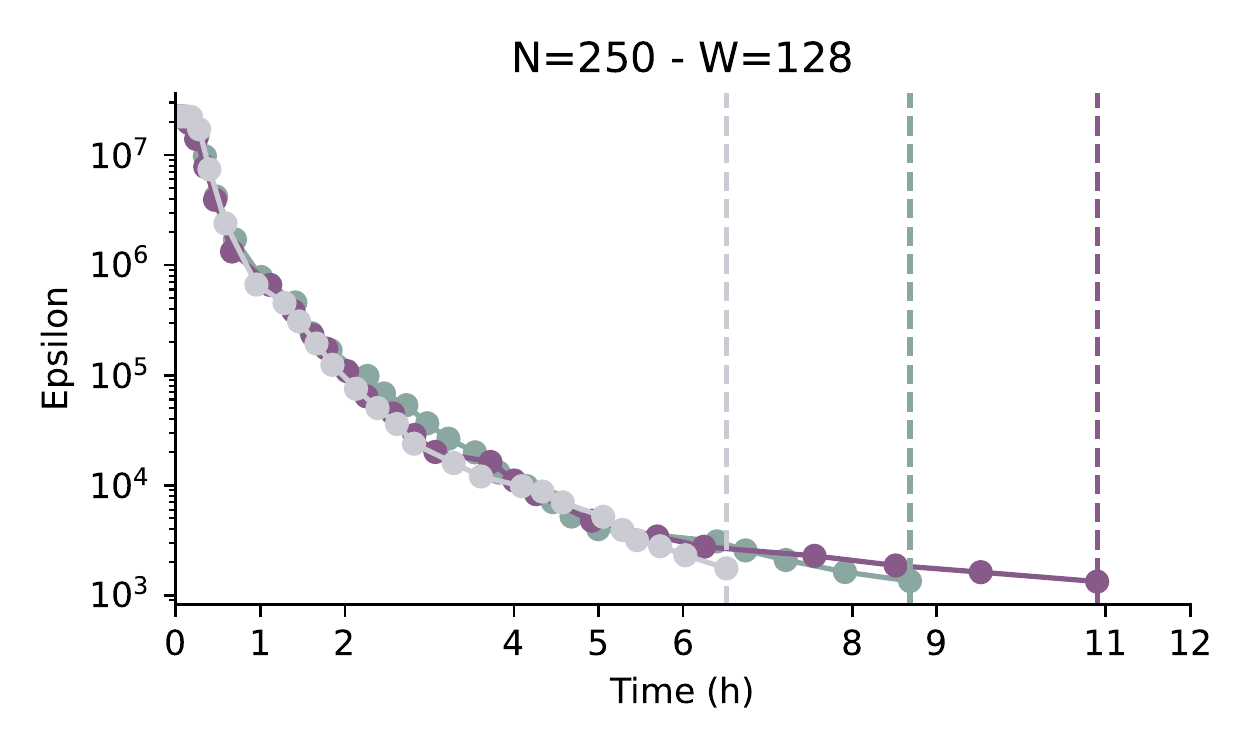}
		\end{minipage}
		\caption{Development of the acceptance threshold over time for the different runs for model M1. These are all the 8 runs we performed.}
		\label{fig:TumorAllEpsOverTime}
	\end{figure}

	In the more extreme cases, it also occurred that the LA scheduling took slightly longer than the corresponding DYN run, but similarly also that the LA run decreased the wall time by a factor of more than 2 when compared to the DYN scheduling based one (see Figure~\ref{fig:TumorAllEpsOverTime}). 
	
	These stronger differences in wall time usually seem to be traceable to one single generation taking vast amounts of time. Those long generations occur in all scheduling variants and exist most likely because the epsilon for that generation was chosen too optimistically.
	
	On average, it seems that acceleration varies based on the population size and the number of workers to be used.
	Over the 8 times we have executed the inference of the tumor model with an adaptive epsilon schedule, we observed a mean acceleration of 21\%, with the median value being 23\% for LA Cur.
	
	\begin{figure}[H]
		\centering
		\includegraphics[width=0.3\textwidth]{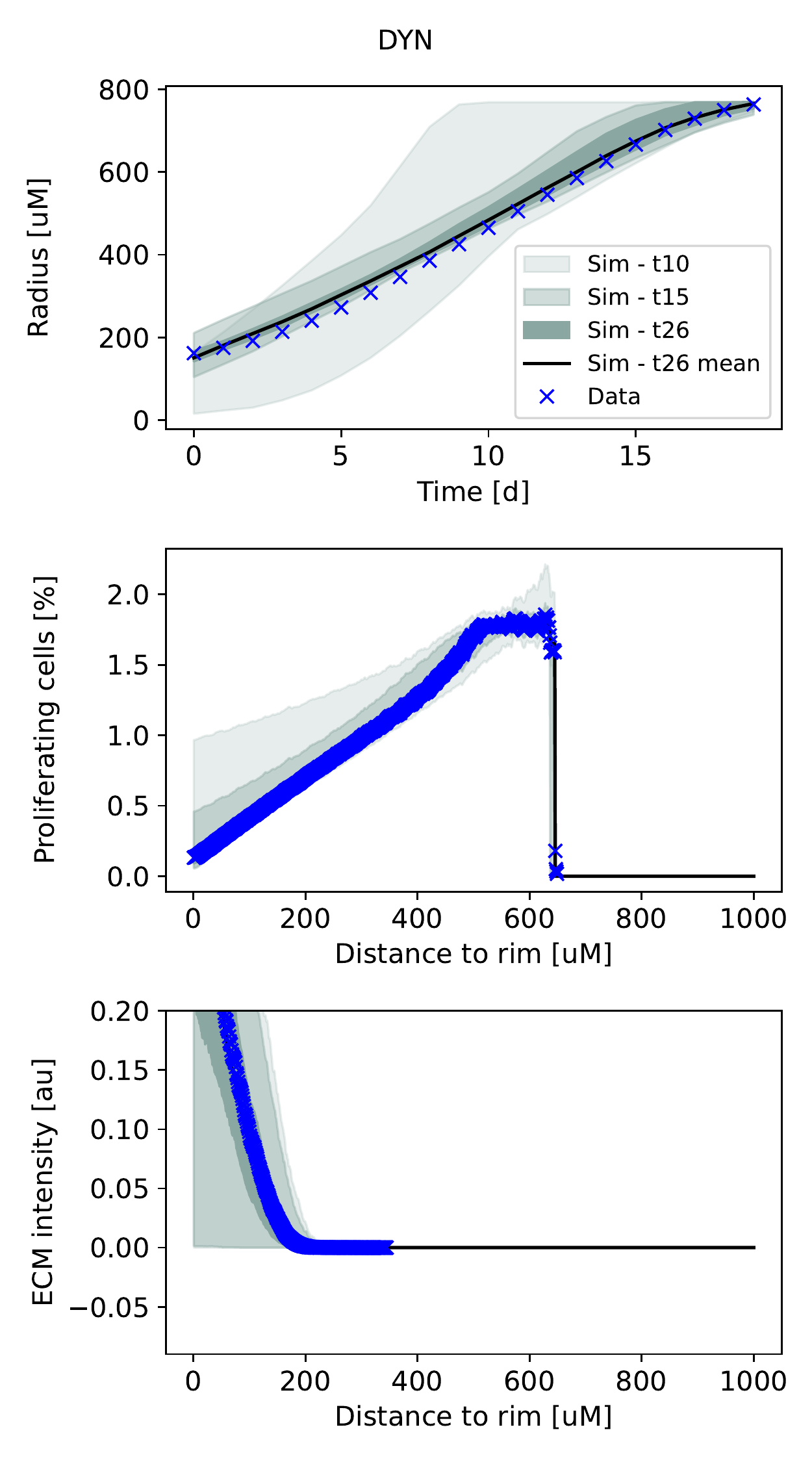}
		\includegraphics[width=0.3\textwidth]{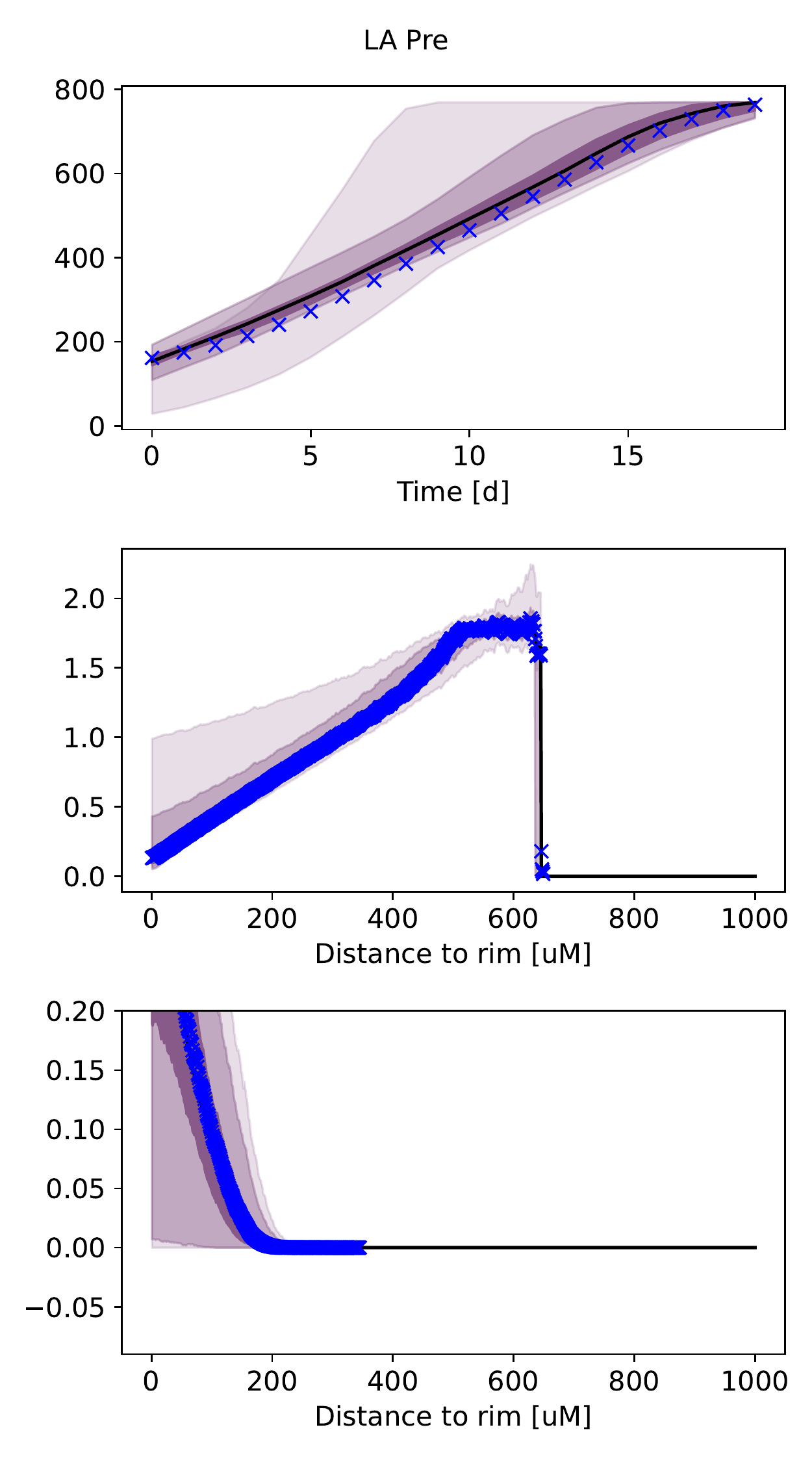}
		\includegraphics[width=0.3\textwidth]{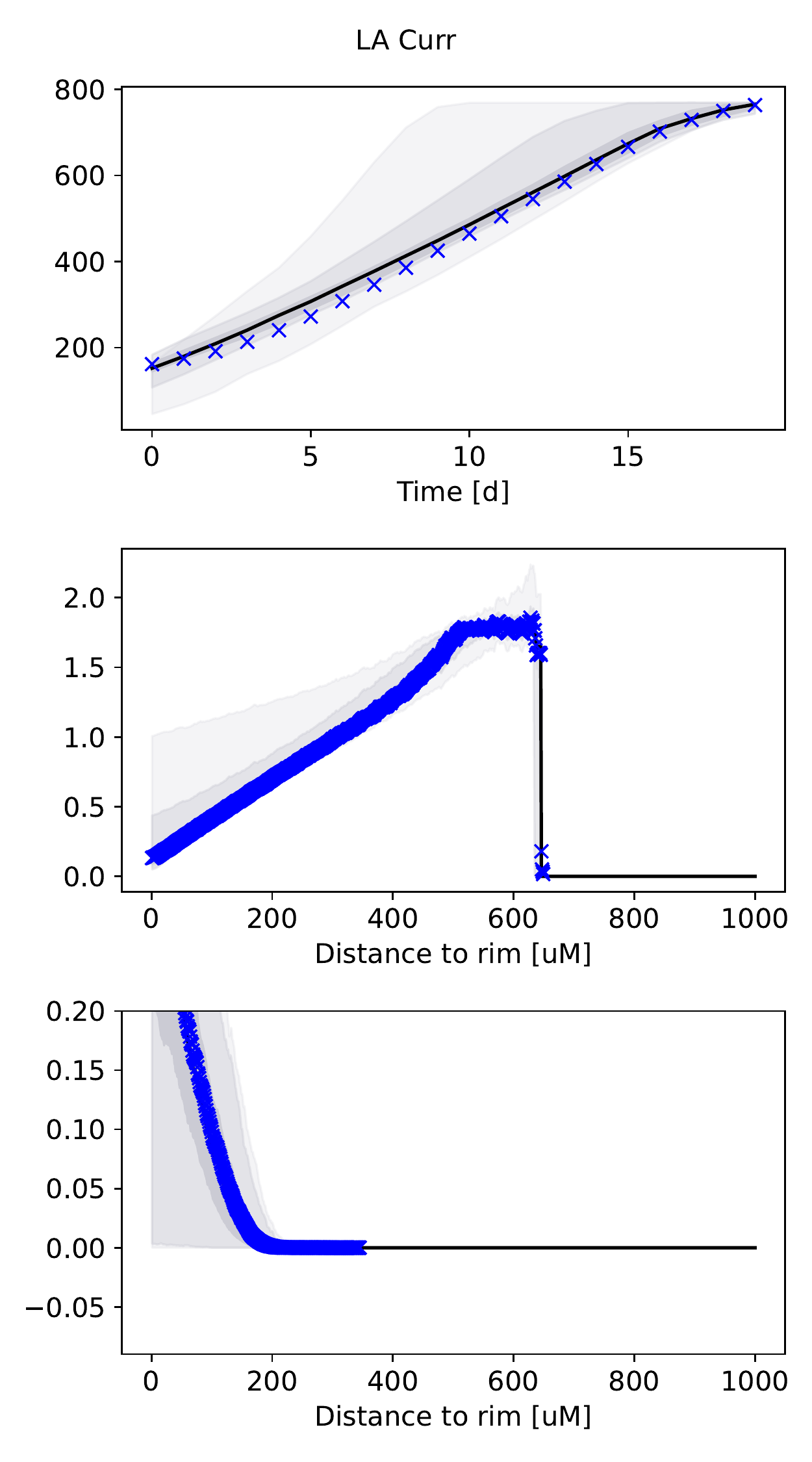}
		\caption{ Comparison of best parameter fit for M1 using DYN (left), LA Pre (middle), and LA Cur (right) for a population size of N= 1000}
		\label{fig:TumorStatEps}
	\end{figure}

	\subsection{(M2) Liver regeneration model}
	
	M2 is a model of YAP regulation by mechanical stimulation through expansion of the bile canaliculi (BC) \cite{MeyerMor2020}. A single realisation of a model varies greatly based on parameter values ranging from 3 to 3,000 seconds. This model has 14 unknown parameters and two observables, namely nuclear YAP and total YAP intensities which were quantified from image tiles covering an entire portal and central vein. The yes-associated protein (YAP) is an activator that activates the Hippo pathway that plays an important role in liver regeneration.
	
	Two sub-models were used to describe the changes of osmotic pressure and the concomitant activation of YAP after PH. The first sub-model is a biophysics-based model to predict the local mechanical stress and apical membrane strain that result from the alteration of osmolyte (bile acid) load in the BC network after partial hepatectomy. It considers the spatial geometry of the BC within the portal and central vein axis of the lobule. Sub-model 2 is a biochemistry-based model that predicts the cellular response of YAP to the local mechanical stress.
	
	\subsubsection*{Results -- Correctness}
	
	Several fitting configurations were done to model (M2) with population sizes ranging from 250 to 1000, with two different amounts of workers, 128 and 256.
	The acceptance criteria for particles were adaptively set to be the 30\% quantile of the previously accepted distances. 
	The run set to be finished after the discrepancy threshold $\varepsilon_t$ falls under some values. The value was set differently based on different configurations ranging from 1.4e+03 to 2.0e+03.
	Similar to model (M1), the three different samplers, DYN, LA Pre, and LA Cur, were executed on similar configurations to ensure a fair comparison.
	
\begin{figure}[H]
		\centering
		\includegraphics[width=1\textwidth]{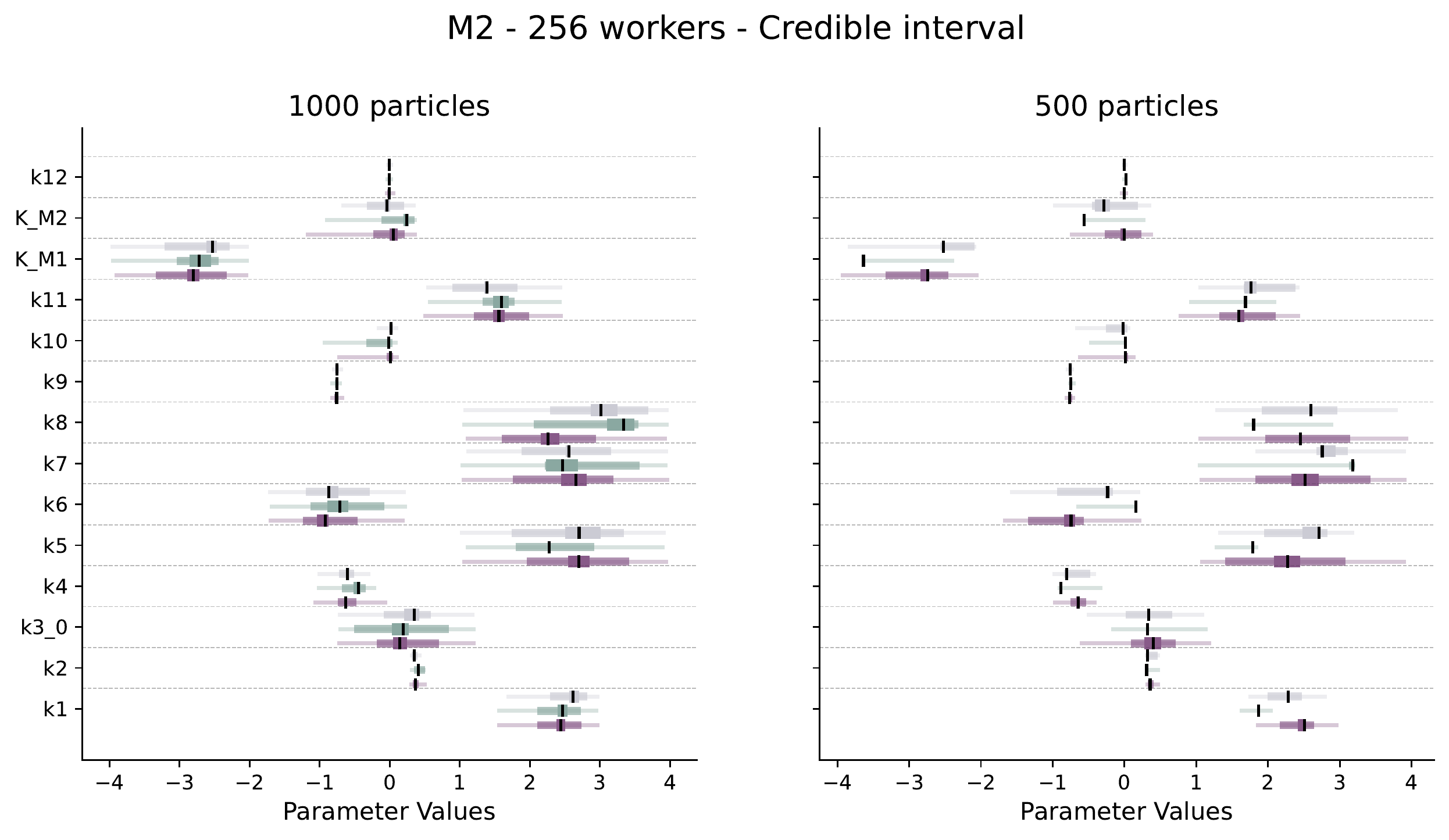}
		\includegraphics[width=1\textwidth]{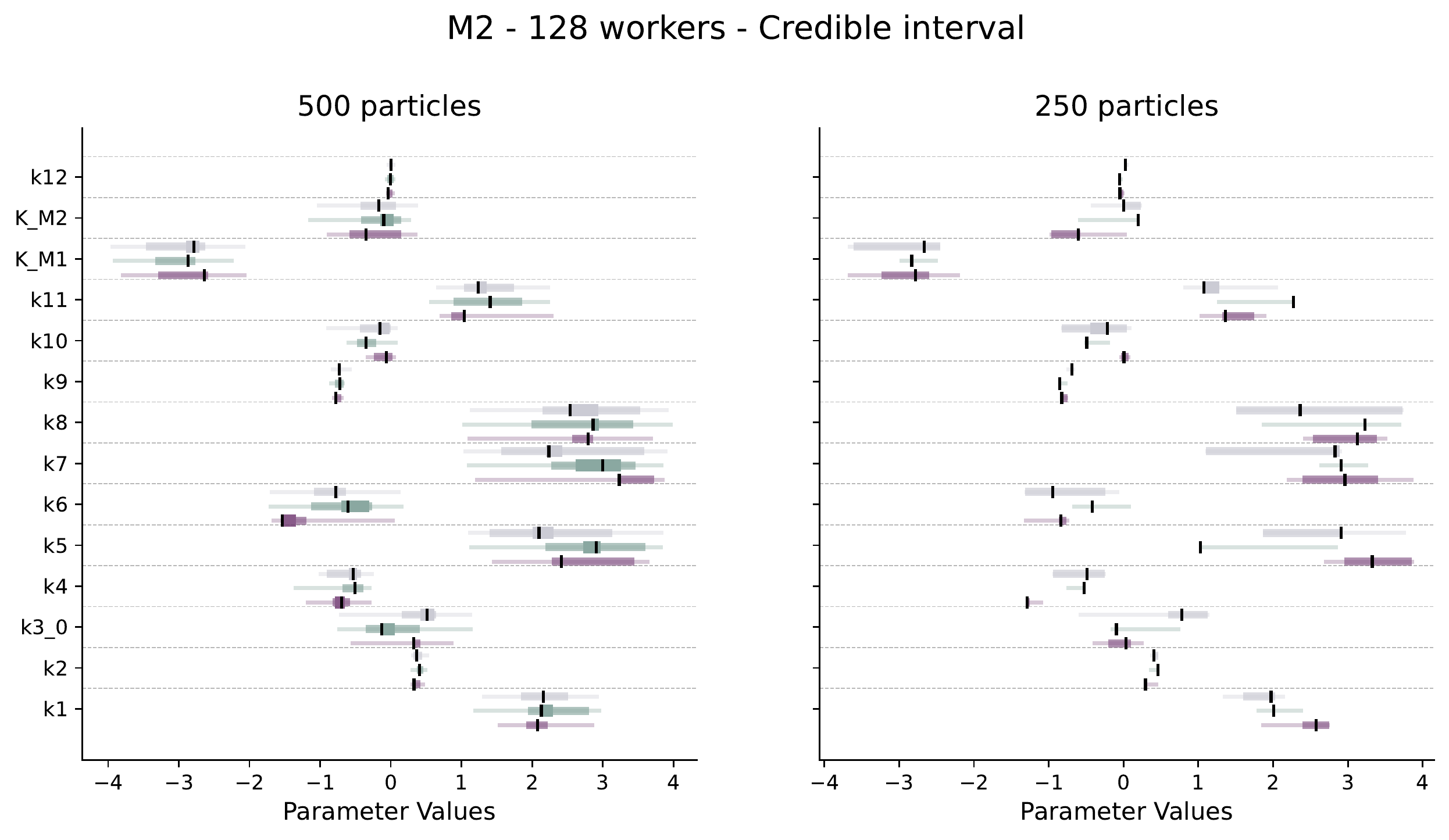}
		
		\caption{The credible interval of the model (M2) with population size 1000, 500, 250 on 128 and 256 workers. A similar maximum simulation time was used here as in Figure \ref{img:M2_time_eps} (900 seconds as maximum simulation time)}
		\label{img:M2_CI_900}
	\end{figure}

\begin{figure}[H]
		\centering
		\includegraphics[width=1\textwidth]{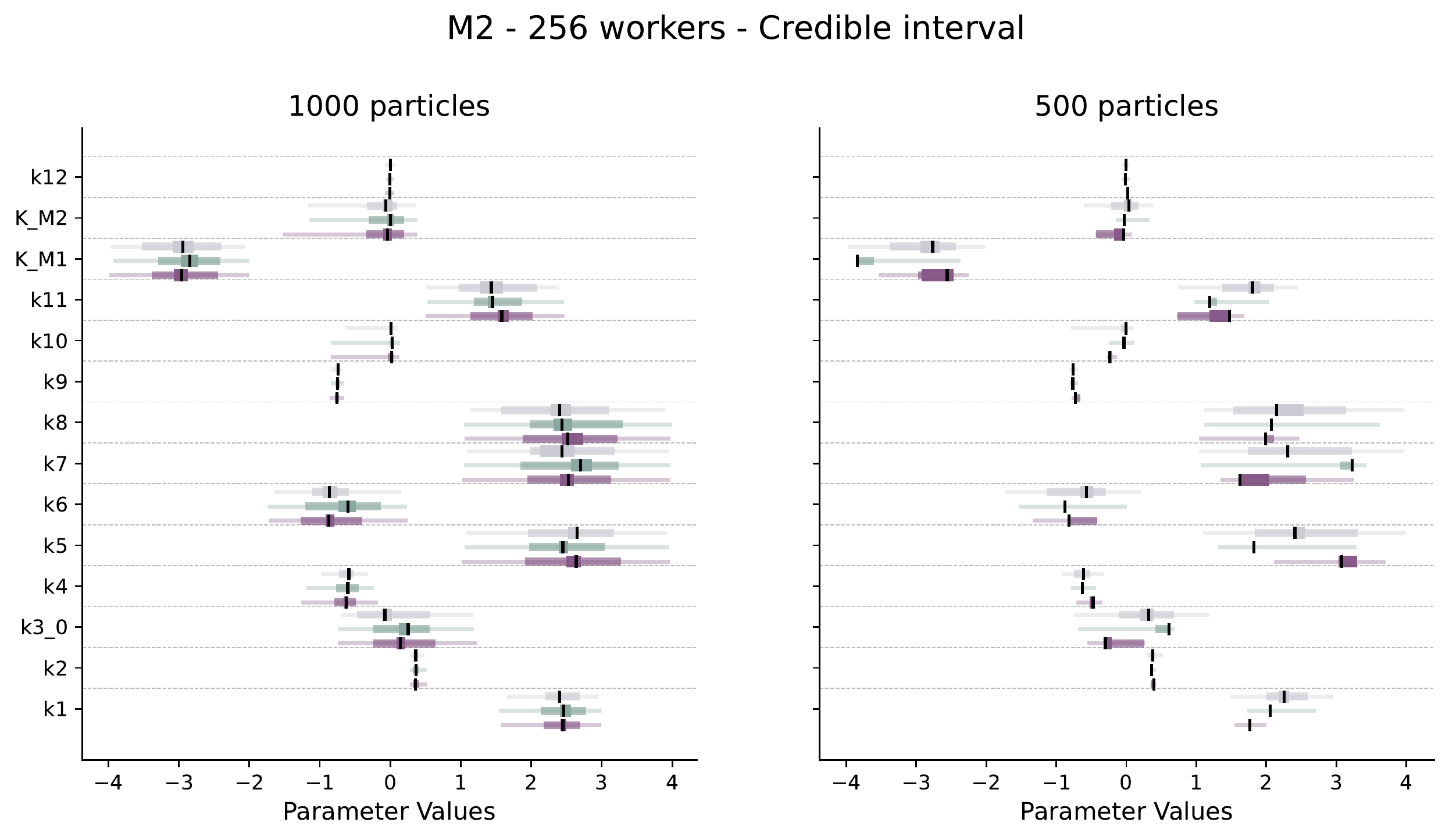}
		\includegraphics[width=1\textwidth]{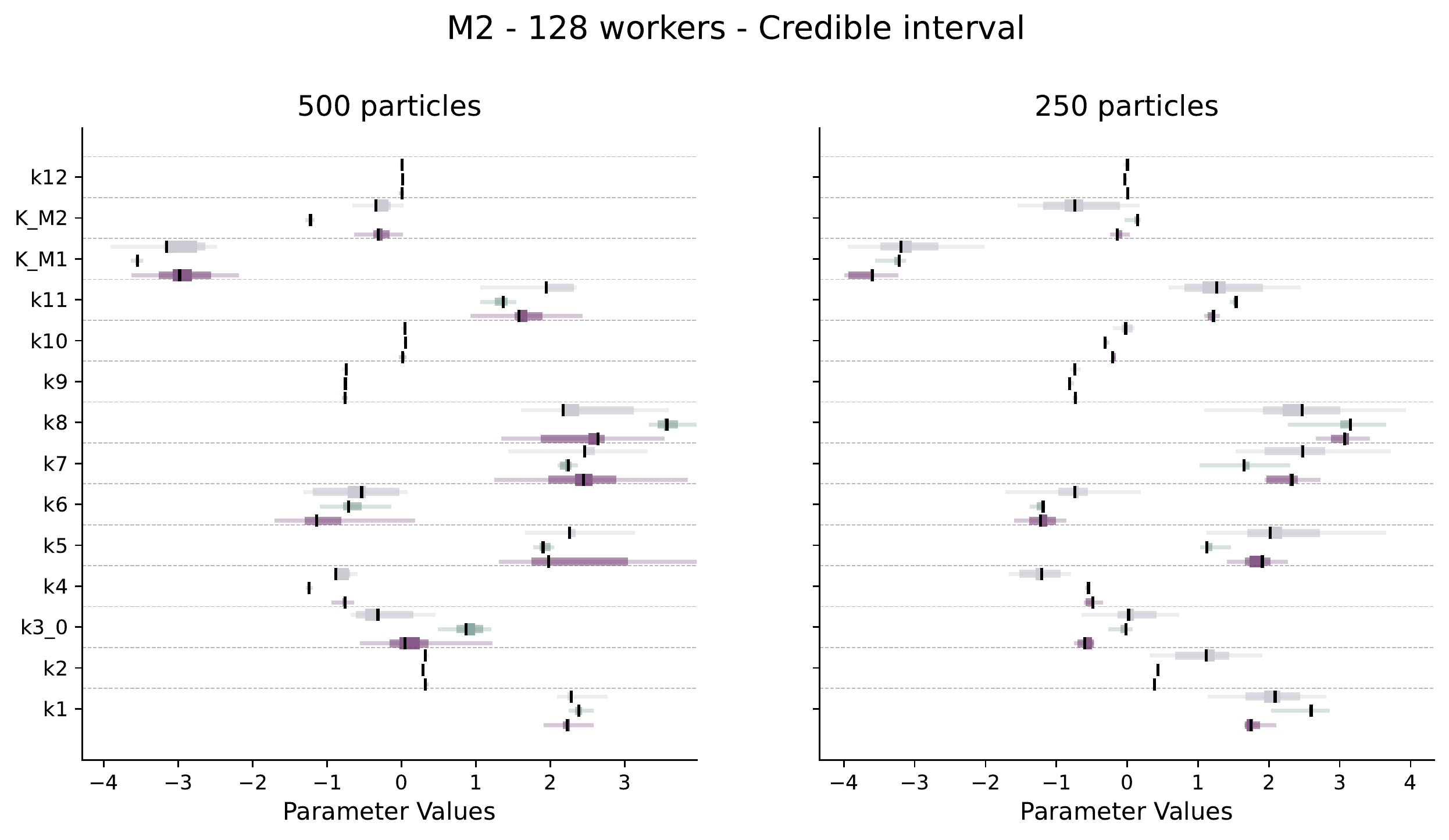}
		
		\caption{The credible interval of the model (M2) with population size 1000, 500, 250 on 128 and 256 workers. A  maximum simulation time was used here as in Figure 6 (1800 seconds as maximum simulation time).}
		\label{img:M2_CI_1800}
	\end{figure}

	Assessing the quality of the result, there were no substantial differences of the results during the run (see Figure~\ref{img:M2_CI_900} and \ref{img:M2_CI_1800} and Figure~\ref{fig:LiverPosteriors}; detailed results of the other runs can be found in the supplementary code).

	\begin{figure}[H]
		\centering
		\includegraphics[width=\textwidth]{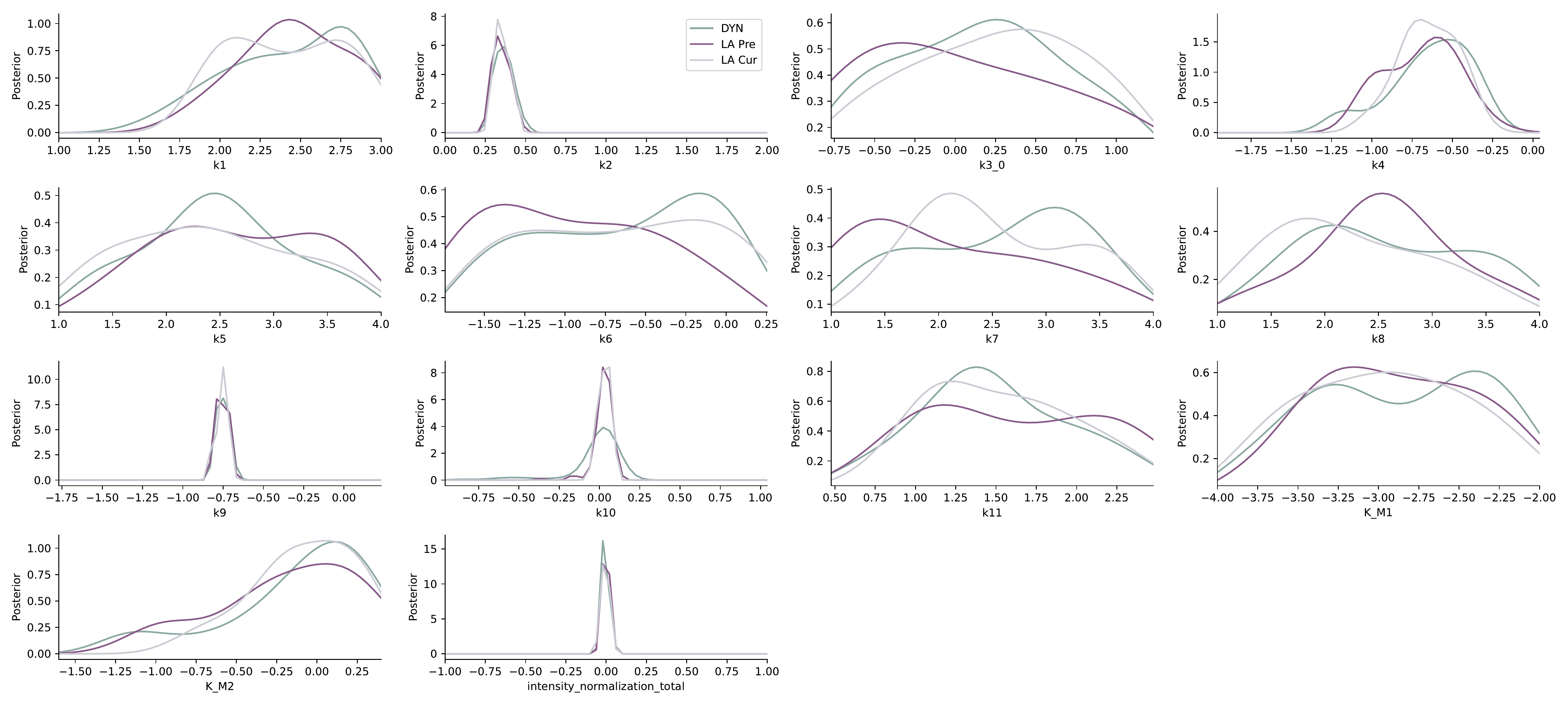}
		\caption{Direct comparison of the posterior distributions for all 14th parameters of (M2) for a run with $N=1000$ on $W=256$ workers using DYN, LA Pre, and LA Cur scheduling. The sharp peaks of the parameters are visibly at the same location, and if the inference yielded a more broader distribution, it does. It did so in all cases..}
		\label{fig:LiverPosteriors}
	\end{figure}
	
\begin{figure}[H]
		\centering
		\includegraphics[width=0.7\textwidth]{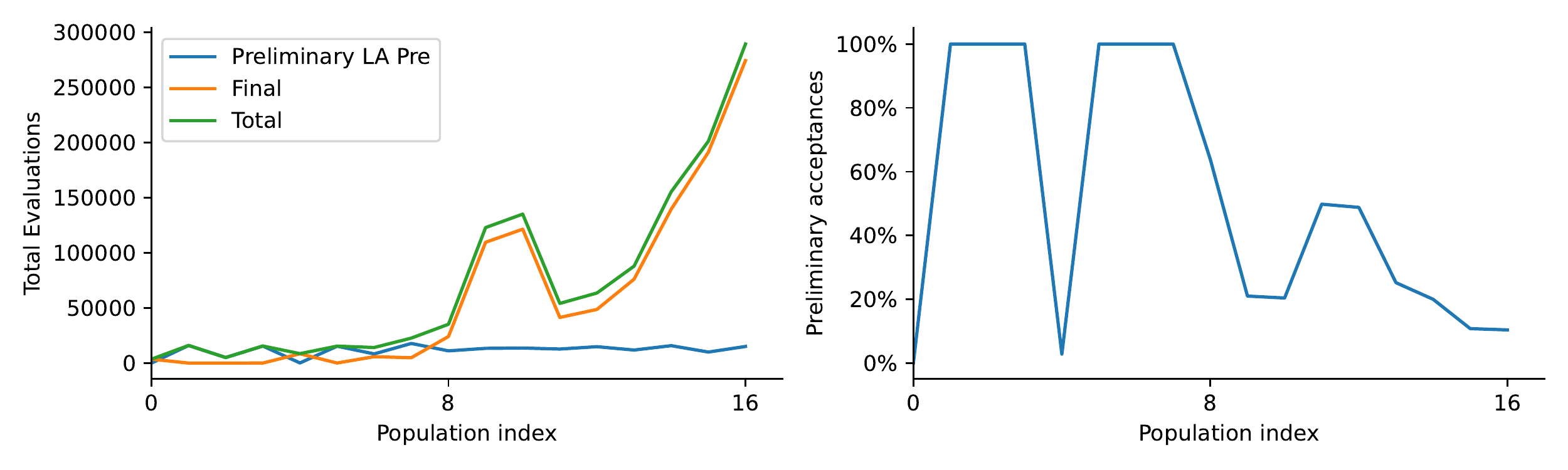}
		\includegraphics[width=0.7\textwidth]{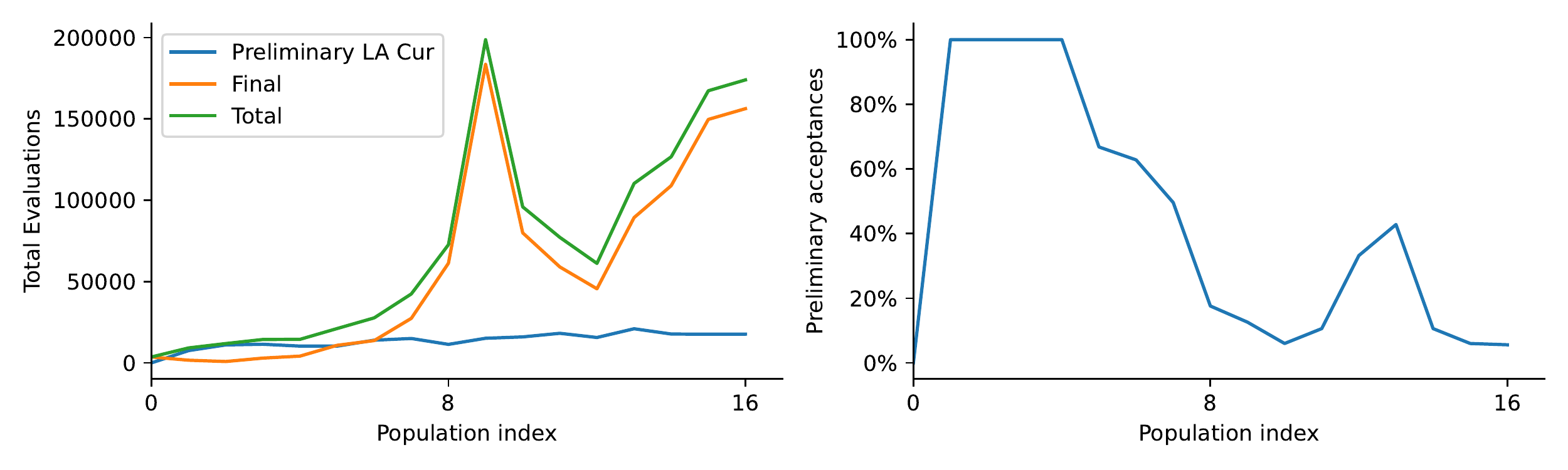}
		
		\caption{Total evaluations and fraction of accepted particles for (M2) based on the preliminary population in each generation in the LA Pre (top) and LA Cur (bottom) runs of the same run as in Figure~\ref{fig:LiverPosteriors}}
		\label{fig:LiverPrelAcc}
	\end{figure}

	Similar to model (M1), the fraction of M2 model preliminary particles decrease over the course of the inference,
	While the fraction of preliminary particles tends to decrease over the course of the inference (see Figure~\ref{fig:LiverPrelAcc}).
	
	The fluctuations of this fraction exist because the acceptance threshold, and thus the acceptance rate, decreases less consistently than when using a static epsilon schedule, as can at least partially be observed in Figure~\ref{fig:LiverAllEpsOverTime}.
	
	As a large fraction of the population is based on the preliminary proposal for a substantial part of the generations, it shows that these preliminaries have a large effect on the LA version of the ABC-SMC algorithm. 
	Nonetheless, the posteriors are consistently similar, indicating that everything works as intended and no bias was introduced.
	
	\begin{figure}[H]
		\centering
		\includegraphics[width=0.5\textwidth]{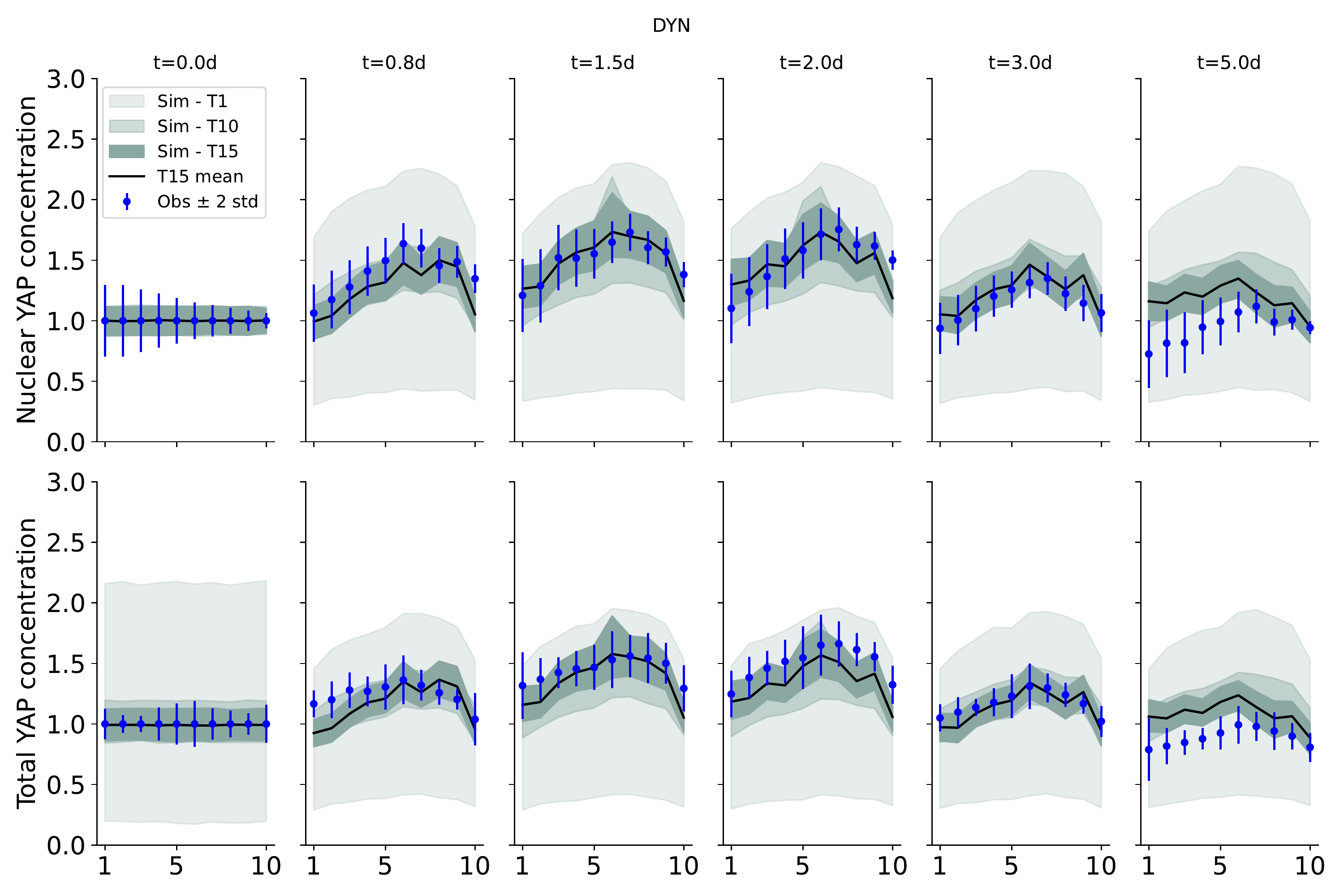}
		\includegraphics[width=0.5\textwidth]{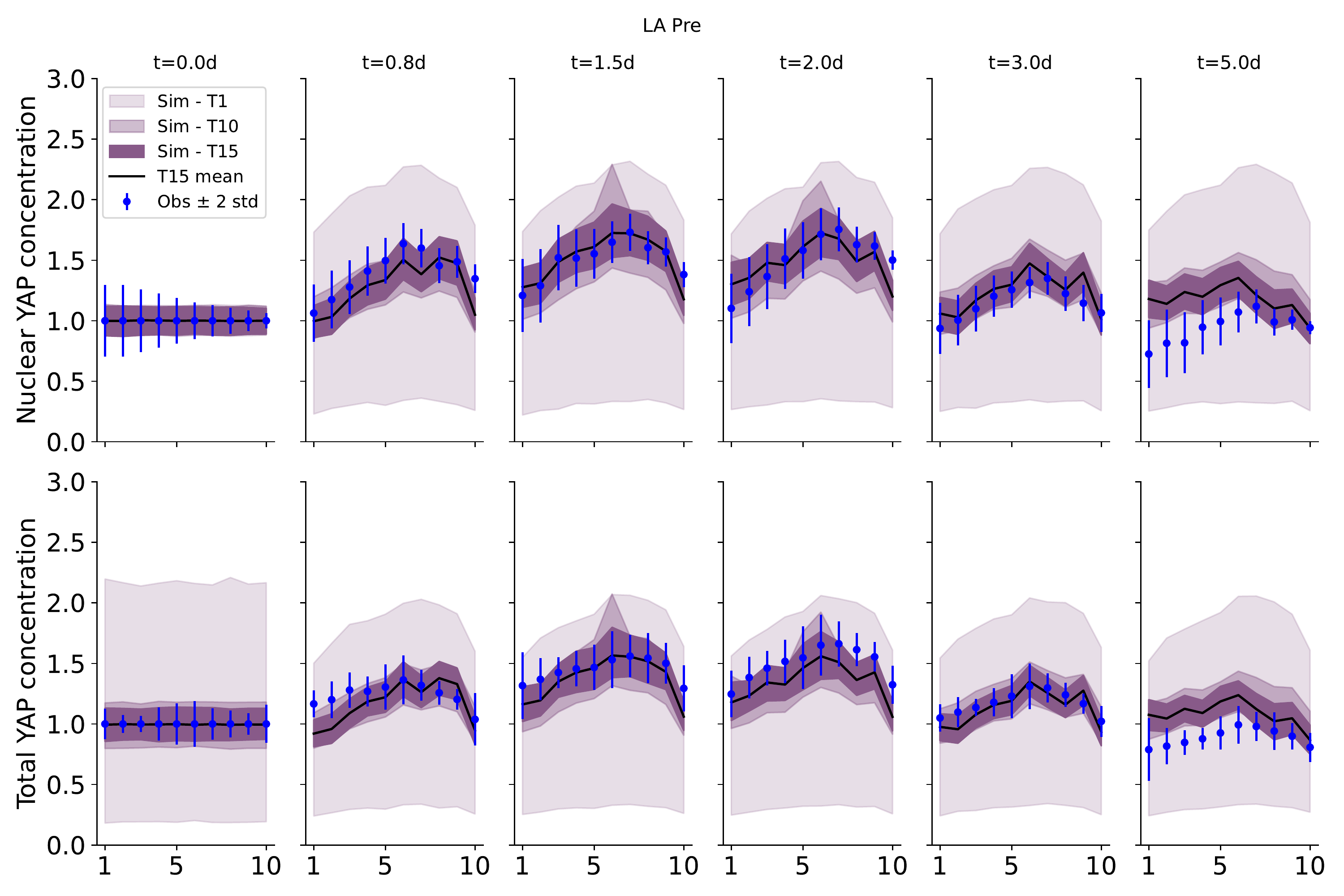}
		\includegraphics[width=0.5\textwidth]{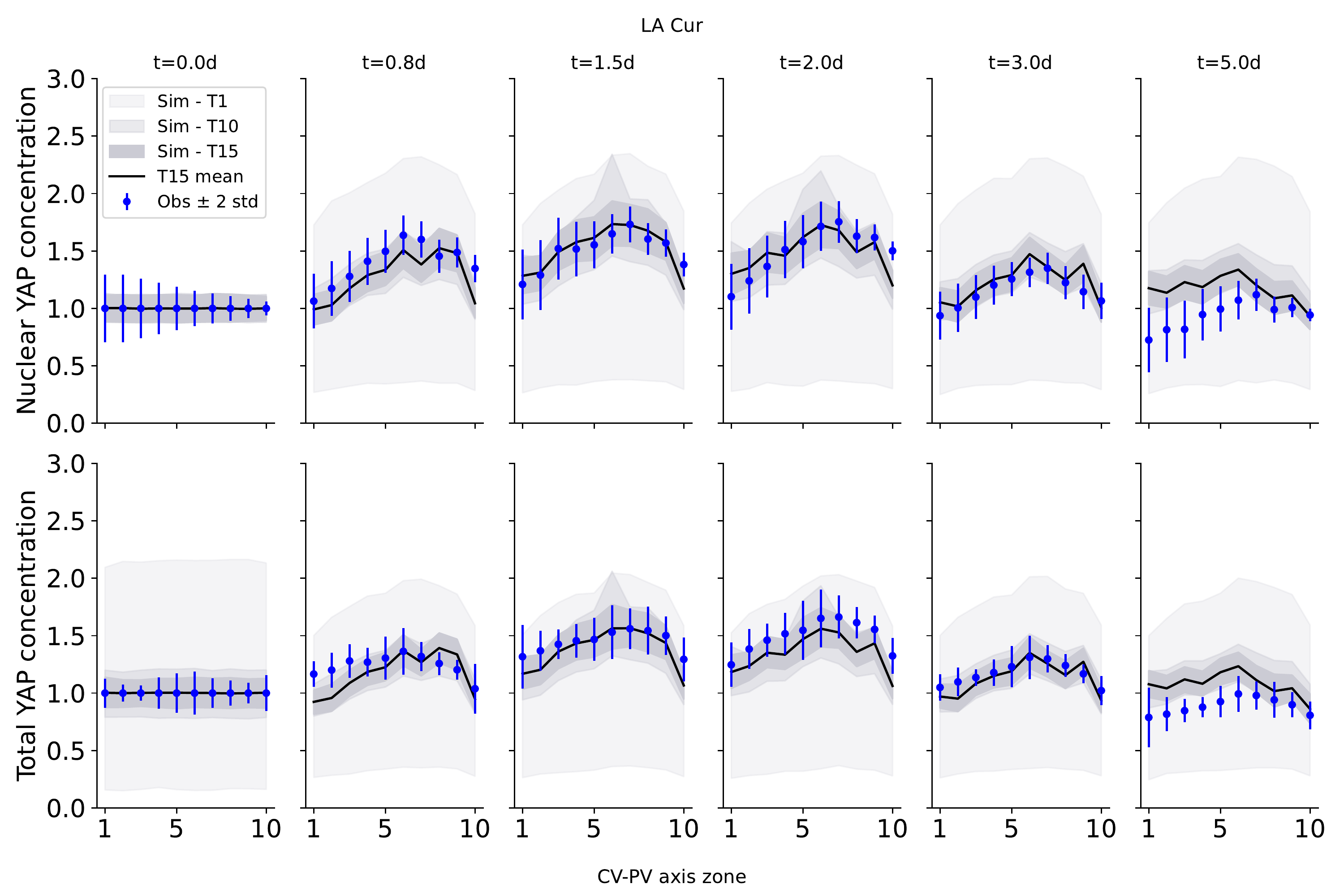}
		\caption{ Comparison of best parameter fit for (M2) using DYN (top), LA Pre (middle), and LA Cur (bottom) for a population size of N= 1000}
		\label{fig:LiverStatEps}
	\end{figure}

	\subsubsection*{Results -- Run-Time}
	\label{sec:LiverRT}
	
\begin{figure}[H]
		\centering
		\includegraphics[width=0.8\textwidth]{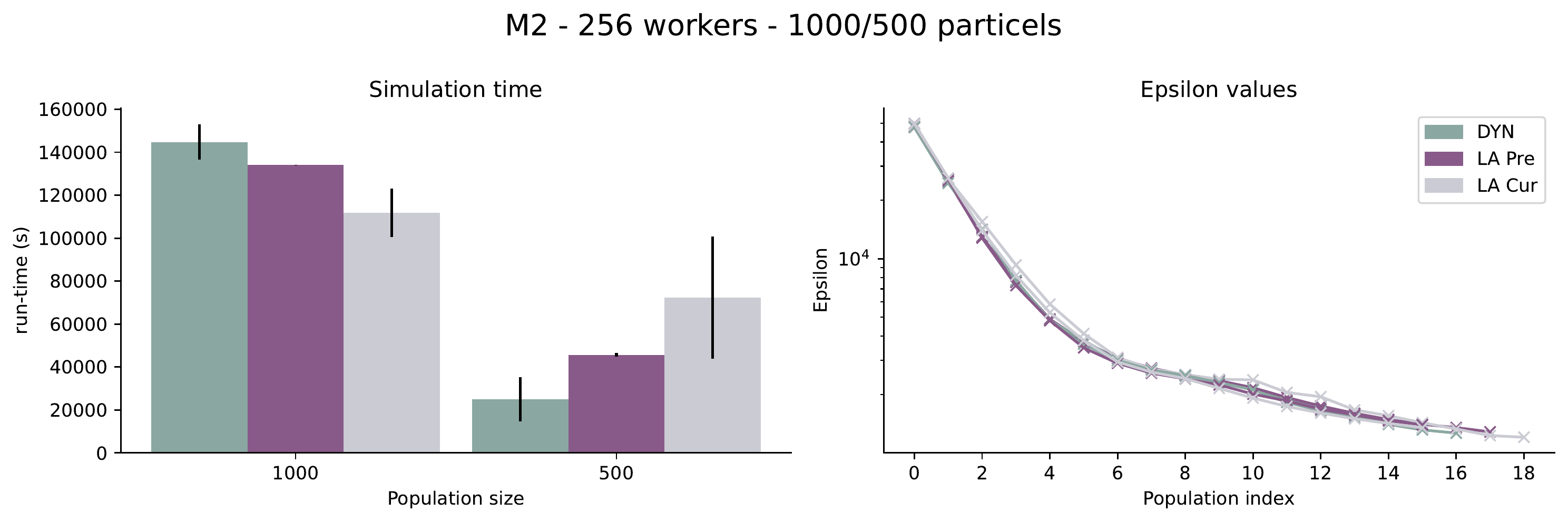}
		\includegraphics[width=0.8\textwidth]{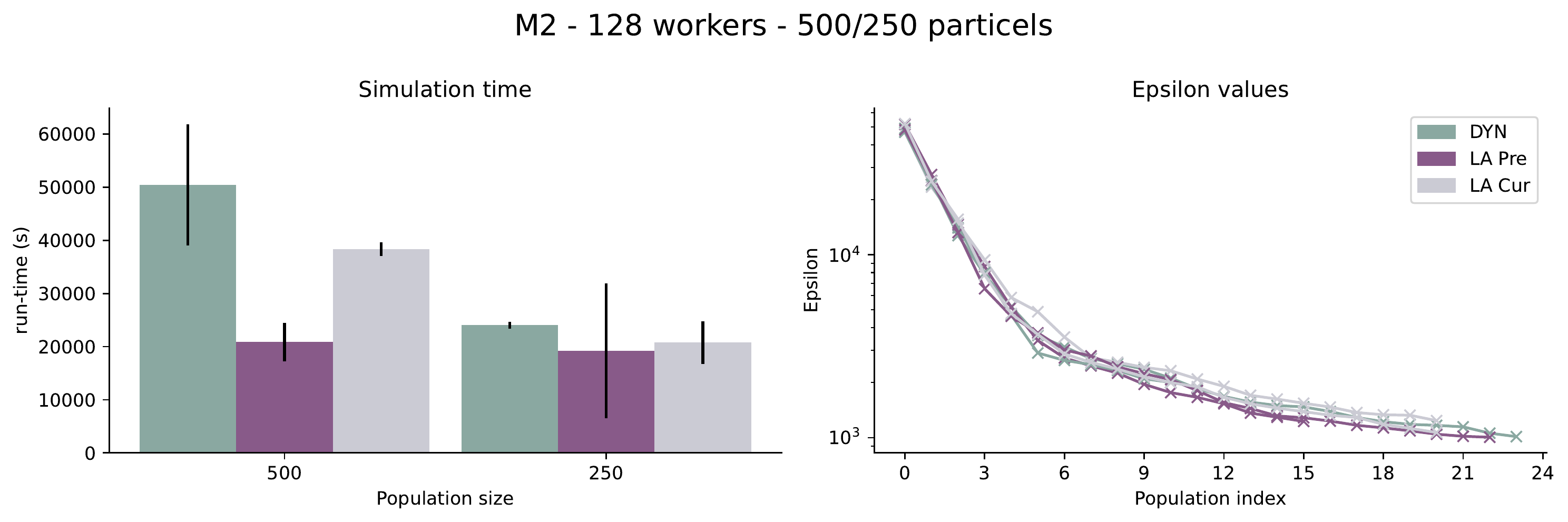}
		
		\caption{The run-time and posterior distributions for 2 different runs of the model M2 with population size 1000, 500, 250 on 128 and 256 workers. In this run, we use a maximum simulation time of 900 seconds, as opposed to 1900 seconds in Figure 5. We can see that decrease in heterogeneity level on simulation time drastically affected the effectiveness of the LA samplers.}
		\label{img:M2_time_eps}
	\end{figure}

	To save computational resources, we employed an early rejection strategy to reject particles based on a maximum run-time for individual simulations not 386 matching the data. However, this will decrease run-time heterogeneity. To assess the affect of that on the speed up of the look-ahead sampler, we fit the model using two different maximum run-time, 15, and 30 minutes. The result of the 30min maximum run-time can be seen in Figure 6. In addition, the 15min maximum run-time is presented in figure \ref{img:M2_time_eps}. Result shows clearly that if heterogeneity increases, one would expect higher speed-up from both LA Pre and LA Cur compared to DYN sampler. 
	
\begin{figure}[H]
		\centering
		\begin{minipage}{0.3\textwidth}
			\includegraphics[width=\textwidth]{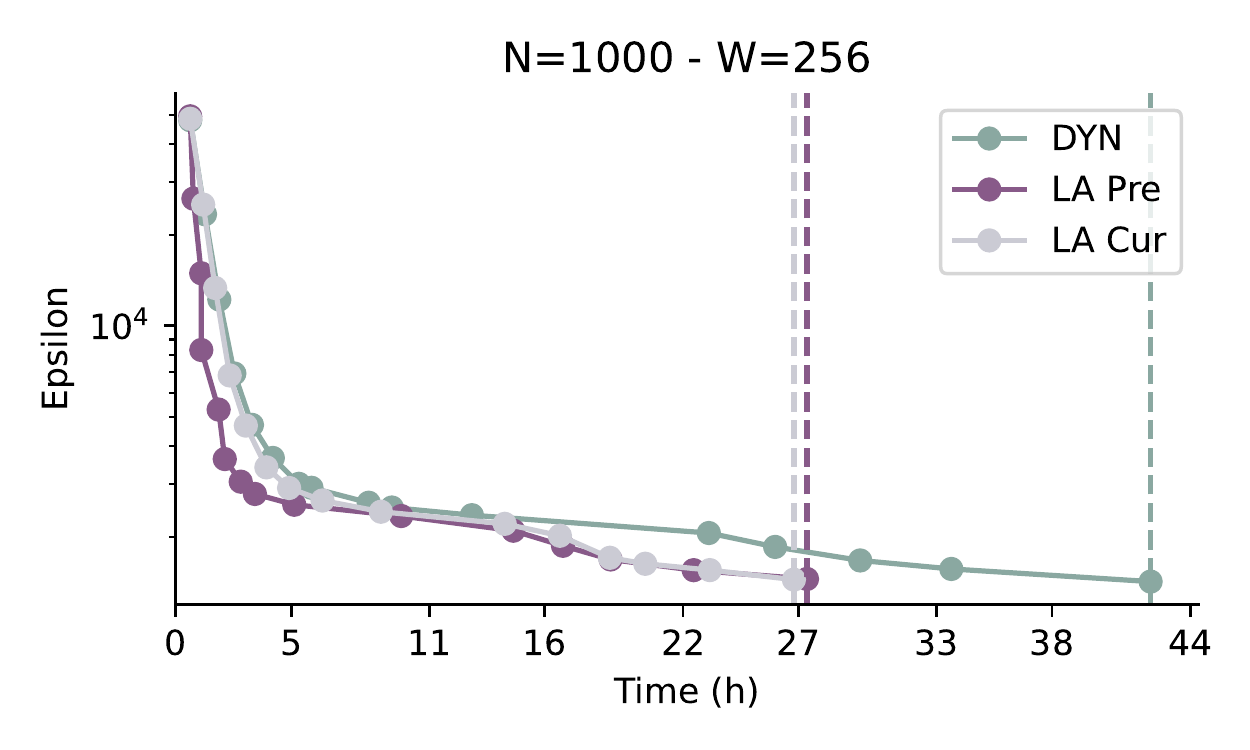}
			\includegraphics[width=\textwidth]{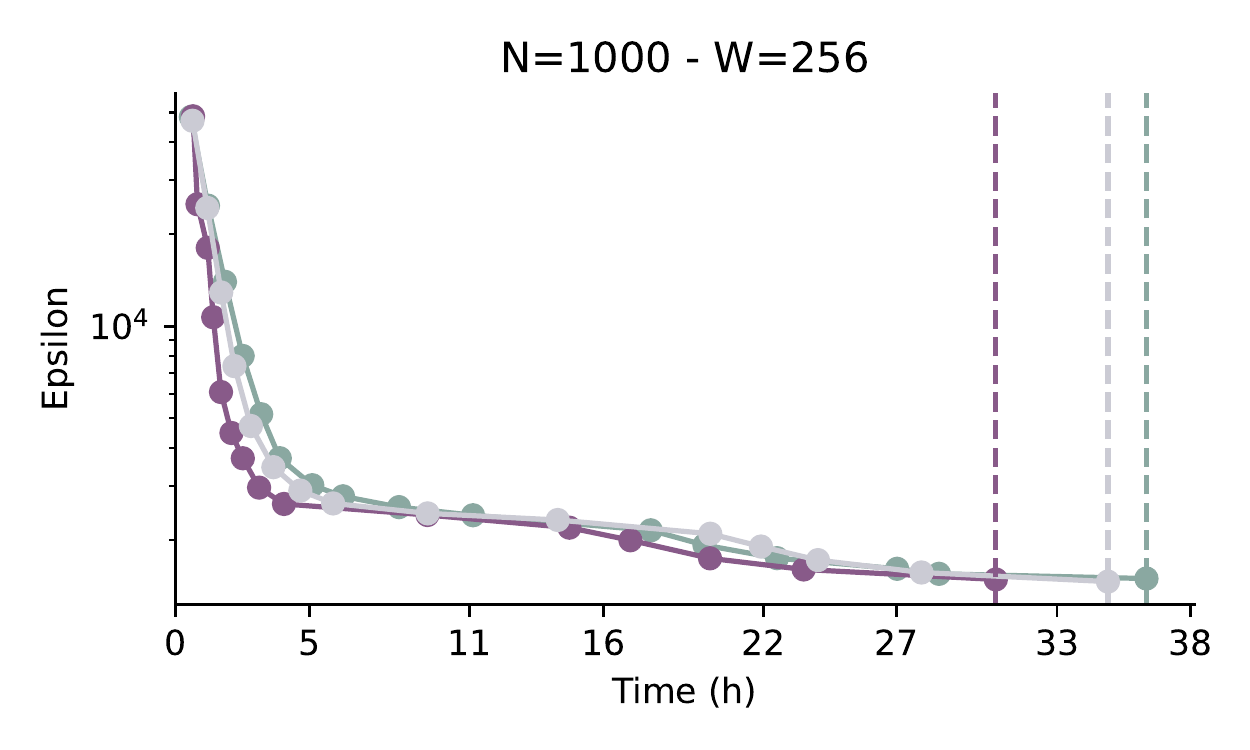}
			\includegraphics[width=\textwidth]{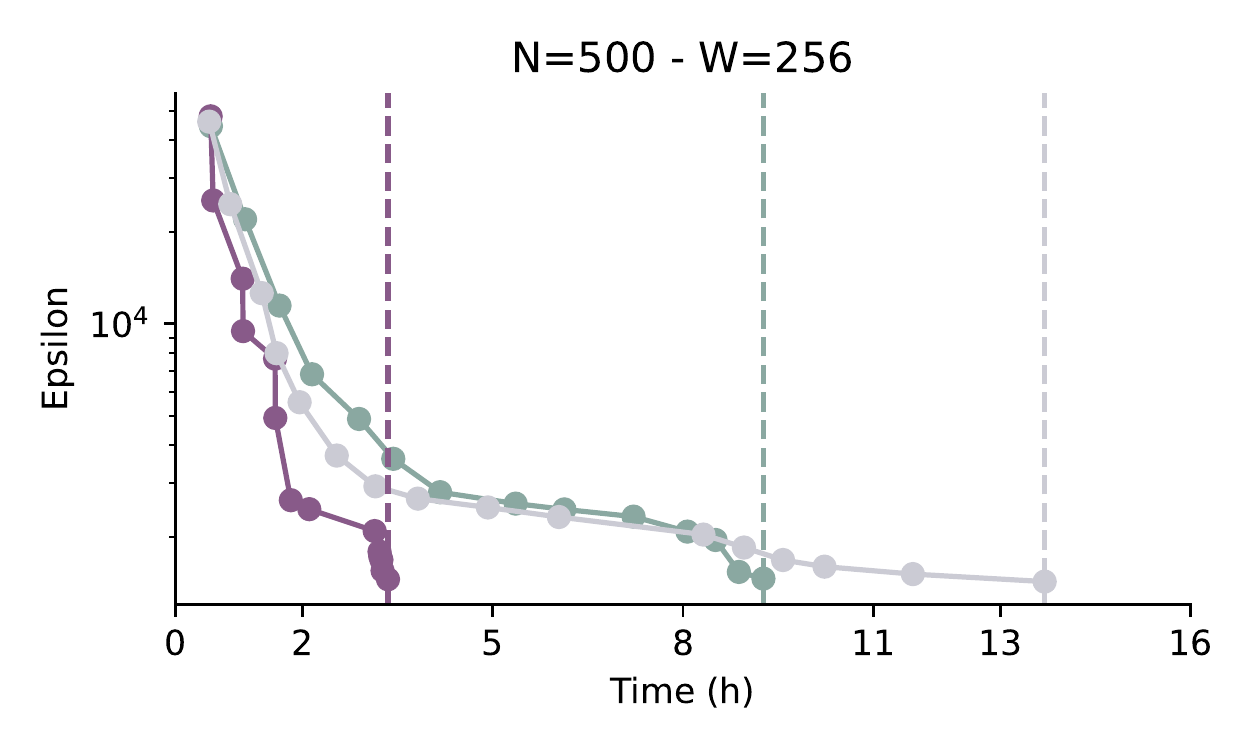}
		\end{minipage}
		\begin{minipage}{0.3\textwidth}
			\includegraphics[width=\textwidth]{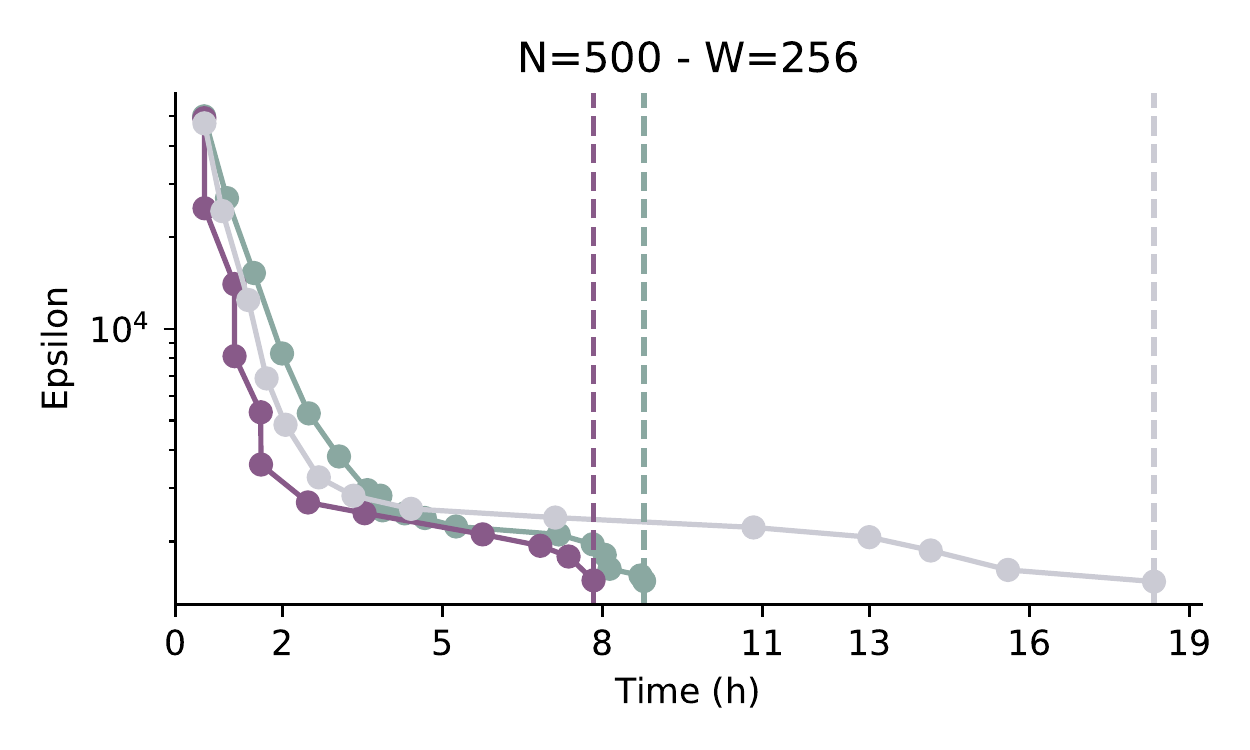}
			\includegraphics[width=\textwidth]{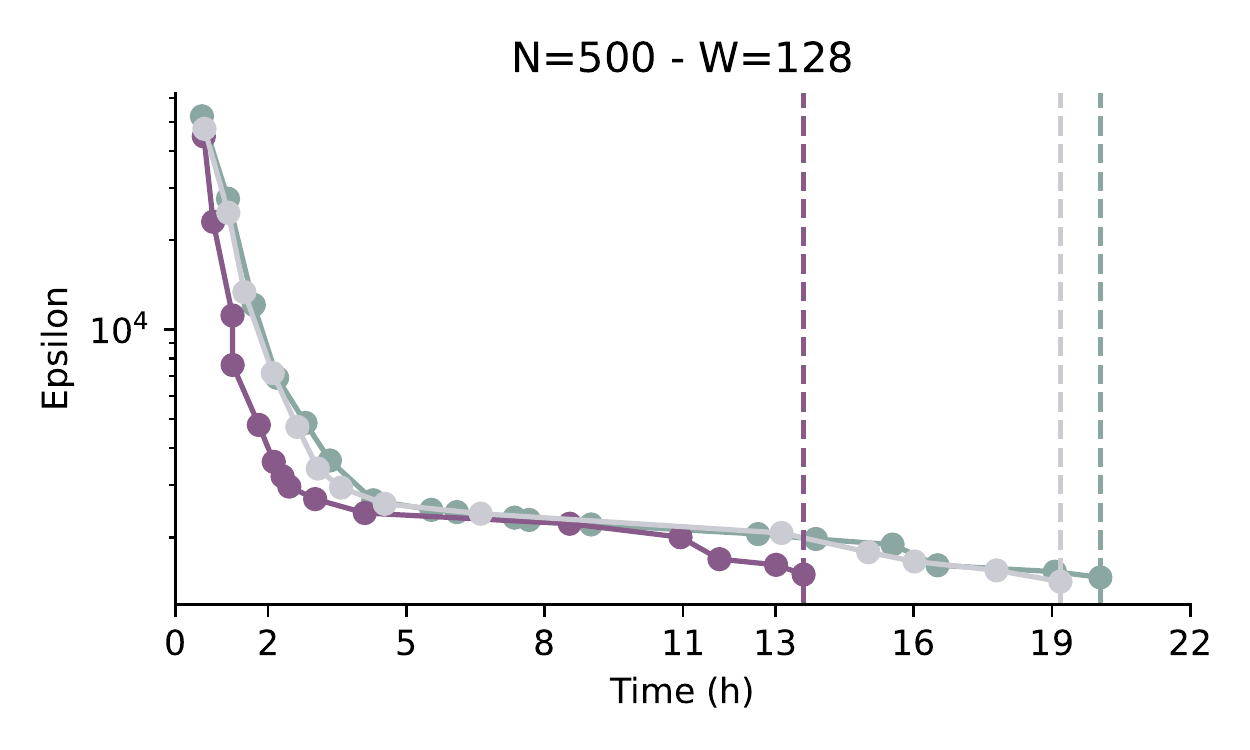}
			\includegraphics[width=\textwidth]{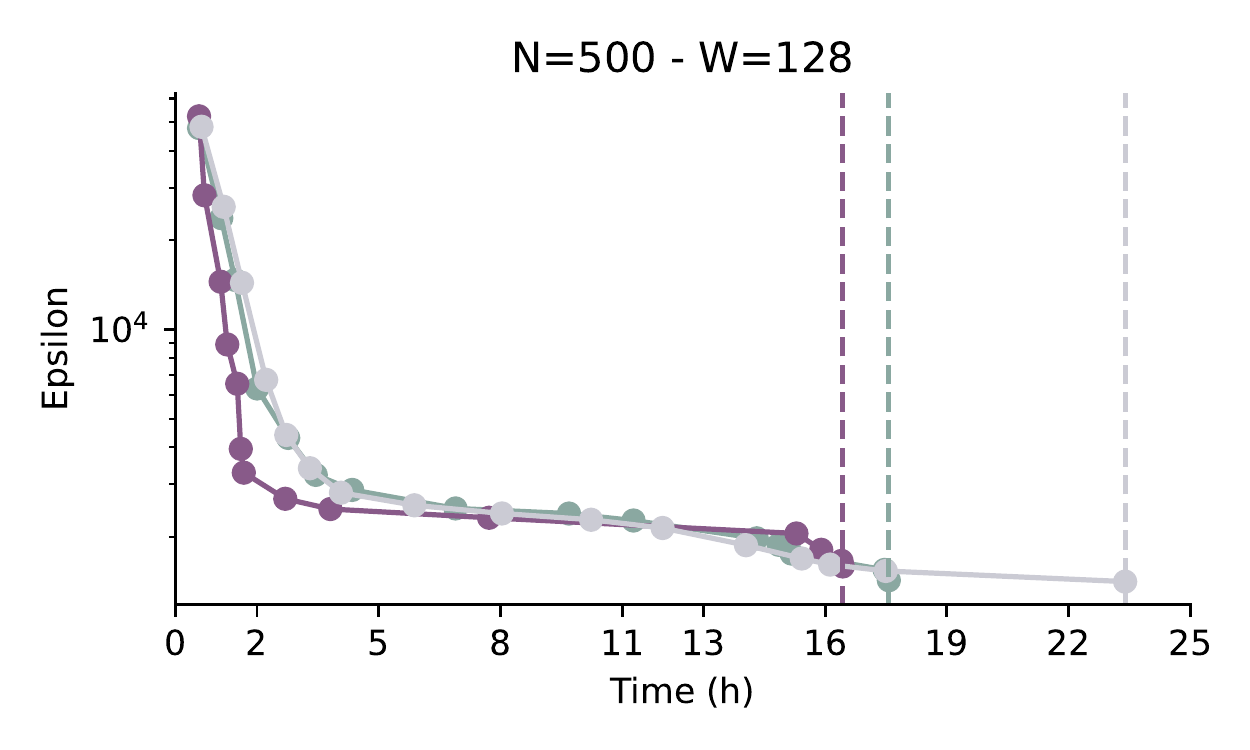}
		\end{minipage}
		\begin{minipage}{0.3\textwidth}
			\includegraphics[width=\textwidth]{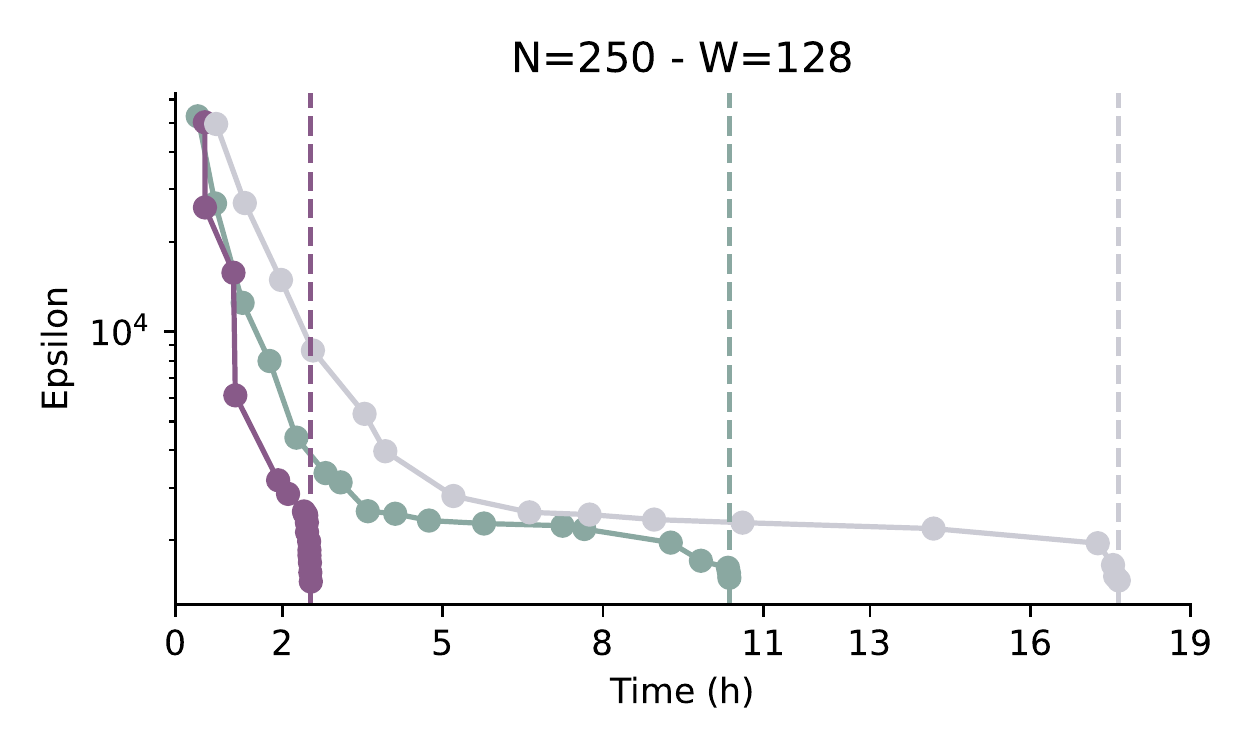}
			\includegraphics[width=\textwidth]{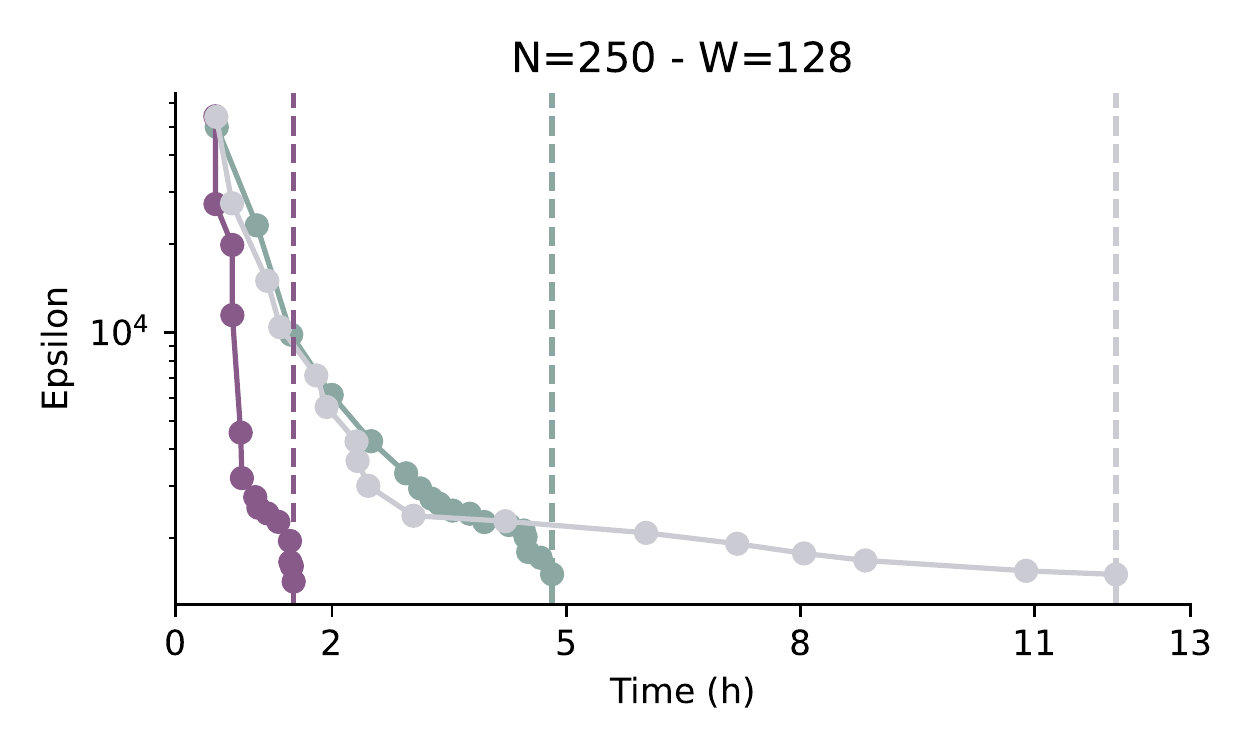}
		\end{minipage}
		\caption{Development of the acceptance threshold over time for the different runs for model M2. These are all the 8 runs we performed.}
		\label{fig:LiverAllEpsOverTime}
	\end{figure}

	Figure~\ref{fig:LiverAllEpsOverTime} shows the development of the epsilon threshold for the 8 different runs. 
	In the more extreme cases, it also occurred that the LA scheduling took slightly longer than the corresponding DYN run, which was observed more in the less heterogeneous runs (e.g. when using 30min as maximum run-time). 
	
	On average, it seems that acceleration varies based on the population size and the number of workers to be used.
	Over the 8 times we have executed the inference of the liver regeneration model with an adaptive epsilon schedule, we observed a mean acceleration of 36\%, with the median value being 31\% for the LA Pre.
	
	\subsubsection{Simulation time variability}
	
\begin{figure}[H]
		\centering
		\includegraphics[width=0.5\textwidth]{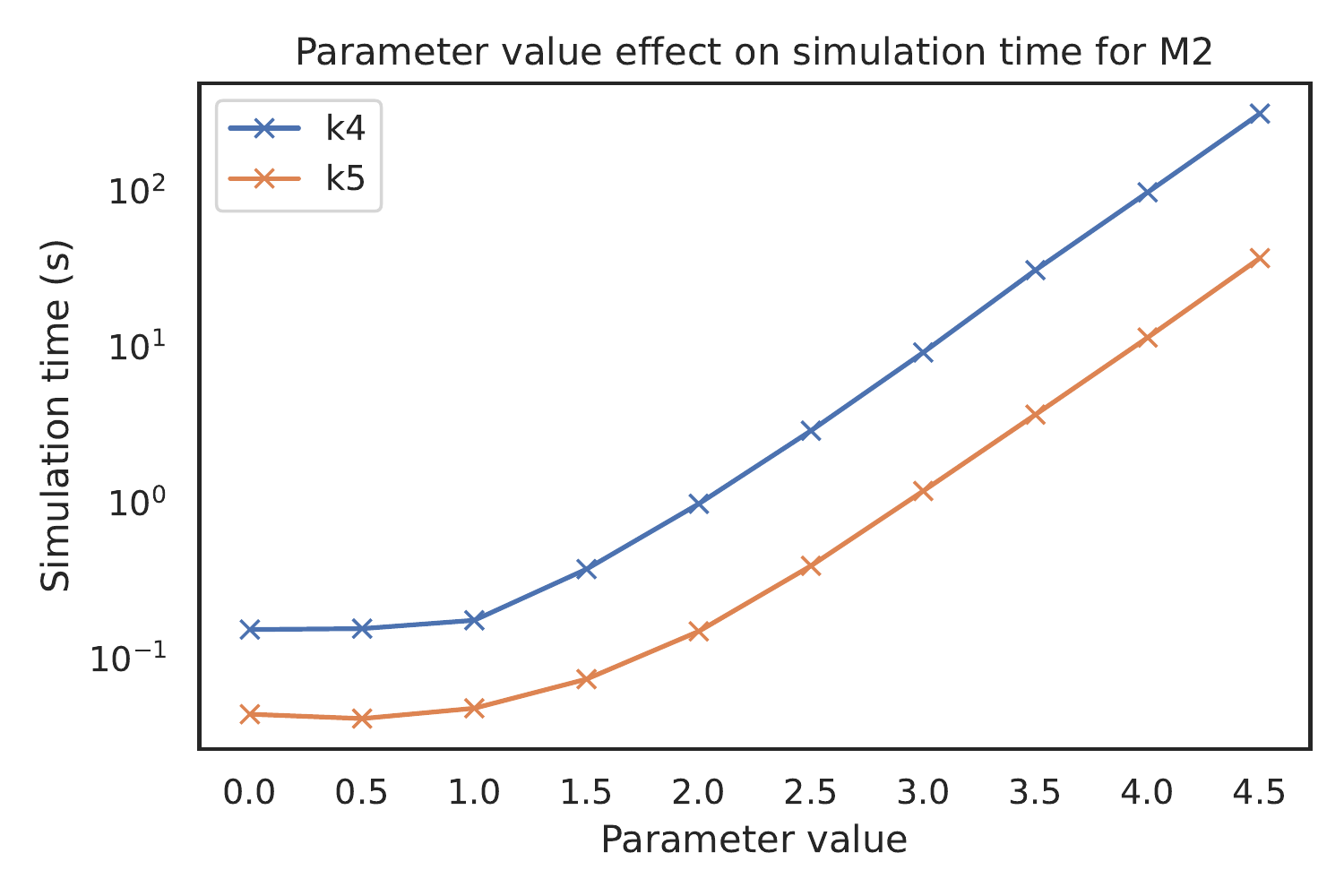}
		\caption{Simulating the liver regeneration model M2 using different values for the parameters k4 and k5 that range from 0 to 4.5 in log10 space}
		\label{fig:sim_vs_par}
	\end{figure}

	The simulation time for a model often relies on the parameter space, as exemplified by model M2. The forward simulation of this model is heavily influenced by the chosen set of parameters. For instance, parameters such as the inactivation rate of YAP (k4) or the activation rate of YAP (k5) both greatly affect the computation time, as demonstrated in Figure \ref{fig:sim_vs_par}.
	
	\bibliographystyle{unsrt}
	\bibliography{Database}